\documentclass{ws-mpla}
\usepackage[super]{cite}
\usepackage{graphicx}
\usepackage{dcolumn}
\usepackage{bm}
\usepackage{epsfig,amsmath}
\usepackage[usenames,dvipsnames]{color}
\usepackage{appendix}
\usepackage{textcomp}
\usepackage{float}
\usepackage{subfigure}
\begin{document}


\catchline{}{}{}{}{}

\title{Mass Independent Area (or Entropy)  and Thermodynamic Volume Products in Conformal Gravity}

\author{Parthapratim Pradhan}

\address{Department of Physics\\ Hiralal Mazumdar Memorial College For Women \\
 Dakshineswar, Kolkata-700035,  India.\\ E-mail: pppradhan77@gmail.com }

\maketitle


\begin{abstract}
In this work we investigate the thermodynamic properties of conformal gravity  in four dimensions. 
We compute the \emph{area(or entropy) functional} relation for this black hole. We consider both 
de-Sitter  (dS) and anti de-Sitter (AdS) cases.  We derive the 
\emph{Cosmic-Censorship-Inequality}  which is an important relation in general relativity that relates 
the total mass of a spacetime to the area of all the black hole horizons. 
Local thermodynamic stability is studied by computing the specific heat. The second order phase transition 
occurs at a certain condition. Various type of second order phase structure has been 
given for various values of  $a$ and the cosmological constant 
$\Lambda$ in the Appendix. When $a=0$, one obtains the result of Schwarzschild-dS and Schwarzschild-AdS cases.
In the limit $aM<<1$, one obtains the result of Grumiller space-time. Where $a$ is non-trivial 
Rindler parameter or Rindler acceleration and $M$ is the mass parameter. The 
\emph{thermodynamic volume functional} relation is derived in the \emph{extended phase space}, where
the cosmological constant treated as a thermodynamic pressure and its conjugate variable as a thermodynamic volume. 
The \emph{mass-independent} area (or entropy) functional relation and thermodynamic volume functional relation that we have 
derived could turn out to be a \emph{universal} quantity.

\keywords{ CG Black Hole, Area product, Entropy product, Cosmic-Censorship-Inequality}
\end{abstract}


\section{Introduction}
It has already been well examined for Kerr-Newman (KN) black hole (BH) possesses event horizon (EH) or outer horizon (OH) and 
Cauchy horizon (CH) or inner horizon (IH) that the mass-independent area product formula should be \cite{ah09}.
\begin{eqnarray}
{\cal A}_{1} {\cal A}_{2} &=& (8\pi)^2\left(J^2+\frac{Q^4}{4}\right) ~.\label{prKN}
\end{eqnarray}
where ${\cal A}_{1}$ and ${\cal A}_{2}$ are area of the EH and CH. $Q$ and $J$ are the charge and 
angular momentum of the BH respectively.

In the limit $Q=0$, one obtains the mass-independent area product formula for Kerr BH. In this two cases, a simple 
area product is sufficient as this can be seen from the exact solution. However, it has been also shown that when 
a BH is perturbed by surrounded matter the same formula does hold \cite{ah09}. Therefore, we can argue that the area product 
formula for Kerr BH and KN BH is a \emph{universal} quantity because it holds independently of the environment 
of the BH. Now if we incorporate the cosmological constant: for example RN-AdS BH \cite{mv13}, the  mass-independent area 
functional relation becomes 
\begin{eqnarray}
f({\cal A}_{1},{\cal A}_{2}) &=&  4\pi Q^2 ~.\label{v2}
\end{eqnarray}
where 
\begin{eqnarray}
f({\cal A}_{1},{\cal A}_{2}) &=& \left[1+\frac{|\Lambda|}{12\pi}({\cal A}_{1}+\sqrt{{\cal A}_{1}{\cal A}_{2}}+{\cal A}_{2})\right]
\sqrt{{\cal A}_{1}{\cal A}_{2}}  ~.\label{v2.1}
\end{eqnarray}
It indicates that it is in fact some complicated function of EH and IHs but it is 
not a simple area product as we have seen in case of Kerr BH or KN BH.  Similarly, it has been extended 
for KN-AdS BH \cite{jh14}:
\begin{eqnarray}
f({\cal A}_{1},{\cal A}_{2}) &=& (8\pi)^2\left(J^2+\frac{Q^4}{4}\right) ~.\label{v3}
\end{eqnarray}
where the function is defined as 
\begin{eqnarray}
f({\cal A}_{1},{\cal A}_{2}) &=& \left[1+\frac{|\Lambda|}{6\pi}({\cal A}_{1}+{\cal A}_{2}+4\pi Q^2)+
\frac{|\Lambda|^2}{144\pi^2}({\cal A}_{1}^2+{\cal A}_{1}{\cal A}_{2}+{\cal A}_{2}^2)\right]
{\cal A}_{1}{\cal A}_{2}  ~.\label{v3.1}
\end{eqnarray}
It implies that it is a mass-independent complicated function of EH and IH that depends 
upon charge, angular momentum and cosmological parameter. Again, it should be remember that 
it is not a simple product of EH and IHs. 

Recently, it  has been shown that for a regular Ay\'{o}n-Beato and Garc\'{i}a (ABG) BH \cite{ppgrg} the
mass-independent relation should read as
\begin{eqnarray}
f({\cal A}_{1},{\cal A}_{2}) &=& 384 \pi^3 q^6  ~.\label{eq}
\end{eqnarray}
where
$$
f({\cal A}_{1},{\cal A}_{2}) = {\cal A}_{1}{\cal A}_{2}({\cal A}_{1}+{\cal A}_{2})+24 \pi q^2 {\cal A}_{1}{\cal A}_{2}-
256 \pi ^4 q^8 (\frac{{\cal A}_{1}+{\cal A}_{2}}{{\cal A}_{1}{\cal A}_{2}})-
$$
$$
\frac{{\cal A}_{1}{\cal A}_{2}}{{\cal A}_{1}+{\cal A}_{2} +4\pi q^2}\left[ ({\cal A}_{1}+{\cal A}_{2})^2 
+24 \pi q^2 ({\cal A}_{1}+{\cal A}_{2})-{\cal A}_{1}{\cal A}_{2}-\frac{256 \pi^4 q^8}
{{\cal A}_{1}{\cal A}_{2}} +176 \pi^2 q^4 \right]
$$
where $q$ is the monopole charge parameter. Once again this is an example of a mass-independent 
complicated function of EH and IH but it is not a simple area product of EH and IHs.  This is in fact
very interesting topic in recent years both in the general relativity community and in the String theory community 
\cite{ah09,cgp11,castro12,chen12,sd12,mv13,don,pp14,pp15} and for the extended phase space 
thermodynamics see the recent work \cite{extend}. 

However in this work, we wish to investigate the \emph{mass-independent entropy functional relation} for conformal gravity (CG) 
holography in four dimensions \cite{grum14}. We also investigate the thermodynamic properties of the said BH. 
We compute the specific heat to determine the local thermodynamic stability of this BH. Furthermore, we have shown graphically  
various types of second order phase transition which occurs at a certain radius of the  EH. Smarr formula 
for this gravity is also calculated. Moreover, we compute the famous Cosmic Censorship Inequality (CCI)  for 
CG BH which is an important relation between the mass and area of the BH  in general theory of relativity 
in the context of naked singularities \cite{rp}.

In 4D, CG is an interesting topic in general relativity (GR) because it is  helpful to solve some problematic 
issues in Einstein's general relativity. It is a re-normalizable theory \cite{stele} of gravity and 
has ghosts, whereas Eintein's GR is ghost free \cite{stele,adler} gravity but 2-loop nonrenormalizable \cite{goroff} . 
For explaining galactic rotation curves without dark matter, Mannheim  \cite{mann} was studied CG 
phenomenologically and it emerges theoretically from twistor string theory \cite{witten9} . 
Also it acts as a counter term  in the AdS/CFT (conformal field theory) correspondence 
\cite{liu,bala}. In the context of quantum gravity, CG has been studied by 't Hooft \cite{hooft}. It has been suggested by 
Maldacena \cite{maldacena} that by imposing appropriate boundary condition \cite{star}, it is possible to remove the ghosts 
and interestingly the author has been shown how Einstein gravity recovered from CG. 

 It has been argued in \cite{capper} due to the conformal anomaly  
it is not possible in general to solve the power-counting renormalizability problem of quantum CG using conformal invariance. 
Because the conformal invariance is not a `good symmetry at the quantum level'. In \cite{mann1}, the author analyzed the 
cosmological perturbation problem by using CG which is connected to the conformal anomaly. In \cite{codello}, the author 
computed the beta functions of higher derivative gravity using one-loop approximation which appears to be  `asymptotically 
safe at a non-Gaussian fixed point rather than perturbatively renormalizable and asymptotically free'.  Inflationary 
cosmology from quantum CG has been studied in \cite{jizba}.

It should be noted that  the most intriguing properties of CG is that the theory depends only 
on the Lorentz angles but not on the distances \cite{grum10}. This implies that the theory 
is invariant under Weyl transformations $g_{\mu\nu}\rightarrow \Omega ^{2}(x)g_{\mu\nu}$ where $g_{\mu\nu}$  is the metric 
tensor and $\Omega(x)$ is a function on space-time.

The action for this theory is given by
\begin{eqnarray}
{\mathcal {I}} &=& \int {\sqrt {-g}} {d}^{4}x C_{\alpha\beta\gamma\delta}C^{\alpha\beta\gamma\delta} ~.\label{ac}
\end{eqnarray}
where $C_{\alpha\beta\gamma\delta}$ is the Weyl tensor. The equation of motion is determined by the Bach 
equation:
\begin{eqnarray}
 2\nabla _{\alpha}\nabla _{\delta}{{C^{\alpha}}_{\beta\gamma }}^{\delta}+{{C^{\alpha}}_{\beta\gamma}}^{\delta}
 R_{\alpha\delta } &=& 0 ~.\label{ac1}
\end{eqnarray}
where $R_{\alpha\beta}$ is the Ricci tensor. This is the basics of CG.

The plan of the paper is as follows. In Sec. (\ref{cgg}), we have described the thermodynamic properties 
of CG BH in four dimensions. Finally, we have given conclusion in Sec. (\ref{dis}).

\section{\label{cgg} Thermodynamic properties of CG BH in Four Dimensions:}
The spherically symmetric solution of the above action in the CG \cite{grum14} is given by
\begin{eqnarray}
ds^2 &= & -{\cal V}(r) dt^2 + \frac{dr^2}{{\cal V}(r)} +r^2 d\Omega_{2}^2 .~\label{mt}
\end{eqnarray}
where,
\begin{eqnarray}
{\cal V}(r) &=&  \sqrt{1-12aM}-\frac{2M}{r}+2ar-\Lambda r^2 .~\label{h1}
\end{eqnarray}
and $d\Omega_{2}^2$ is the metric on unit sphere in two dimension. In the limit $a=0$, one 
obtains the Schwarzschild-AdS space-time. It should be noted that in the limit $aM<<1$, one has 
the Grumiler space-time \cite{grum10}. Where the author proposed that the model 
for gravity at large distances. Approximately, the values of $\Lambda$, $a$ and $M$ (in Planck units) are 
$\Lambda \approx 10^{-123}$, $a \approx 10^{-61}$ and $M \approx 10^{38} M_{*}$, where $M_{*}=1$ for the sun, therefore 
it indeeds $aM \approx 10^{-23} M_{*} \ll 1$ for all BHs or galaxies in our Universe \cite{grum10}.

The  Eq. (\ref{mt}) and Eq. (\ref{h1}) are called Mannheim-Kazanas-Riegert (MKR) solution \cite{mann,rieg84} 
as described in \cite{grum14}.  In terms of length scale $\ell$ in Einstein gravity, the Eq. (\ref{mt}) and 
Eq. (\ref{h1}) are related to the cosmological constant as $\Lambda=\frac{3}{\ell^2}\rho$ with $\rho=-1$ for 
AdS case and $\rho=+1$ for dS case. For mathematical simplicity, let us now put $\Lambda=\frac{1}{\ell^2}$ for 
$\Lambda>0$.
The BH horizons computed by the condition ${\cal V}(r)=0$ i.e. 
\begin{eqnarray}
 r^3 -2a\ell^2r^2-\sqrt{1-12aM}\ell^2r+ 2M\ell^2 &=& 0 ~.\label{h2}
\end{eqnarray}

To determine the roots we apply  Vieta's theorem \cite{vit}, one obtains
\begin{eqnarray}
\sum_{i=1}^{3} r_{i} &=& 2a\ell^2  ~.\label{eq1}\\
\sum_{1\leq i<j\leq 3} r_{i}r_{j} &=& -\sqrt{1-12aM}\ell^2 ~.\label{eq2}\\
\prod_{i=1}^{3} r_{i} &=& - 2M\ell^2 ~.\label{eq3}
\end{eqnarray}
It should be mentioned that $i=1$, $i=2$ and $i=3$ denotes EH, IH and 
cosmological horizon respectively.
The entropy \cite{grum14} of the BH using Wald's approach is given by 
\begin{eqnarray}
{\cal S}_{i}  &=& \frac{{\cal A}_{i}}{4\ell^2} ~.\label{eq5}
\end{eqnarray}
where the area of the BH reads 
\begin{eqnarray}
{\cal A}_{i}  &=& 4\pi r_{i}^2 ~.\label{eq6}
\end{eqnarray}
Interestingly, the entropy satisfied the area law despite the fact that CG is a higher derivative 
gravity.

The BH temperature is calculated to be 
\begin{eqnarray}
T_{i} &=& \frac{{\cal V}'(r)}{4\pi}= 
\frac{1}{4\pi r_{i}} \left(\sqrt{1-3a^2r_{i}^2+\frac{6a}{\ell^2}r_{i}^3}+ar_{i}-3 \frac{r_{i}^2}{\ell^2}\right)
~. \label{eq7}
\end{eqnarray} 
It indicates that the product of Hawking temperature does depend on the mass parameter.

The mass parameter could be expressed as in terms of area:
\begin{eqnarray}
M  &=& \sqrt{\frac{{\cal A}_{i}}{16\pi}}\left[\sqrt{1-3a^2\left(\frac{{\cal A}_{i}}{4\pi}\right)
+\left(\frac{6a}{\ell^2}\right)\left(\frac{{\cal A}_{i}}{4\pi}\right)^{3/2}}-\left(a\sqrt{\frac{{\cal A}_{i}}{4\pi}}
+\left(\frac{{\cal A}_{i}}{4\pi\ell^2}\right)\right)\right] ~.\label{h3}
\end{eqnarray}
This is an exact \emph{mass-area} equality in CG.

\subsection{Mass-independent Area (or Entropy) Functional Form:}
Here, we shall derive exact mass independent quantity in terms of two BH physical horizons. To do this we first 
eliminate the mass parameter between Eq. (\ref{eq2}) and Eq. (\ref{eq3}), one obtains in terms of $\Lambda$
\begin{eqnarray}
\Lambda^2\left(r_{1}r_{2}+r_{1}r_{3}+r_{2}r_{3} \right)^2 &=& 1+6a\Lambda r_{1}r_{2}r_{3} ~.\label{eq8}
\end{eqnarray}
Now we can eliminate the third root between Eq. (\ref{eq8}) and Eq. (\ref{eq1}) then we find 
\begin{eqnarray}
\Lambda^2\left[r_{1}r_{2}+\frac{2a}{\Lambda}(r_{1}+r_{2})-(r_{1}+r_{2})^2 \right]^2 &=& 
1+12a^2 r_{1}r_{2}-6a\Lambda r_{1}r_{2}(r_{1}+r_{2}) ~.\label{eq9}
\end{eqnarray}
If we are restricted to  two physical horizons by eliminating third root and if we computed in 
terms of area then we find the mass-independent area functional relation:
\begin{eqnarray}
f({\cal A}_{1},{\cal A}_{2}) &=& 16 \pi^2 ~.\label{eq10}
\end{eqnarray}
where the function $f({\cal A}_{1},{\cal A}_{2})$ is given by
$$
f({\cal A}_{1},{\cal A}_{2}) = \left[\Lambda \sqrt{{\cal A}_{1}{\cal A}_{2}}
+2a(\sqrt{4\pi{\cal A}_{1}}+\sqrt{4\pi{\cal A}_{2}})-\Lambda(\sqrt{{\cal A}_{1}}+\sqrt{{\cal A}_{2}})^2\right]^2
$$
\begin{eqnarray}
+6a\Lambda \sqrt{4\pi {\cal A}_{1}{\cal A}_{2}} (\sqrt{{\cal A}_{1}}+\sqrt{{\cal A}_{2}})
-48\pi a^2\sqrt{{\cal A}_{1}{\cal A}_{2}} ~.\label{eq11}
\end{eqnarray}
This is indeed a function of OH area and IH area but it is quite \emph{mass-independent}. 
This is one of the key result of this work.

In terms of three horizons, the area functional form should be 
\begin{eqnarray}
f({\cal A}_{1},{\cal A}_{2},{\cal A}_{3}) &=& 16 \pi^2 ~.\label{eq12}
\end{eqnarray}
where the function $f({\cal A}_{1},{\cal A}_{2},{\cal A}_{3})$ should be 
\begin{eqnarray}
f({\cal A}_{1},{\cal A}_{2},{\cal A}_{3}) &=& 
\Lambda^2 \left(\sqrt{{\cal A}_{1}{\cal A}_{2}}+\sqrt{{\cal A}_{2}{\cal A}_{3}}+\sqrt{{\cal A}_{3}{\cal A}_{1}}\right)^2
-6a\Lambda \sqrt{4\pi {\cal A}_{1}{\cal A}_{2}{\cal A}_{3}}~.\label{eq13}
\end{eqnarray}
This is again explicitly \emph{mass-independent}. But it is not a simple area products as in Kerr BH.

\subsection{Specific Heat:}
The specific heat is only an indicator of stability of the BH. To determine this stability 
we can compute the specific heat as
\begin{eqnarray}
C_{i} &=& \frac{\partial{M}}{\partial T_{i}} .~\label{eq14}
\end{eqnarray}
which is found to be for this BH
\begin{eqnarray}
C_{i} &=& 2\pi r_{i}^2\left[\frac{15a\Lambda r_{i}^3-6a^2r_{i}^2-(2ar_{i}+3\Lambda r_{i}^2)
\sqrt{6a\Lambda r_{i}^3-3a^2r_{i}^2+1}+1}
{3a\Lambda r_{i}^3-3\Lambda r_{i}^2\sqrt{6a\Lambda r_{i}^3-3a^2r_{i}^2+1}-1 } \right] .~\label{eq15}
\end{eqnarray}
It should be noted that 
\begin{eqnarray}
\prod_{i=1}^{3} C_{i}
\end{eqnarray}
does depend on the mass parameter. So it is not a mass-independent quantity.
The second order phase transitions occur at 
\begin{eqnarray}
1-3a^2r_{i}^2+6a\Lambda r_{i}^3 &=& \left(\frac{3a\Lambda r_{i}^3-1}{3\Lambda r_{i}^2} \right)^2.~\label{eq16}
\end{eqnarray}

In appendix A, we have plotted the specific heat with EH radius which demonstrates the second order phase 
transition for various values of $a$ and $\Lambda$.

Now we consider the AdS case. In this case we have to put $-\Lambda=\frac{1}{\ell^2}$. We find the 
Killing horizons equation:
\begin{eqnarray}
 r^3 +2a\ell^2r^2+\sqrt{1-12aM}\ell^2r-2M\ell^2 &=& 0 ~.\label{eq17}
\end{eqnarray}
In this case, the roots of the cubic equation are
\begin{eqnarray}
\sum_{i=1}^{3} r_{i} &=& -2a\ell^2  ~.\label{eq18}\\
\sum_{1\leq i<j\leq 3} r_{i}r_{j} &=& \sqrt{1-12aM}\ell^2 ~.\label{eq19}\\
\prod_{i=1}^{3} r_{i} &=&  2M\ell^2 ~.\label{eq20}
\end{eqnarray}
Eliminating mass parameter, we obtain the following expression:
\begin{eqnarray}
\left(r_{1}r_{2}+r_{1}r_{3}+r_{2}r_{3} \right)^2 &=& \ell^2\left(\ell^2-6a r_{1}r_{2}r_{3}\right) ~.\label{eq21}
\end{eqnarray}
Eliminating third root, we get the following radial expression:
\begin{eqnarray}
\left[r_{1}r_{2}-2a\ell^2(r_{1}+r_{2})-(r_{1}+r_{2})^2 \right]^2-6a\ell^2 r_{1}r_{2}(r_{1}+r_{2})-12a^2\ell^4 r_{1}r_{2}
&=& \ell^4 ~.\label{eq22}
\end{eqnarray}

In terms of two physical horizons by eliminating third root and  in 
terms of area one can find the mass-independent area functional 
relation:
\begin{eqnarray}
f({\cal A}_{1},{\cal A}_{2}) &=& 16 \pi^2\ell^4 ~.\label{eq23}
\end{eqnarray}
where the function $f({\cal A}_{1},{\cal A}_{2})$ is given by
$$
f({\cal A}_{1},{\cal A}_{2}) = \left[(\sqrt{{\cal A}_{1}}+\sqrt{{\cal A}_{2}})^2
+2a\ell^2(\sqrt{4\pi{\cal A}_{1}}+\sqrt{4\pi{\cal A}_{2}})-\sqrt{{\cal A}_{1}{\cal A}_{2}}\right]^2
$$
\begin{eqnarray}
-6 a \ell^2 \sqrt{4\pi {\cal A}_{1}{\cal A}_{2}} (\sqrt{{\cal A}_{1}}+\sqrt{{\cal A}_{2}})
-48 \pi a^2 \ell^4 \sqrt{{\cal A}_{1}{\cal A}_{2}} ~.\label{eq24}
\end{eqnarray}
Again, this is explicitly mass-independent but $\Lambda$ dependent. But it is not straightforward
as a simple product of two BH horizons.

In the AdS case, the BH temperature is given by 
\begin{eqnarray}
T_{i} &=& 
\frac{1}{4\pi r_{i}} \left(\sqrt{1-3a^2r_{i}^2-\frac{6a}{\ell^2}r_{i}^3}+ar_{i}+3\frac{r_{i}^2}{\ell^2}\right)
~. \label{eq25}
\end{eqnarray} 
The mass parameter\footnote{It may be noted that actually the non-trivial mass parameter for CG BH 
\cite{grum14} becomes ${\cal M} = \frac{M}{\ell^2}-\frac{a}{6}\left(1-\sqrt{1-12aM} \right)$.} in terms of the 
area could be written as
\begin{eqnarray}
M  &=& \sqrt{\frac{{\cal A}_{i}}{16\pi}}\left[\sqrt{1-3a^2\left(\frac{{\cal A}_{i}}{4\pi}\right)
-\left(\frac{6a}{\ell^2}\right)\left(\frac{{\cal A}_{i}}{4\pi}\right)^{3/2}}-\left(a\sqrt{\frac{{\cal A}_{i}}{4\pi}}
-\left(\frac{{\cal A}_{i}}{4\pi\ell^2}\right)\right)\right] ~.\label{eq26}
\end{eqnarray}

In 1973, Penrose \cite{rp} derived an important relation between the total mass of the spacetime and 
an area of the horizon is so called \emph{Cosmic-Censorship-Inequality} (CCI) \cite{gibb05} regarding the 
``cosmic censorship conjecture'' (See \cite{bray,bray1,jang,rg,gibb99}) which is an important topic in GR.
For CG BH, this inequality\footnote{If we take into account the non-trivial mass parameter 
in CG, then the CCI for CG BH should be ${\cal M} \geq \frac{M}{\ell^2}-\frac{a}{6}\left(1-\sqrt{1-12aM} \right)$, 
where $M$ is defined in Eq. (\ref{eq26}).} 
becomes 
\begin{eqnarray}
M  & \geq & \sqrt{\frac{{\cal A}_{i}}{16\pi}}\left[\sqrt{1-3a^2\left(\frac{{\cal A}_{i}}{4\pi}\right)
-\left(\frac{6a}{\ell^2}\right)\left(\frac{{\cal A}_{i}}{4\pi}\right)^{3/2}}-\left(a\sqrt{\frac{{\cal A}_{i}}{4\pi}}
-\left(\frac{{\cal A}_{i}}{4\pi\ell^2}\right)\right)\right] ~.\label{pia}
\end{eqnarray}
In the limit $a=0$, one obtains the \emph{Cosmic-Censorship-Inequality} for Schwarzschild-AdS BH:
\begin{eqnarray}
M  & \geq & \sqrt{\frac{{\cal A}_{i}}{16\pi}}\left[1 +\frac{{\cal A}_{i}}{4\pi\ell^2}\right] ~.\label{pia1}
\end{eqnarray}
In the limit $a=0$ and $\ell \rightarrow \infty$, one obtains the famous \emph{Cosmic-Censorship-Inequality} for 
Schwarzschild spacetime:
\begin{eqnarray}
M  &\geq&  \sqrt{\frac{{\cal A}}{16\pi}} ~.\label{scpi}
\end{eqnarray}
The physical significance of CCI is that it suggests the lower bound on the mass (or energy) for 
a time-symmetric initial data which satisfied the non-linear Einstein equations with cosmological constant, and which 
is also fulfilled the dominant energy condition and which has no naked singularities. 

Now if we consider the cosmological constant as a thermodynamic pressure i.e. $\Lambda=-\frac{1}{\ell^2}=-8\pi P$ 
(in geometric units where $G_{N}=\hbar=c=k_{B}=1$ ) and 
its conjugate variable as a thermodynamic volume i.e. $V_{i}=\frac{4}{3} \pi r_{i}^3$ then the ADM mass of an AdS BH could be 
treated as the enthalpy of the space-time i.e. $M=H=U+PV$. Where $U$ is the thermal energy of the system \cite{kastor09}.
Therefore in the extended phase space the first law of BH thermodynamics becomes 
\begin{eqnarray}
dH &=& T_{i} d{\cal S}_{i}+ V_{i} dP + \Upsilon_{i} da ~. \label{eq27}
\end{eqnarray}
where, 
\begin{eqnarray}
\Upsilon_{i} &=& \left(\frac{\partial H}{\partial a} \right)_{{\cal A}_{i},P}=-\left(\frac{{\cal A}_{i}}{8\pi}\right)
\left[1+\sqrt{\frac{{\cal A}_{i}}{4\pi}}\frac{\left(3a-24\pi P\sqrt{\frac{{\cal A}_{i}}{4\pi}}\right)}
{\sqrt{1-3a^2\left(\frac{{\cal A}_{i}}{4\pi}\right)
+48\pi P a\left(\frac{{\cal A}_{i}}{4\pi}\right)^{3/2}}}\right] ~. \label{e28}
\end{eqnarray}
is defined to be a physical quantity conjugate to the parameter $a$.

It may be noted that when cosmological constant should be treated as constant then the Eq. (\ref{eq27}) 
reduces to the original first law of BH thermodynamics in the non-extended phase space. Since we are now 
in the extended phase space therefore we could find another important relation  in terms of thermodynamic
volume:
\begin{eqnarray}
\sum_{i=1}^{3} {V_{i}}^{\frac{1}{3}} &=& -\left(\frac{4\pi}{3}\right)^{\frac{1}{3}}\frac{a}{2\pi P}  ~.\label{eq28}\\
\sum_{1\leq i<j\leq 3}(V_{i}V_{j})^{\frac{1}{3}}  &=& \left(\frac{4\pi}{3}\right)^{\frac{2}{3}}\frac{\sqrt{1-12aM}}{8\pi P} 
~.\label{eq29}\\
\prod_{i=1}^{3} V_{i}^{\frac{1}{3}} &=& \frac{M}{3P} ~.\label{eq30}
\end{eqnarray}

Now eliminating mass(or enthalpy) parameter we find the mass(or enthalpy) independent \emph{thermodynamic volume functional} 
relation in terms of two physical horizons:
\begin{eqnarray}
f(V_{1},V_{2}) &=& 1 ~.\label{eq31}
\end{eqnarray}
where the function $f(V_{1},V_{2})$ is given by
$$
f(V_{1},V_{2}) = \left[8\pi P\left(\frac{3}{4\pi}\right)^{\frac{2}{3}} \left(V_{1}V_{2}\right)^{\frac{1}{3}}
-2a\left(\frac{3}{4\pi}\right)^{\frac{1}{3}}\left(V_{1}^{\frac{1}{3}}+V_{2}^{\frac{1}{3}}\right)
-\left(\frac{3}{4\pi}\right)^{\frac{2}{3}}\left(V_{1}^{\frac{1}{3}}+V_{2}^{\frac{1}{3}}\right)^2\right]^2
$$
\begin{eqnarray}
-36aP\left(V_{1}V_{2}\right)^{\frac{1}{3}}\left(V_{1}^{\frac{1}{3}}+V_{2}^{\frac{1}{3}}\right)
-12a^2\left(\frac{3}{4\pi}\right)^{\frac{2}{3}}\left(V_{1}V_{2}\right)^{\frac{1}{3}}~.\label{eq32}
\end{eqnarray}
This is indeed a \emph{mass-independent} functional relation in terms of \emph{thermodynamic volume}. Thus we may 
conclude that the entropy functional relation and thermodynamic volume functional relation that we have obtained 
in this work could turn out to be \emph{universal} i.e. hold in a larger contexts.

Since we have considered the extended phase space, therefore there are two types of 
specific heat. The specific heat at constant thermodynamic volume and the specific heat at constant thermodynamic 
pressure. They should read as 
\begin{eqnarray}
\left( C_{V} \right)_{i} &=& T_{i} \left(\frac{\partial S_{i}}{\partial T_{i}}\right)_{V} 
.~\label{cv}
\end{eqnarray}
and
\begin{eqnarray}
\left( C_{P} \right)_{i} &=& T_{i} \left(\frac{\partial S_{i}}{\partial T_{i}}\right)_{P}
.~\label{cp}
\end{eqnarray} 
Since the entropy ${\cal S}_{i}$ is independent of $T_{i}$ thus one obtains 
\begin{eqnarray}
\left( C_{V} \right)_{i} &=& 0  .~\label{cv1}
\end{eqnarray}
and for AdS case, one should derive 
\begin{eqnarray}
(C_{P})_{i} &=& -2\pi r_{i}^2\left[\frac{1-120\pi P ar_{i}^3-6a^2r_{i}^2-(2ar_{i}-24\pi Pr_{i}^2)
\sqrt{1-3a^2r_{i}^2-48 \pi P ar_{i}^3}}
{1+24 \pi ar_{i}^3-24\pi Pr_{i}^2\sqrt{1-3a^2r_{i}^2-48 \pi Pa r_{i}^3}} \right] .~\label{eq33}
\end{eqnarray}
The second order phase transition point should occur at 
\begin{eqnarray}
1-3a^2r_{i}^2-48 \pi P ar_{i}^3 &=& \left(\frac{1+24 \pi ar_{i}^3}{24\pi Pr_{i}^2}\right)^2 .~\label{eq34}
\end{eqnarray}
In the limit $a=0$, we get the result for Schwarzschild-AdS BH. We have plotted the variation of specific heat with 
EH radius which displayed second order phase transitions for various values of $a$ and $\Lambda$ in 
Appendix section.

\begin{figure}[h]
\begin{center}
 \subfigure[]{
 \includegraphics[width=2.1in,angle=0]{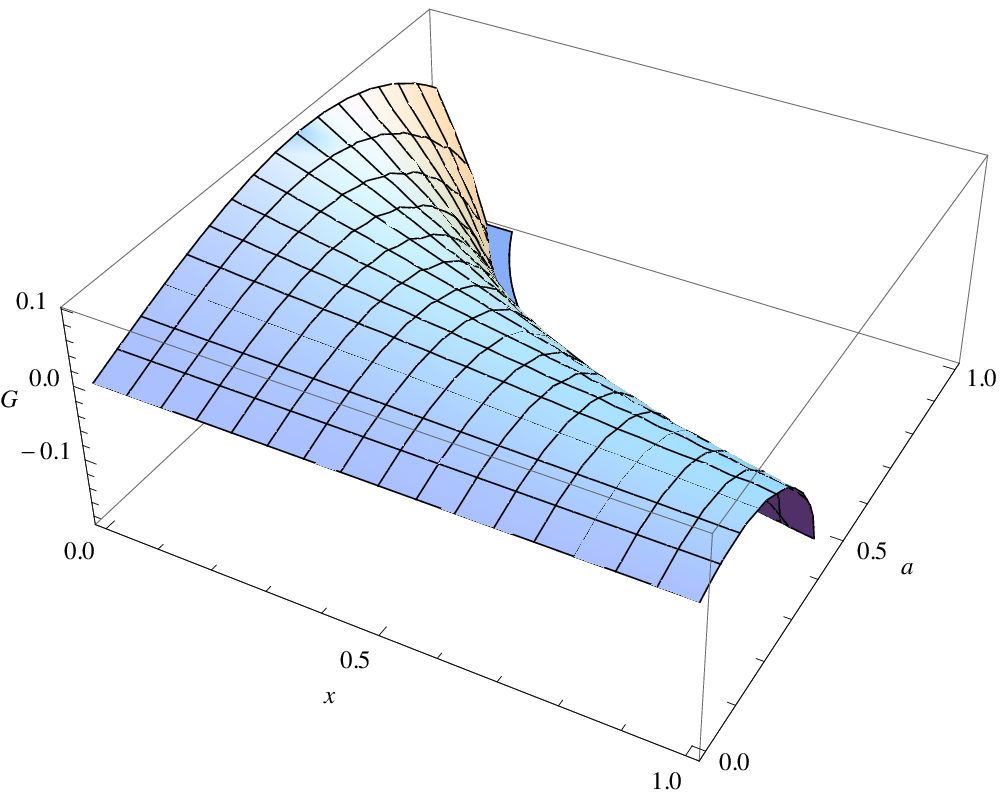}}
 \subfigure[]{
 \includegraphics[width=2.1in,angle=0]{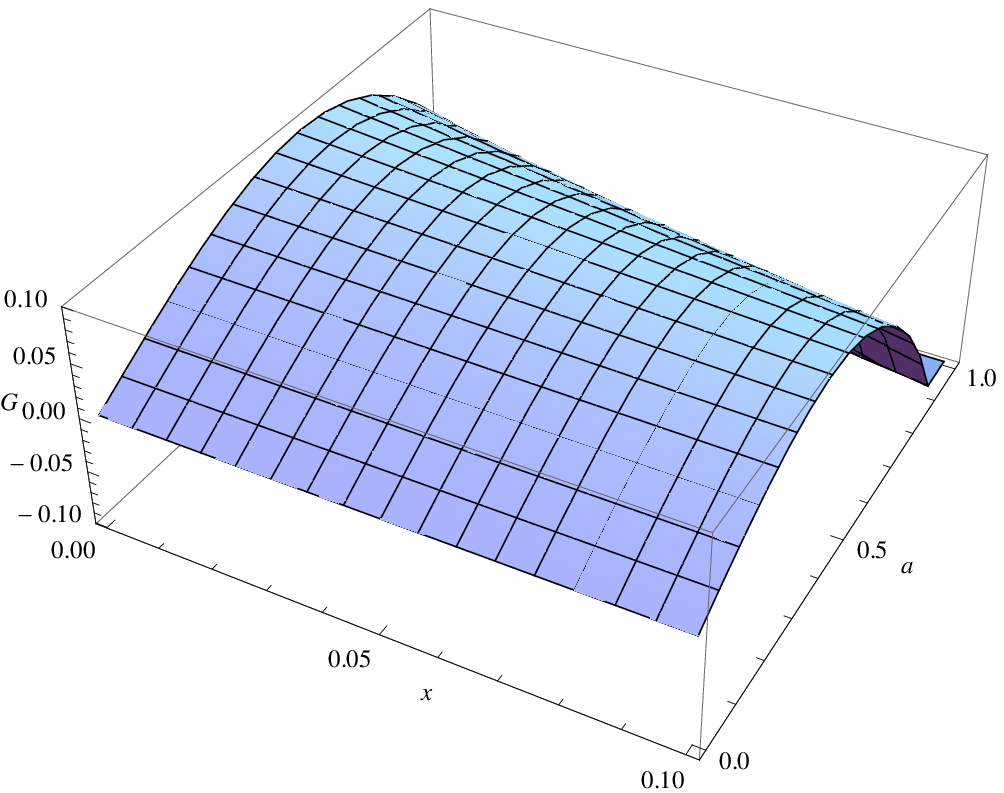}}
  \caption{\label{fgg}\textit{The figure depicts the variation  of $G$  with $r_{+}$ and $a$, Here $x=r_{+}$ and $\ell=1$. }}
\end{center}
\end{figure}

Now one can define the Gibbs free energy in the extended phase space as
\begin{eqnarray}
 G_{i} &=& H-T_{i}S_{i}=\frac{r_{i}}{4}\left[\sqrt{1-3a^2r_{i}^2-48\pi P a r_{i}^3}-3ar_{i}-8\pi Pr_{i}^2\right]
 ~. \label{cg6}
\end{eqnarray}
It also should be noted that the 
\begin{eqnarray}
\prod_{i=1}^{3} G_{i}
\end{eqnarray}
does depend on the mass parameter. Hence it is not a mass-independent quantity. We have plotted the Gibbs free energy 
in the Fig. \ref{fgg}.

\section{\label{dis} Conclusion:}
This work deals with the thermodynamic properties of CG holography in four dimensions. We have 
derived the \emph{mass-independent} area functional relation that could turn out to be a \emph{universal}
quantity. We have considered both the $\Lambda>0$ and $\Lambda<0$ cases. We have also derived the specific heat to 
study the stability of the said BH. The second order phase transition occurs at a certain condition. Various 
types of phase structure has been given for various values of $a$ and $\Lambda$. In the limit $a=0$, one obtains
the result of Schwarzschild-dS and Schwarzschild-AdS BH. In the limit $aM<<1$ one has the results of Grumiler 
space-time. Smarr formula has been studied.  Famous \emph{Cosmic-Censorship-Inequality} has been derived which relates the
total mass of the spacetime and area of the all horizons. The first law of thermodynamics also has been studied in the 
extended phase space. The extended phase space means here the cosmological constant should be treated as a thermodynamic 
pressure and its conjugate variable as thermodynamic volume. Finally, the \emph{mass-independent} volume 
functional relation has been derived. In summary, the \emph{mass-independent area functional} 
relation and \emph{mass-independent volume functional} relation that we have derived could turn out to be a
\emph{universal} quantity.


\appendix

\section{Plot of Specific Heat with $r_{+}$:}

In the appendix section, we have plotted the variation of specific heat with $r_{+}$ for various values of 
$a$ and $\Lambda$.

\begin{figure}[h]
\begin{center}
\subfigure[]{
\includegraphics[width=2.1in,angle=0]{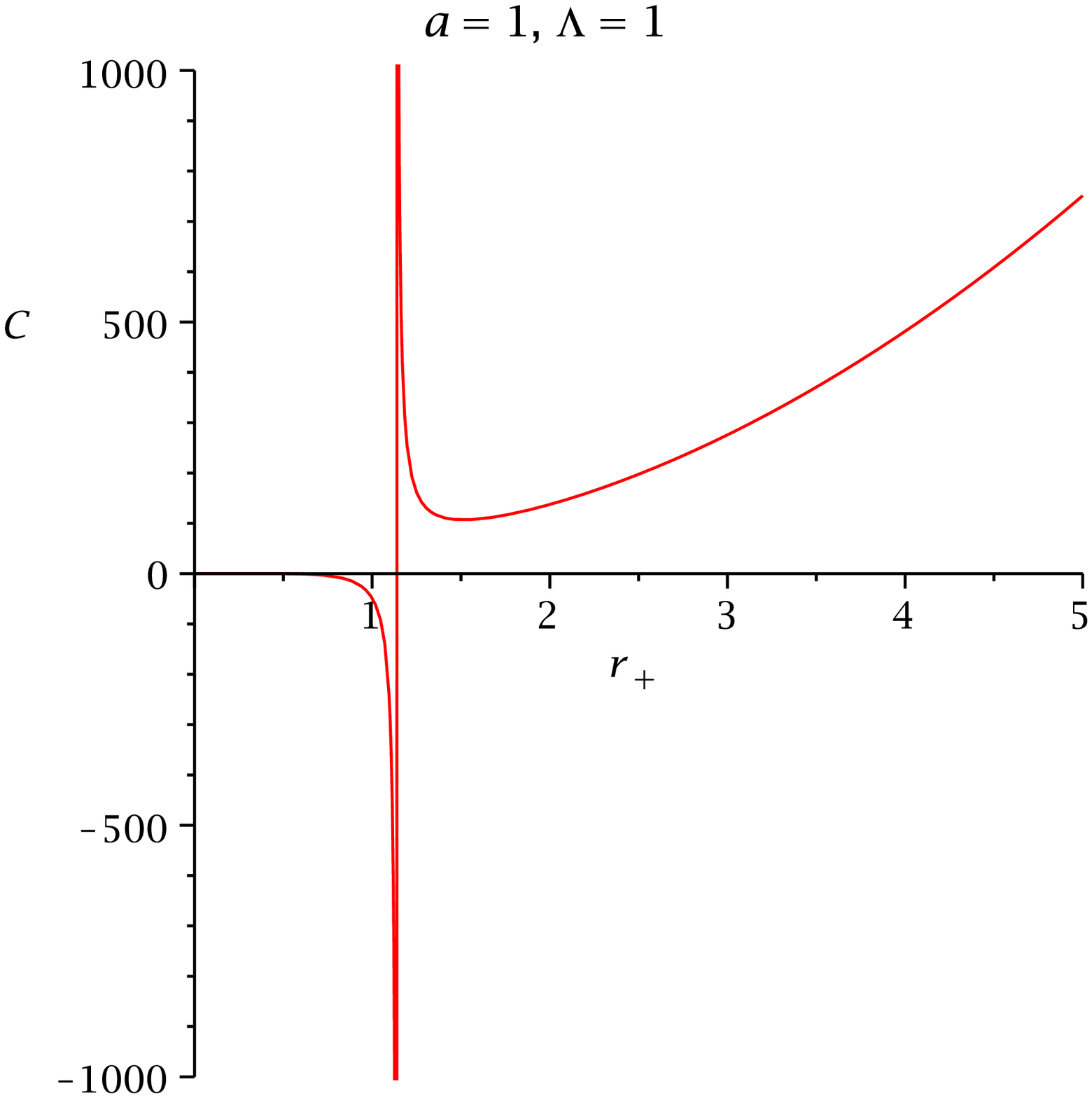}} 
\subfigure[]{
 \includegraphics[width=2.1in,angle=0]{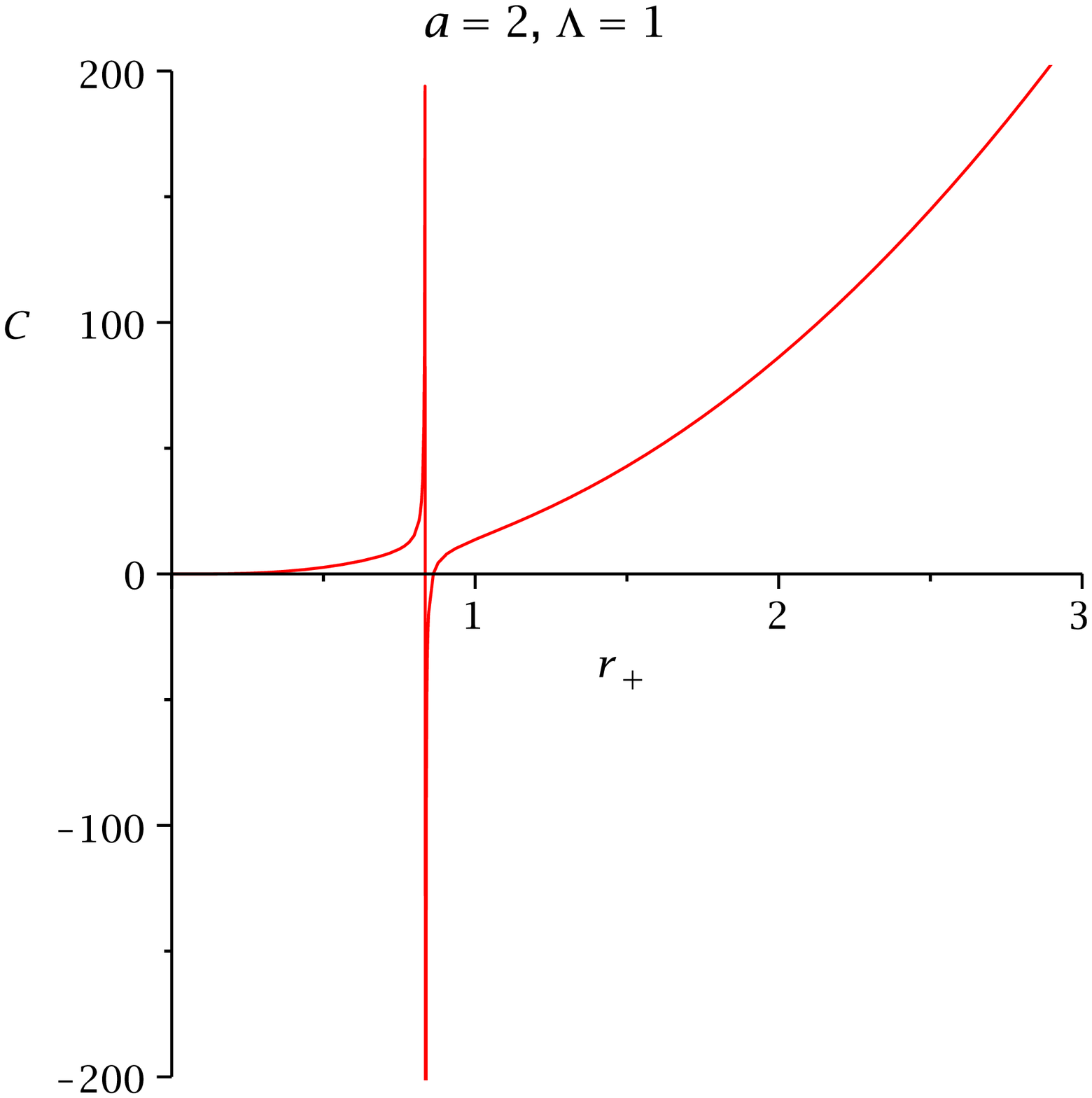}}
 \subfigure[]{
 \includegraphics[width=2.1in,angle=0]{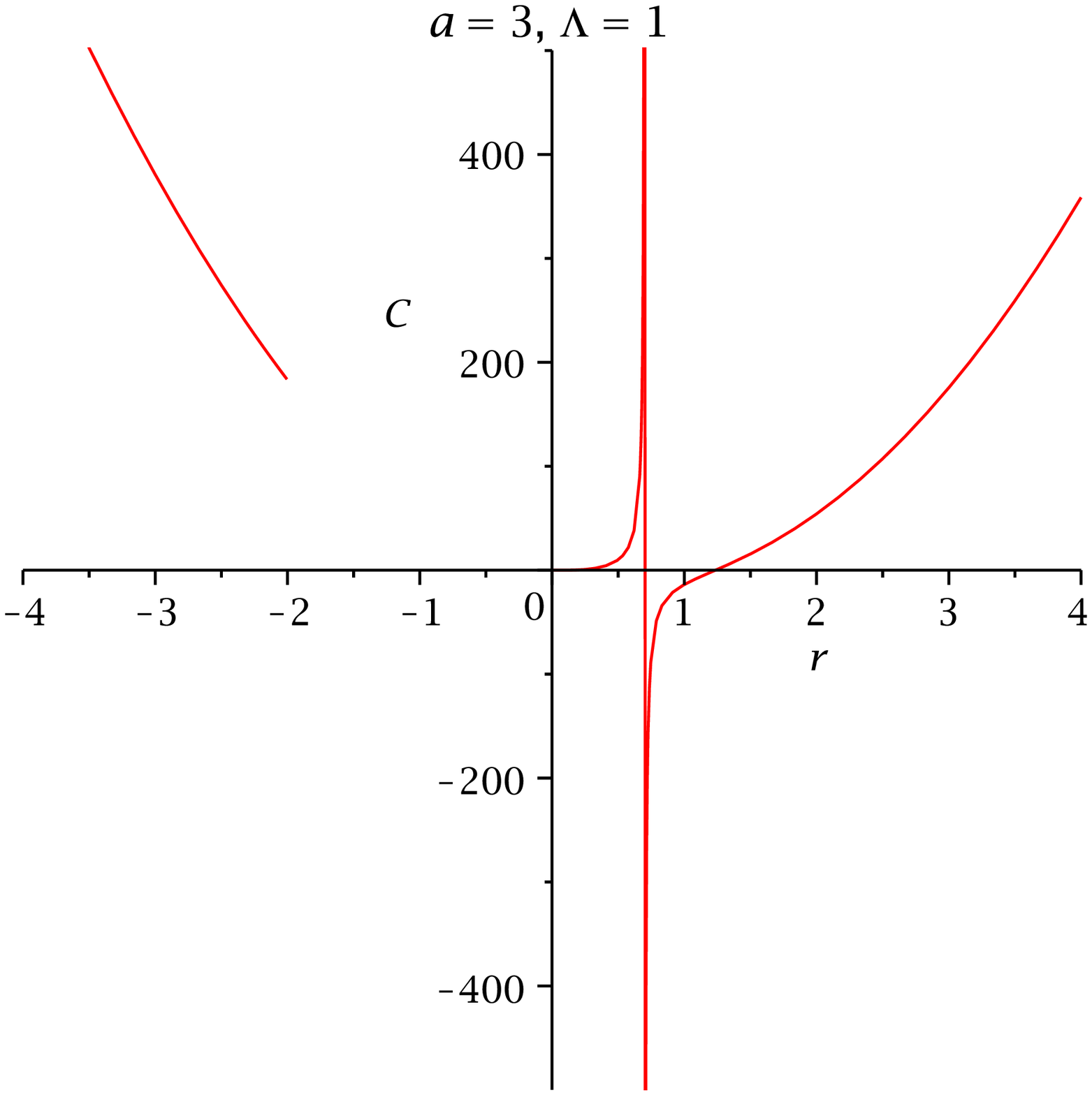}}
 \subfigure[ ]{
 \includegraphics[width=2.1in,angle=0]{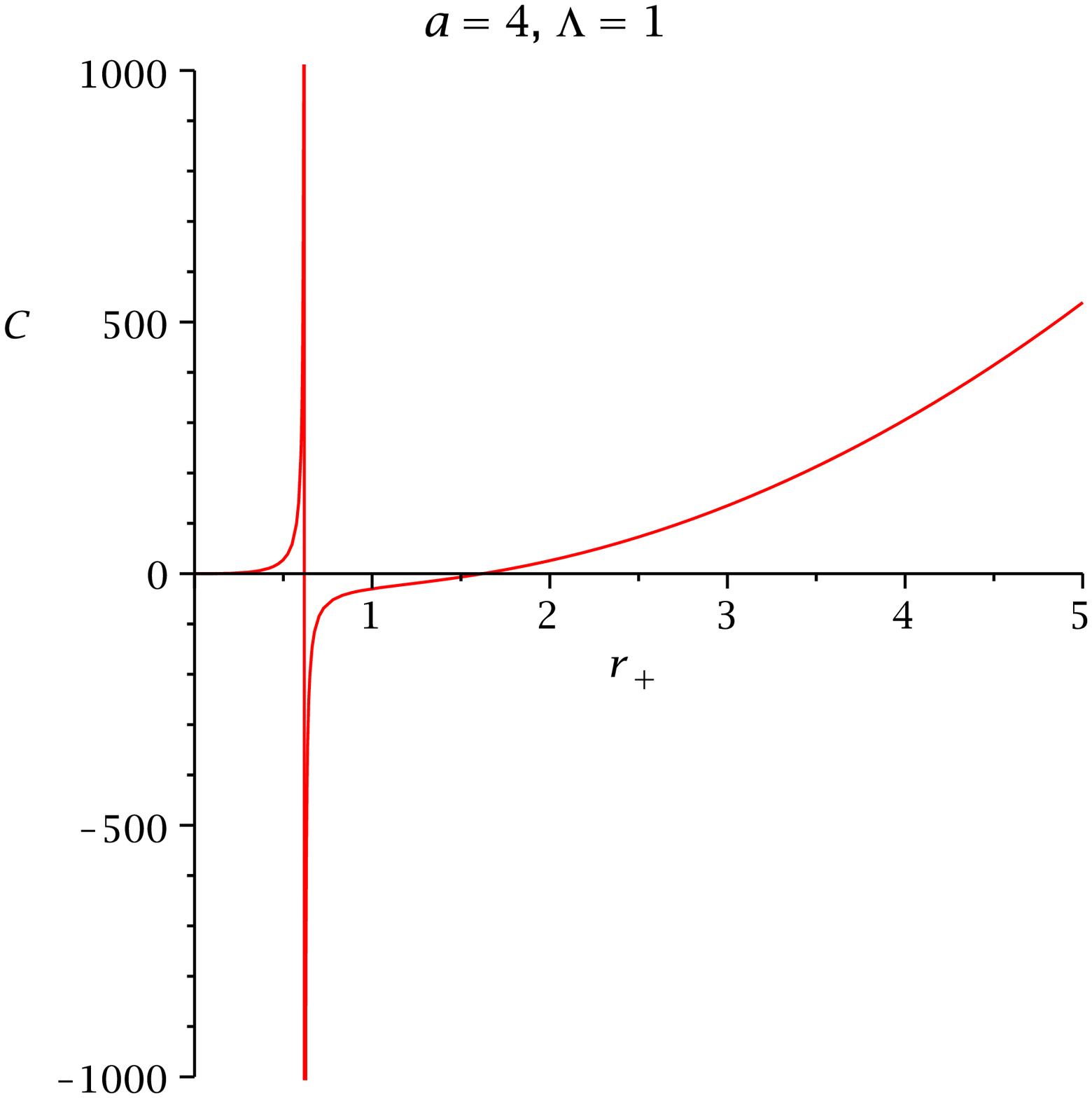}}
 \subfigure[]{
 \includegraphics[width=2.1in,angle=0]{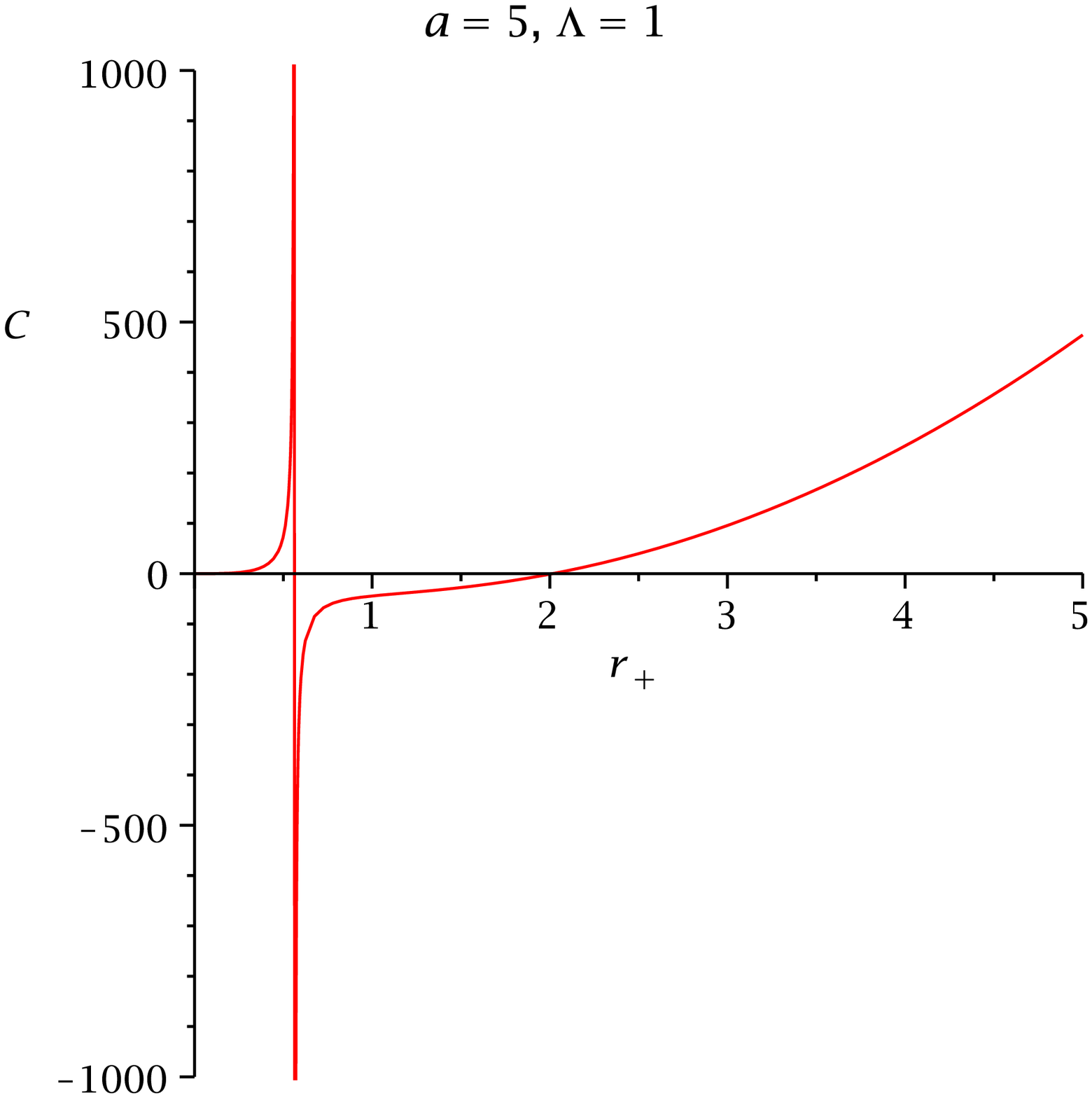}}
 \subfigure[]{
 \includegraphics[width=2.1in,angle=0]{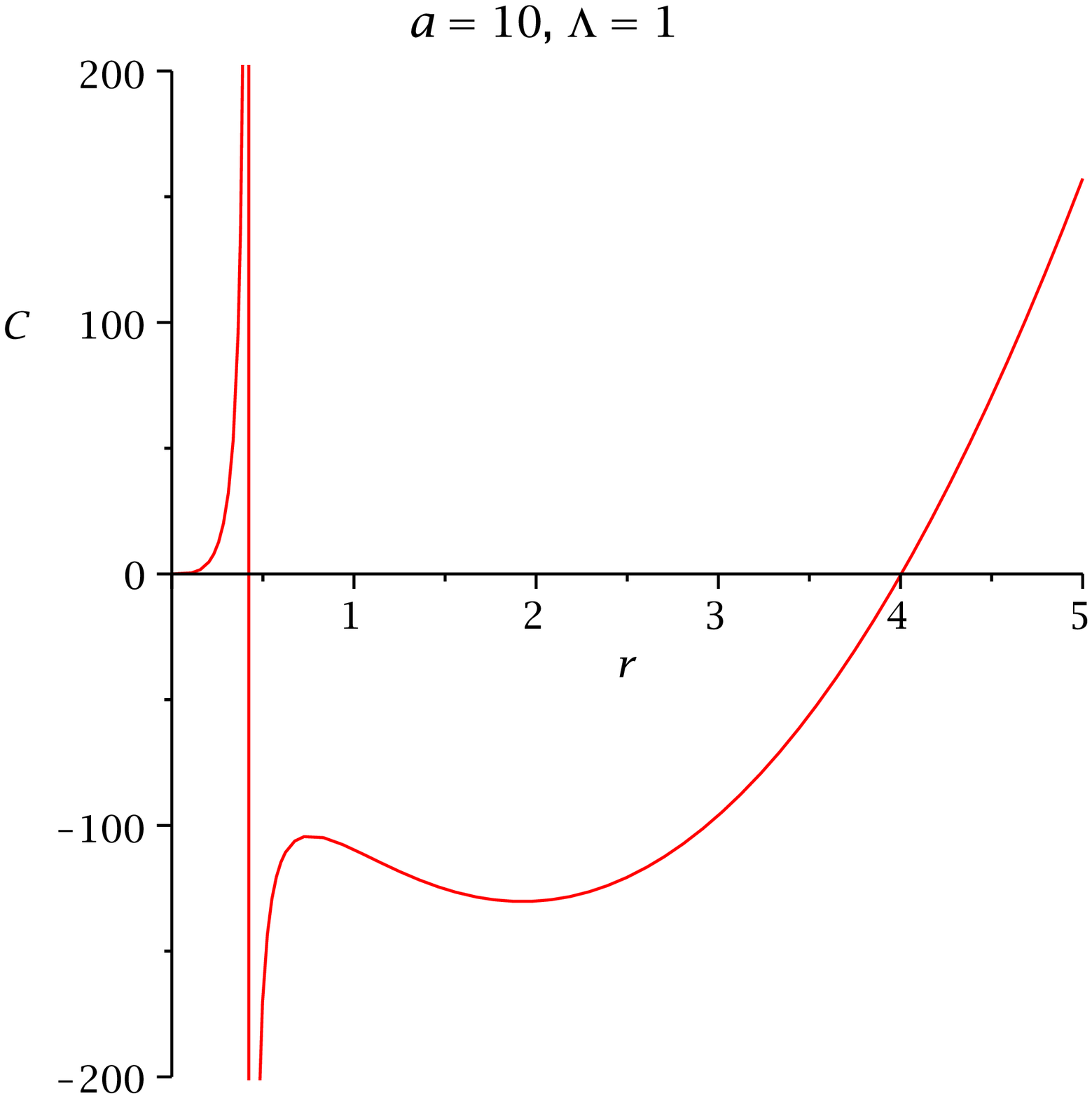}}
 
 \caption{\label{fg1}\textit{The figure shows the variation  of $C$  with $r_{+}$ }}
\end{center}
\end{figure}

\begin{figure}[h]
\begin{center}
\subfigure[]{
\includegraphics[width=2.1in,angle=0]{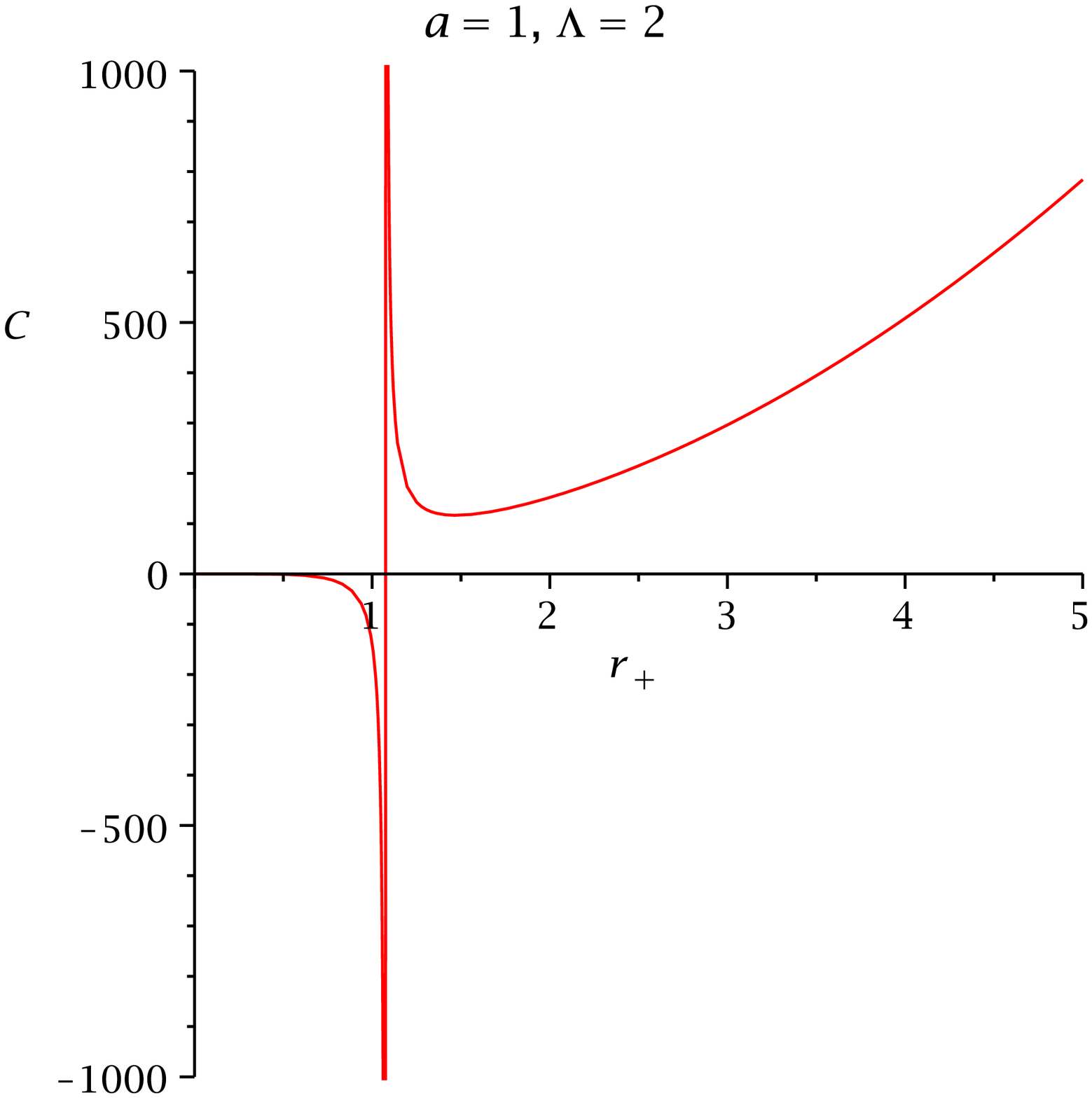}} 
\subfigure[]{
 \includegraphics[width=2.1in,angle=0]{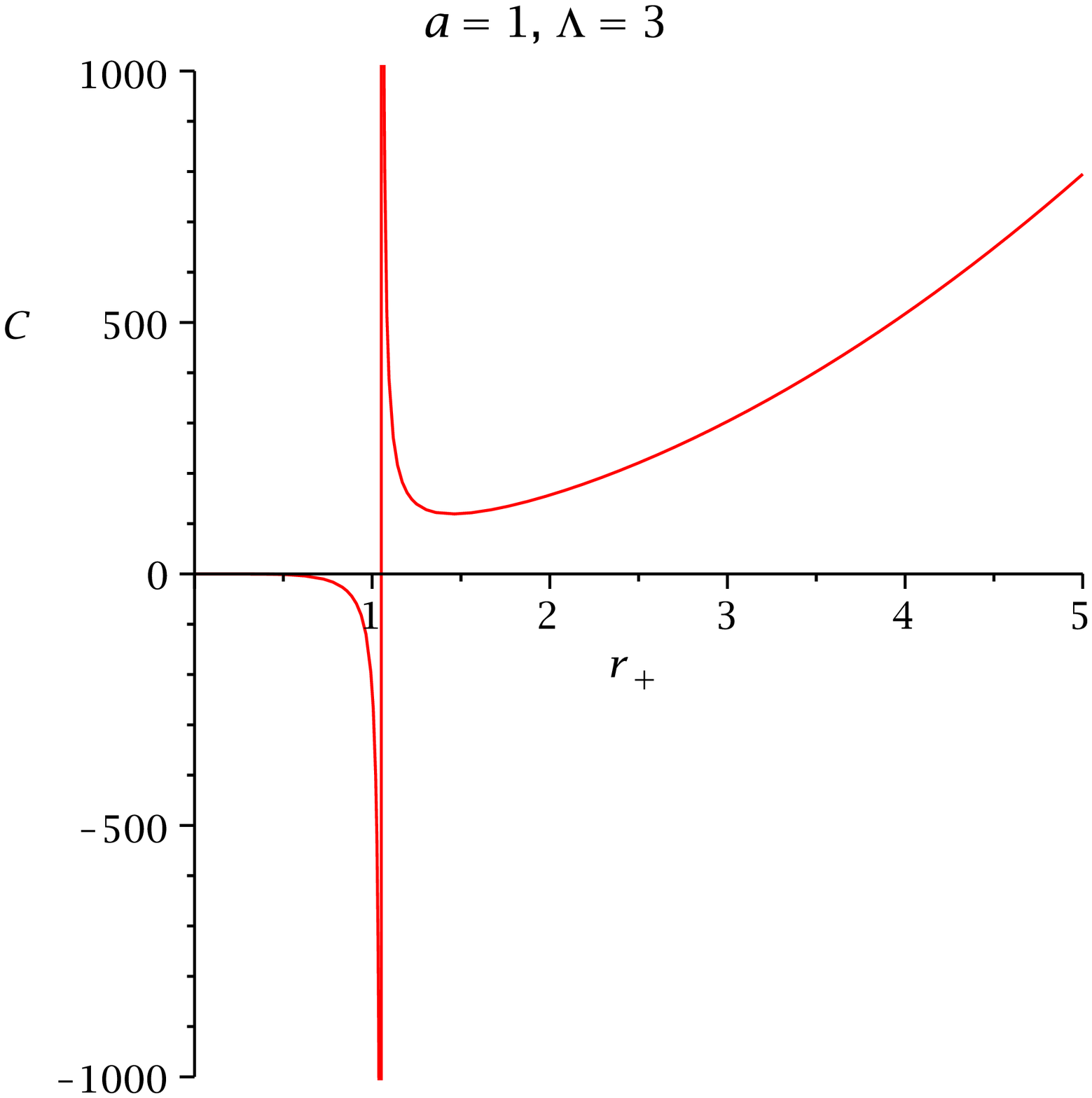}}
\subfigure[]{
 \includegraphics[width=2.1in,angle=0]{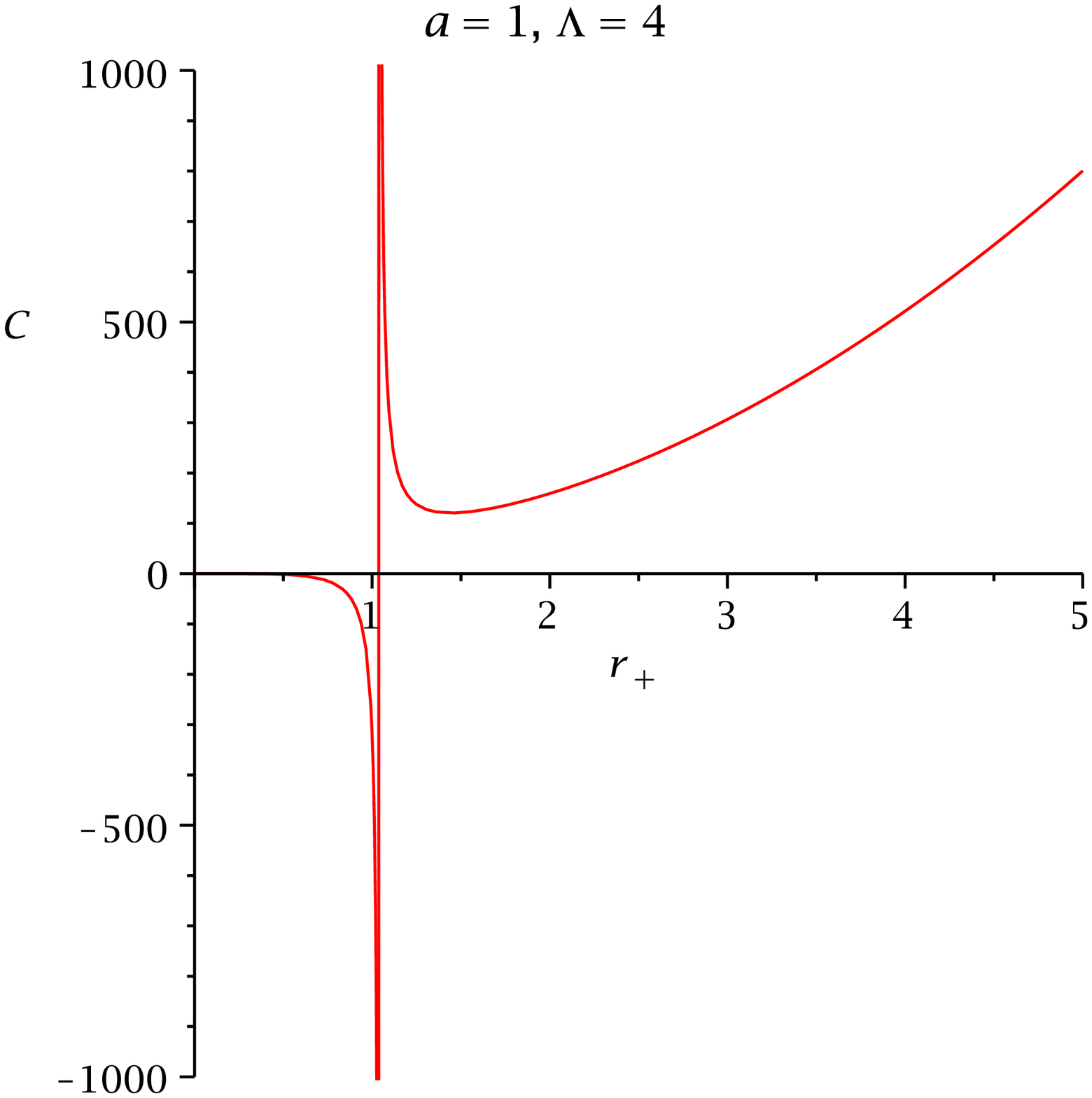}}
 \subfigure[ ]{
 \includegraphics[width=2.1in,angle=0]{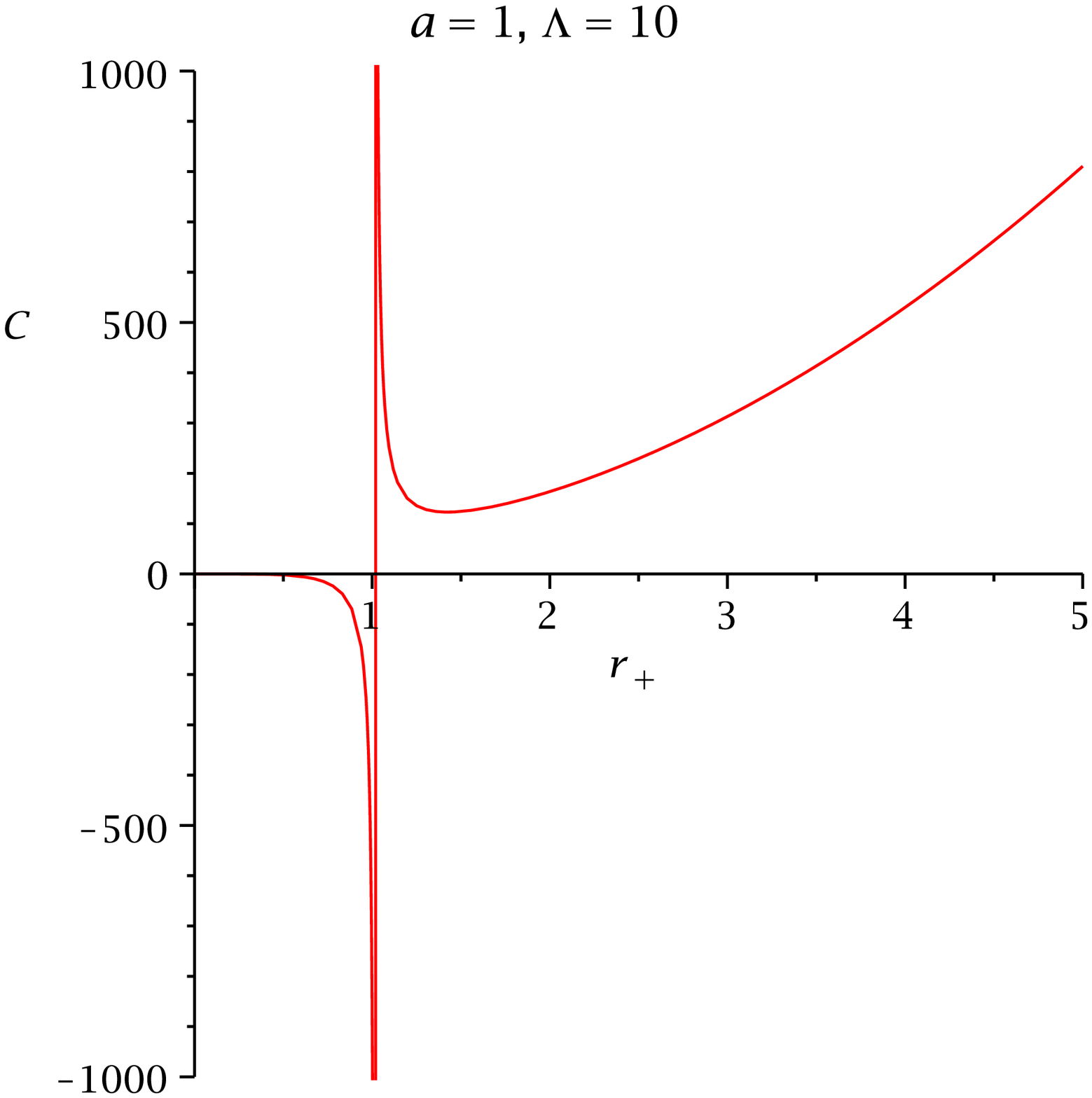}}
 \subfigure[]{
 \includegraphics[width=2.1in,angle=0]{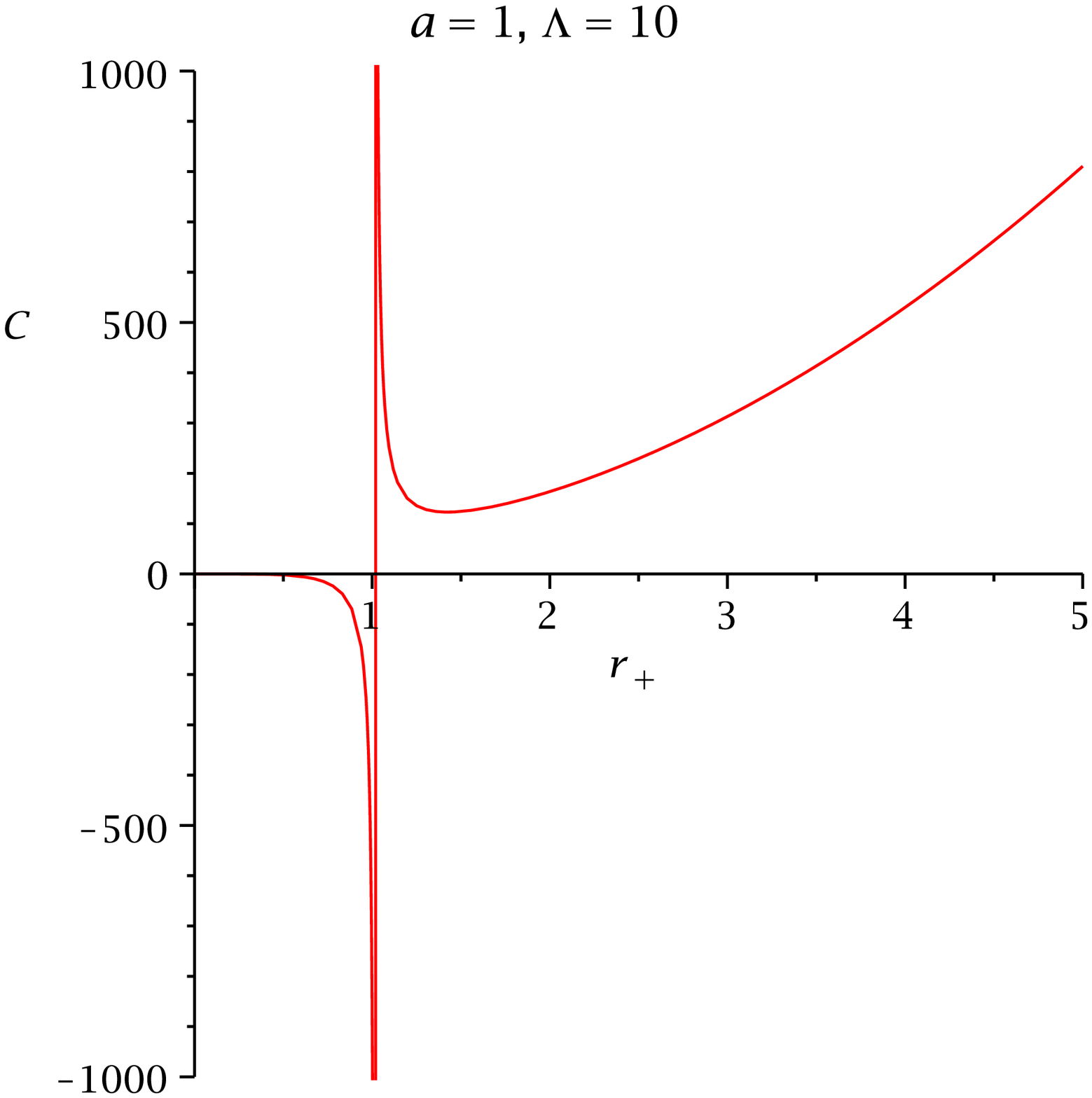}}
 \subfigure[]{
 \includegraphics[width=2.1in,angle=0]{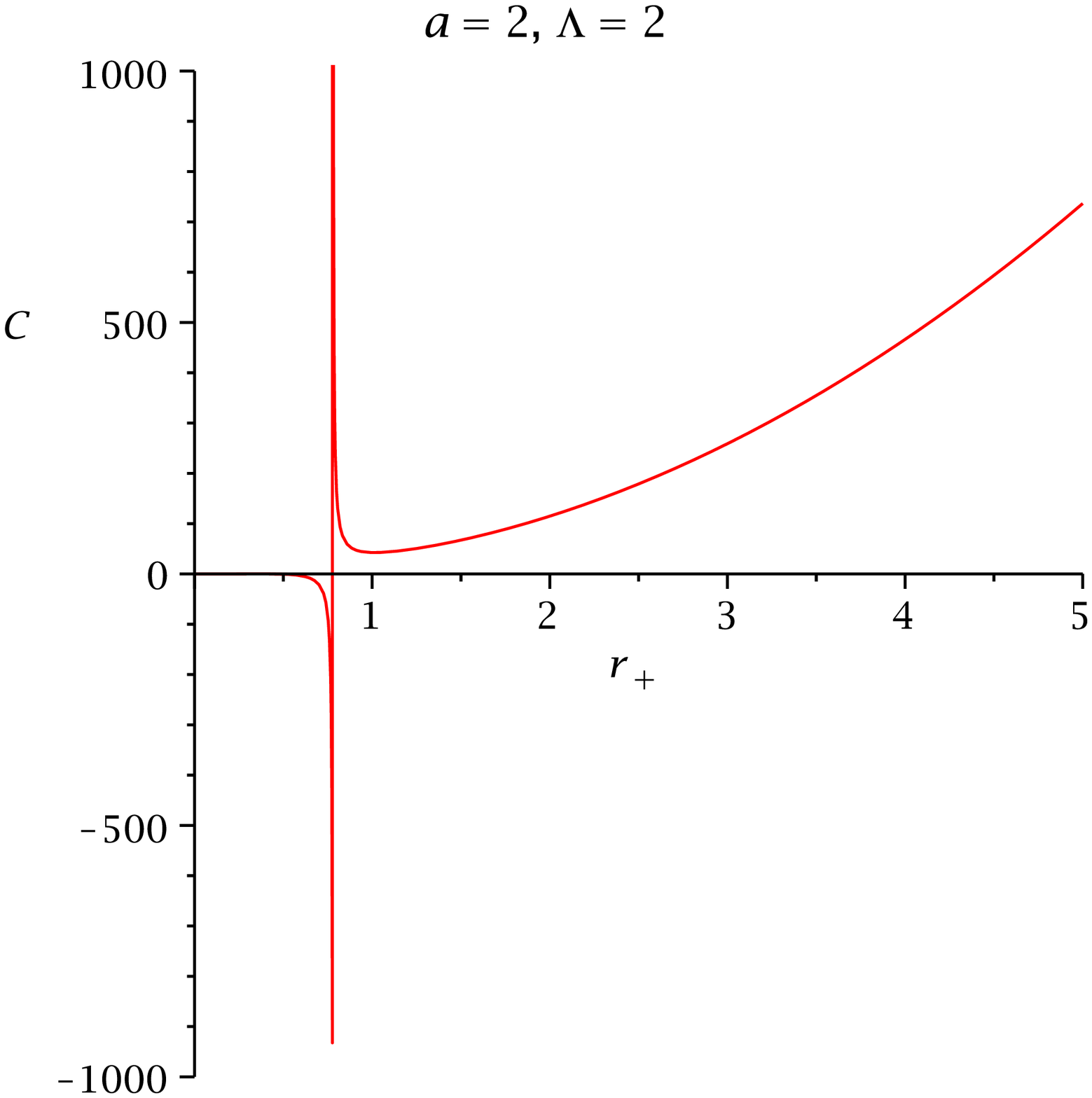}}
 \caption{\label{fg2}\textit{The figure shows the variation  of $C$  with $r_{+}$ }}
\end{center}
\end{figure}

\begin{figure}[h]
\begin{center}
\subfigure[ ]{
 \includegraphics[width=2.1in,angle=0]{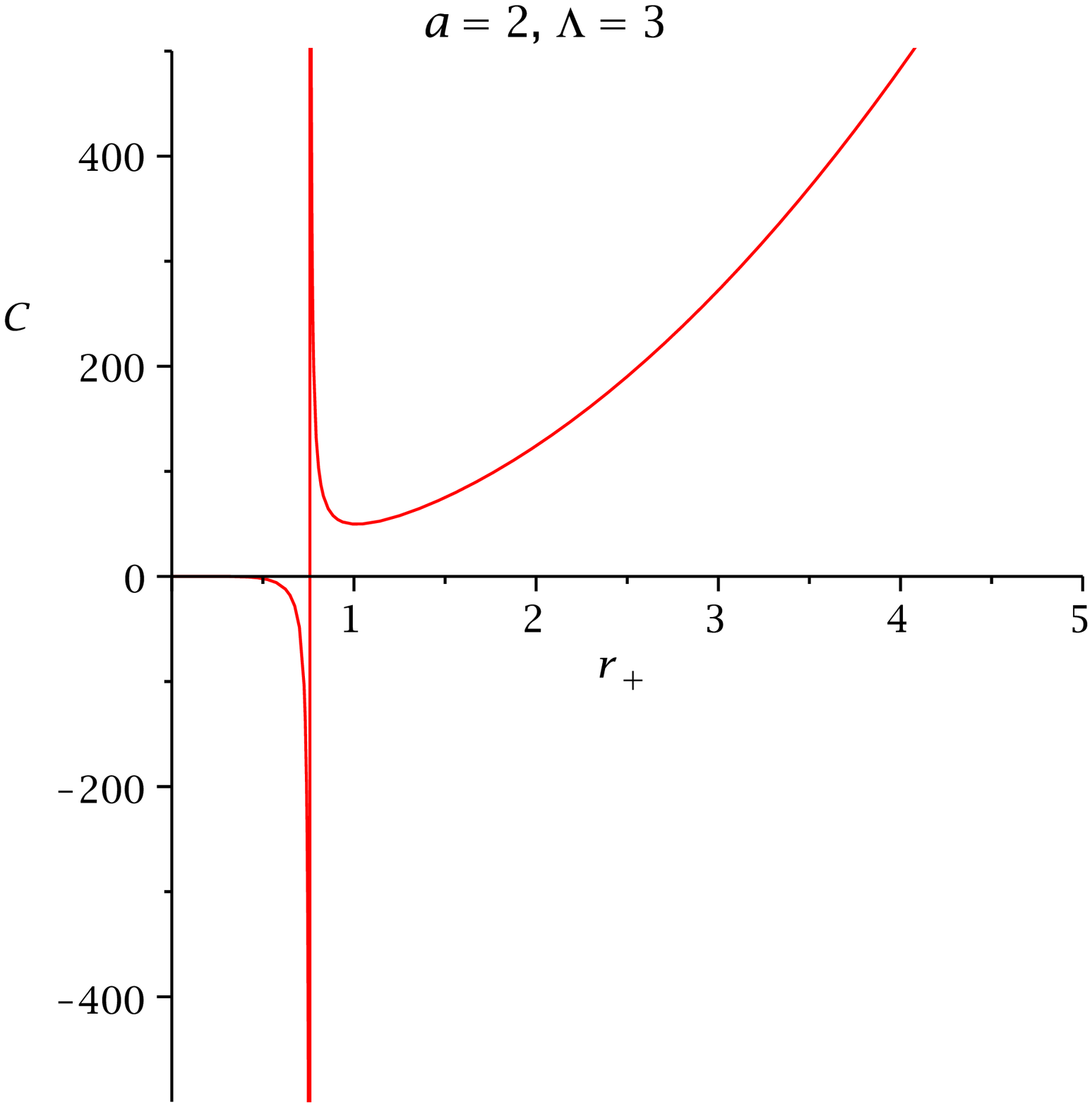}}
 \subfigure[]{
 \includegraphics[width=2.1in,angle=0]{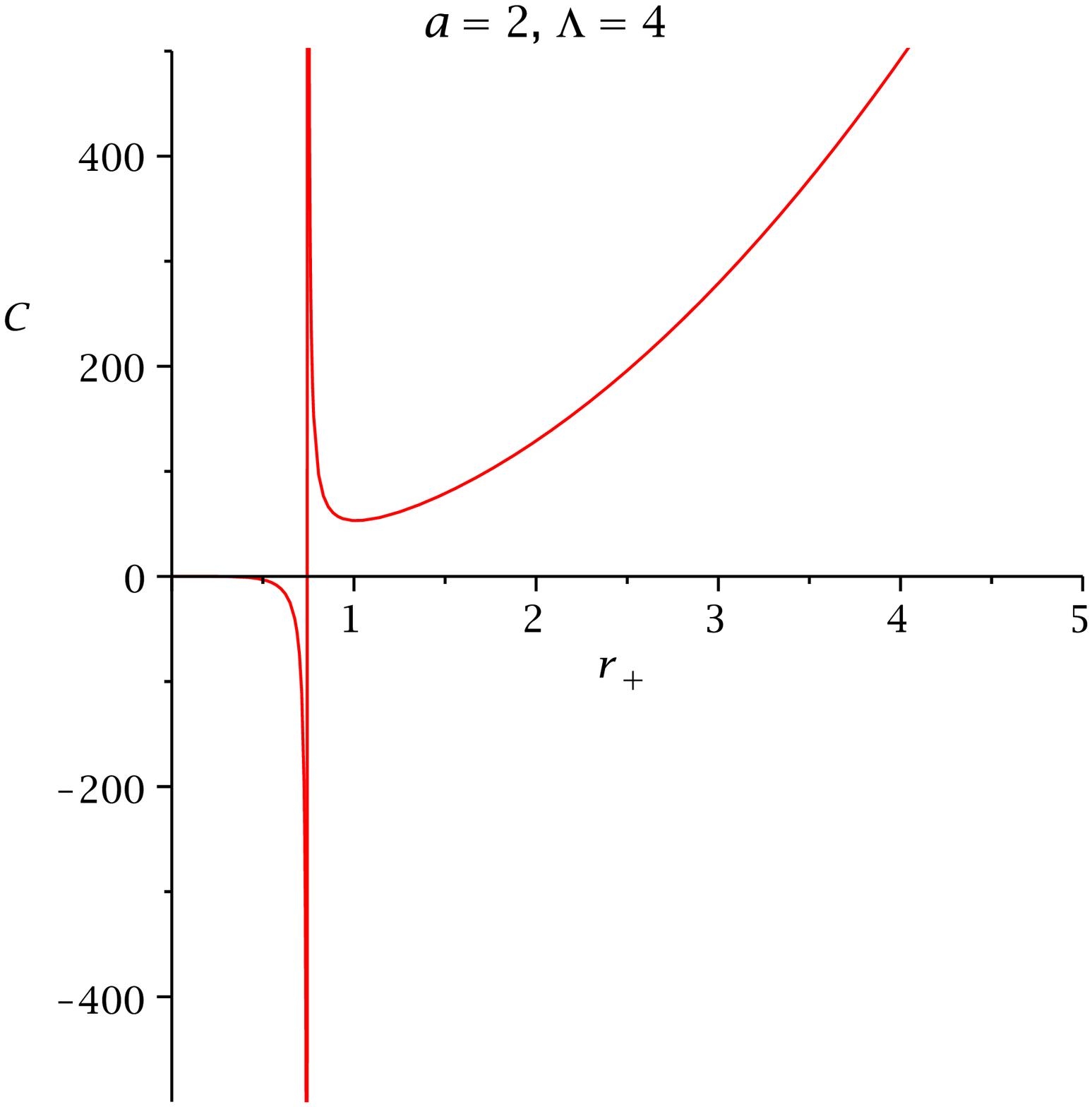}}
 \subfigure[]{
 \includegraphics[width=2.1in,angle=0]{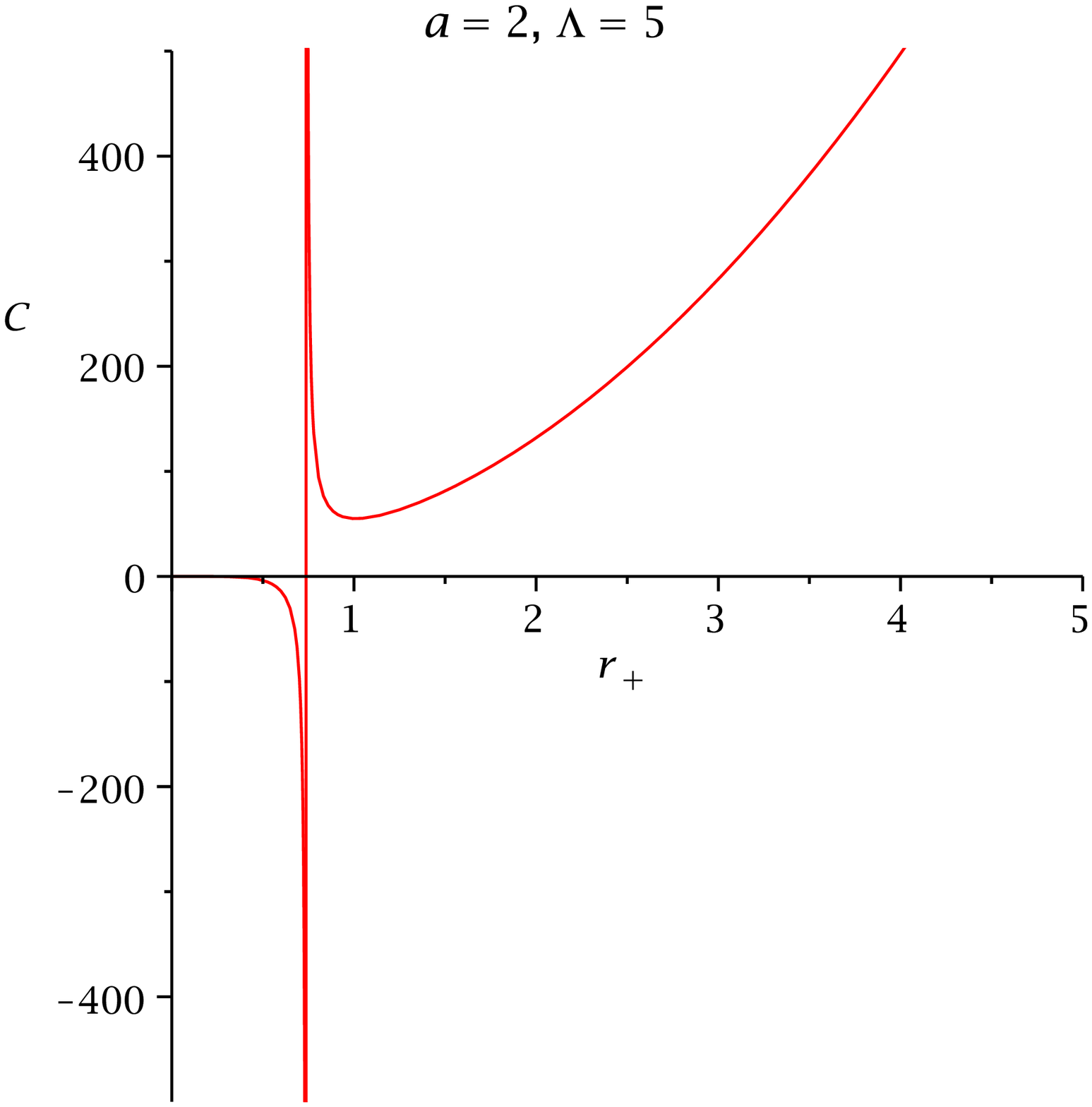}}
 \subfigure[]{
\includegraphics[width=2.1in,angle=0]{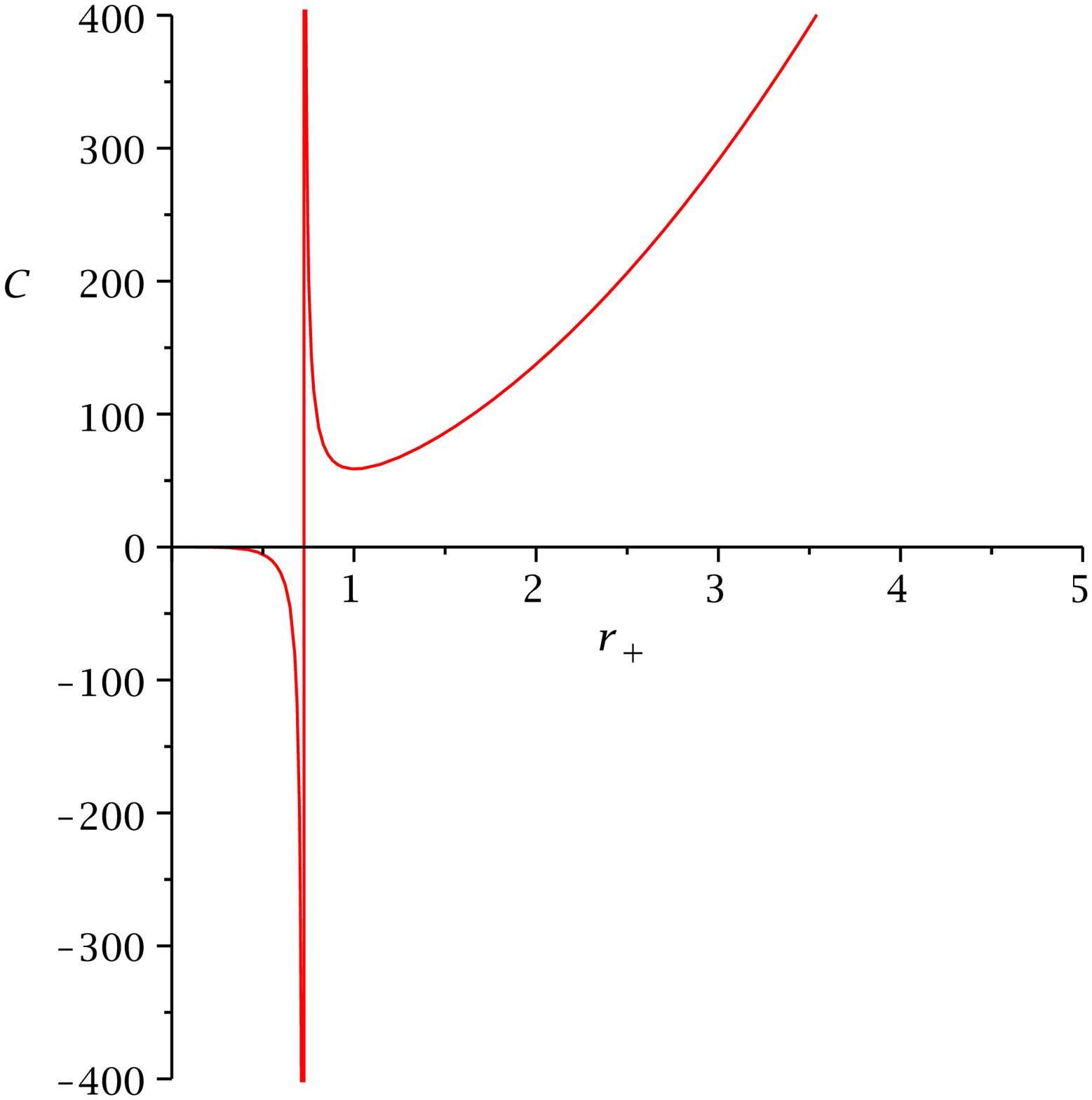}} 
\subfigure[]{
 \includegraphics[width=2.1in,angle=0]{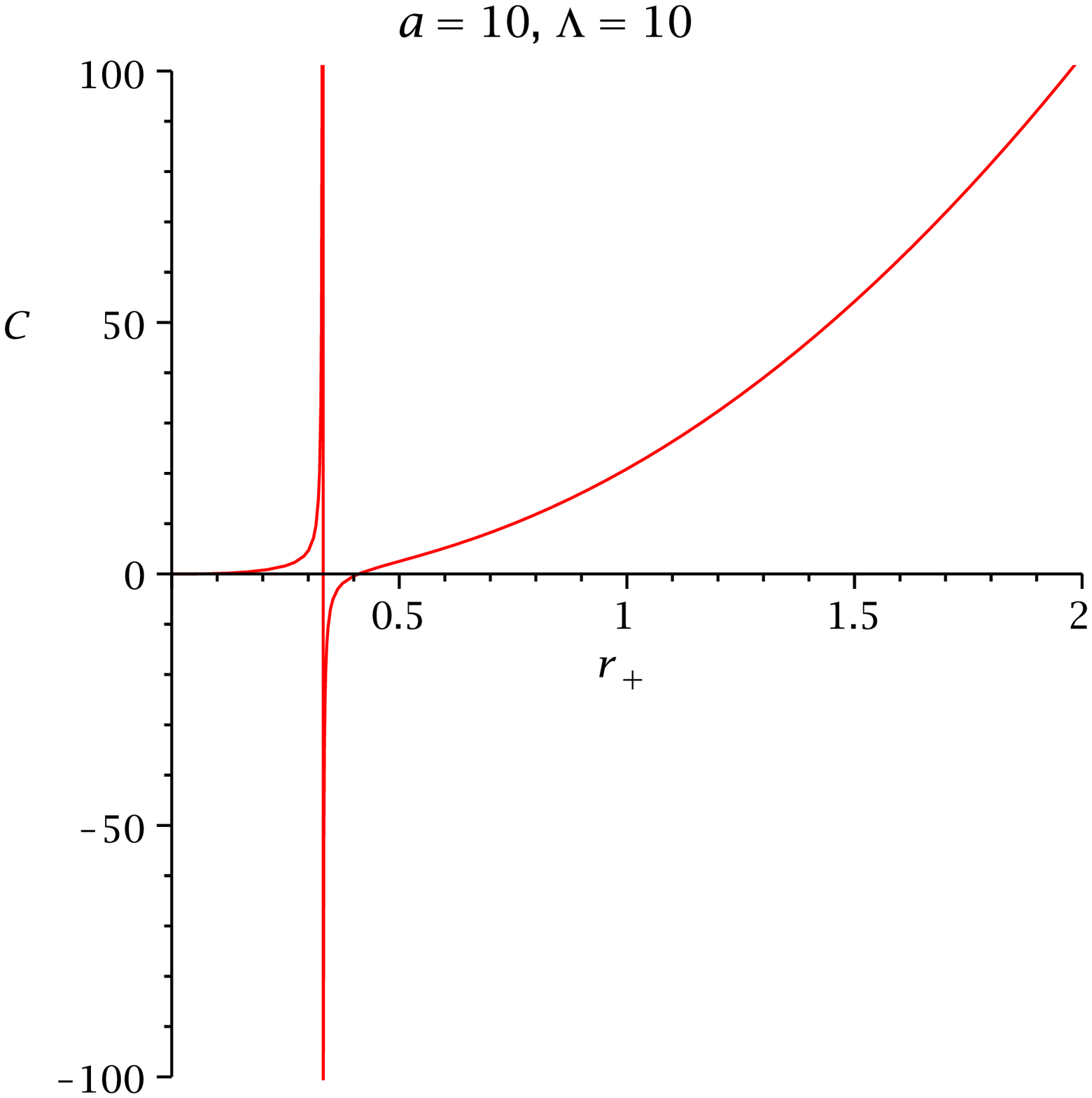}}
 \subfigure[]{
 \includegraphics[width=2.1in,angle=0]{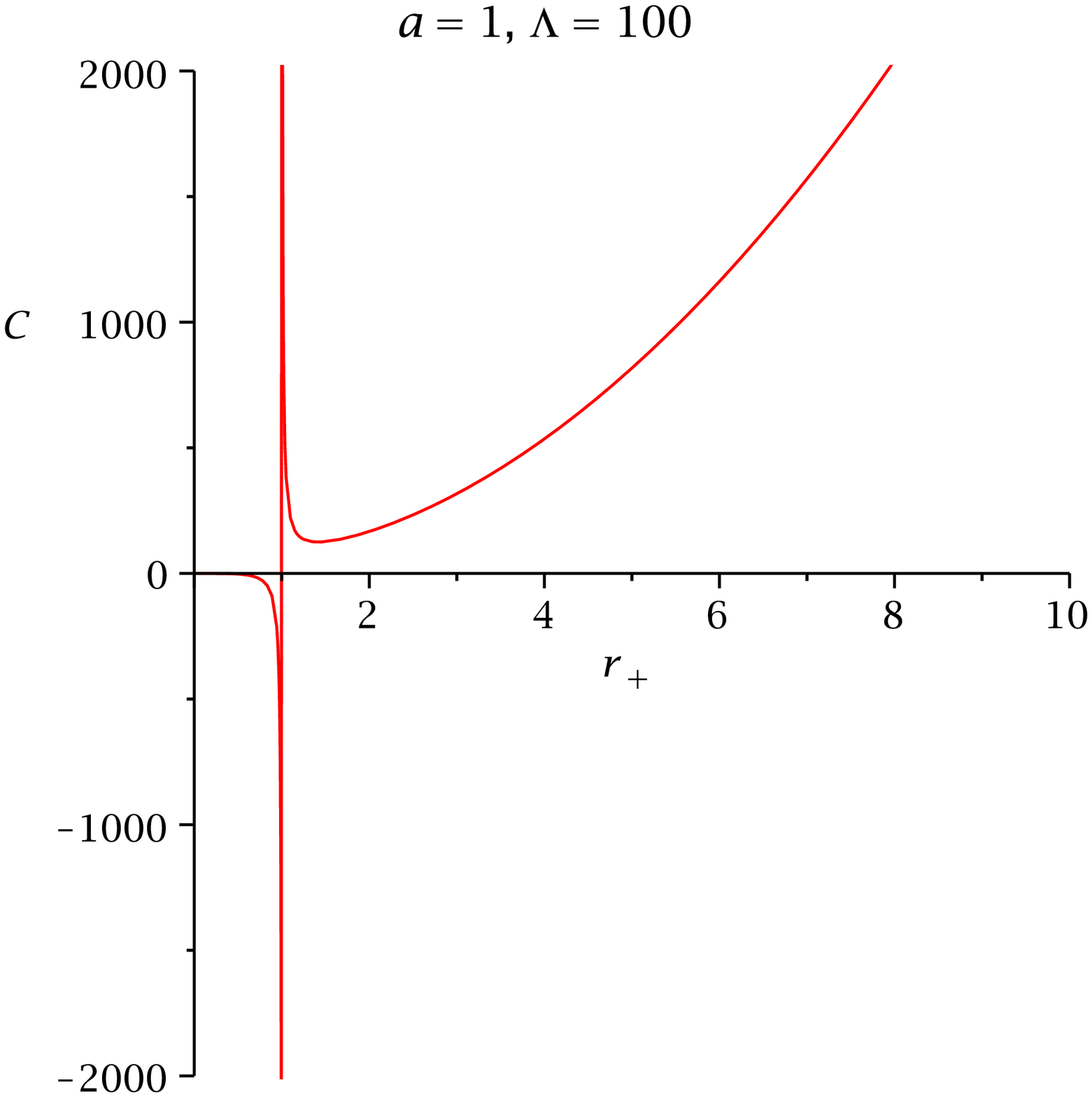}}
   \caption{\label{fg3}\textit{The figure depicts the variation  of $C$  with $r_{+}$ }}
\end{center}
\end{figure}

\begin{figure}[h]
\begin{center}
 \subfigure[ ]{
 \includegraphics[width=2.1in,angle=0]{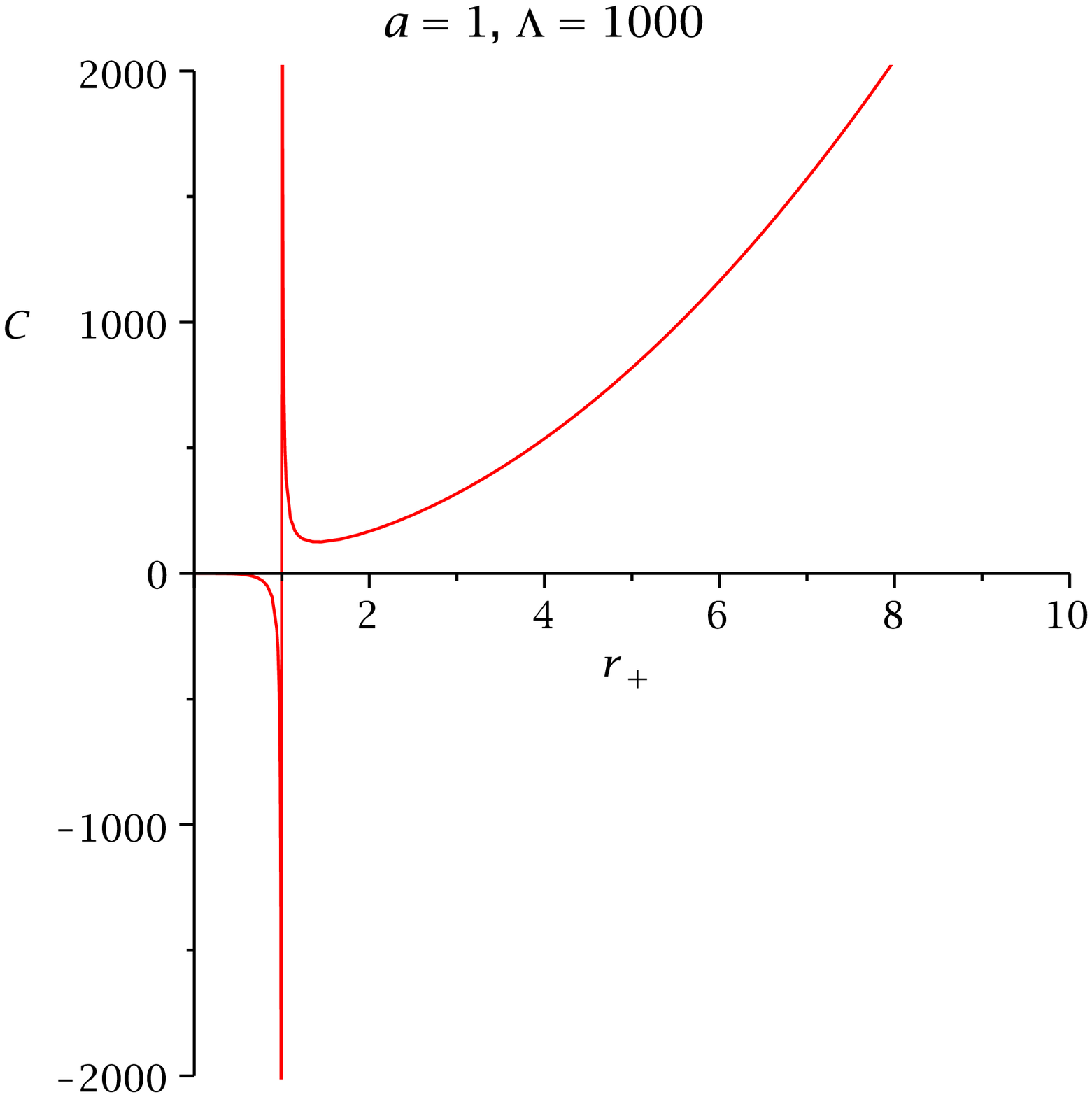}}
 \subfigure[]{
 \includegraphics[width=2.1in,angle=0]{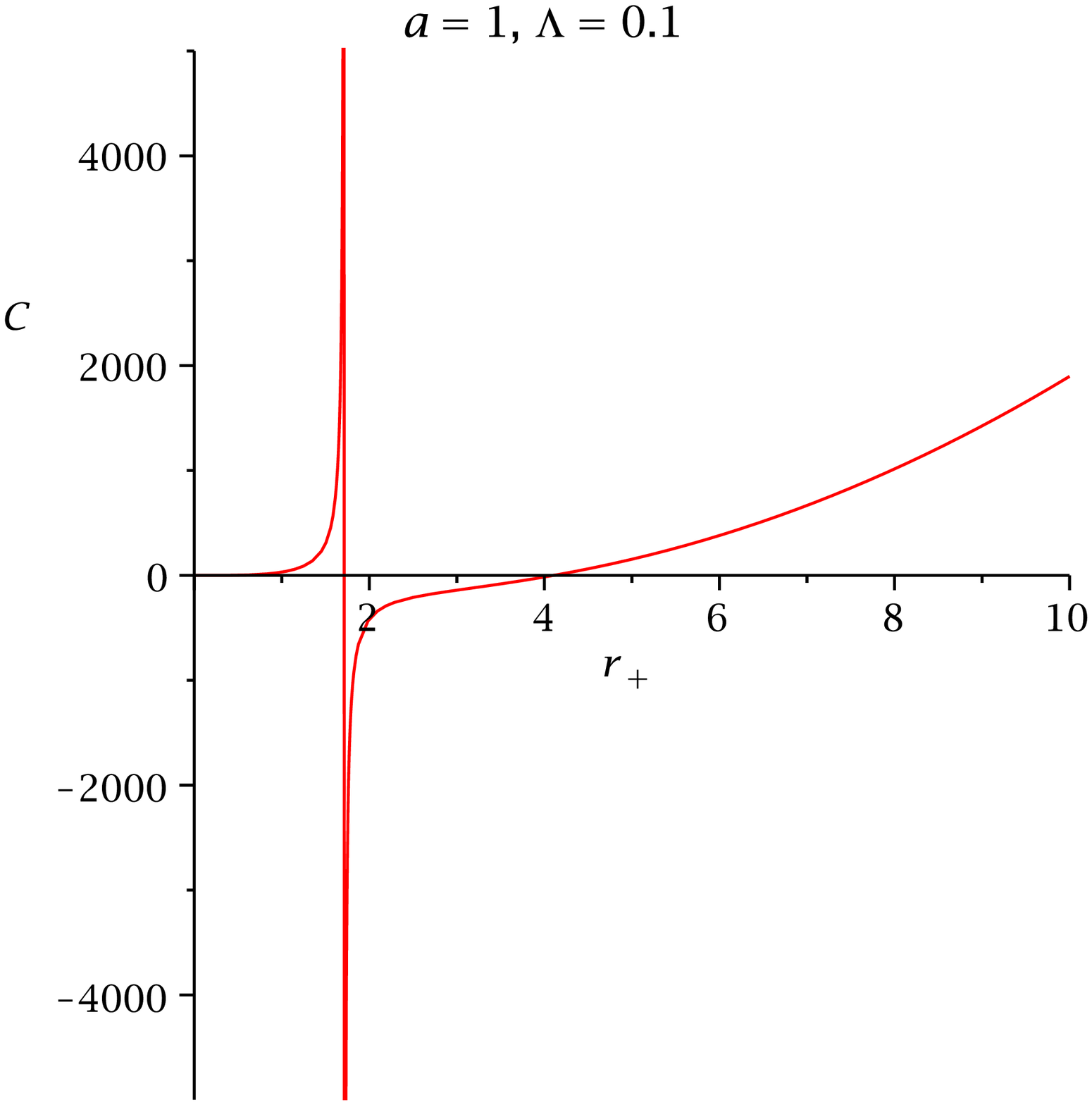}}
 \subfigure[]{
 \includegraphics[width=2.1in,angle=0]{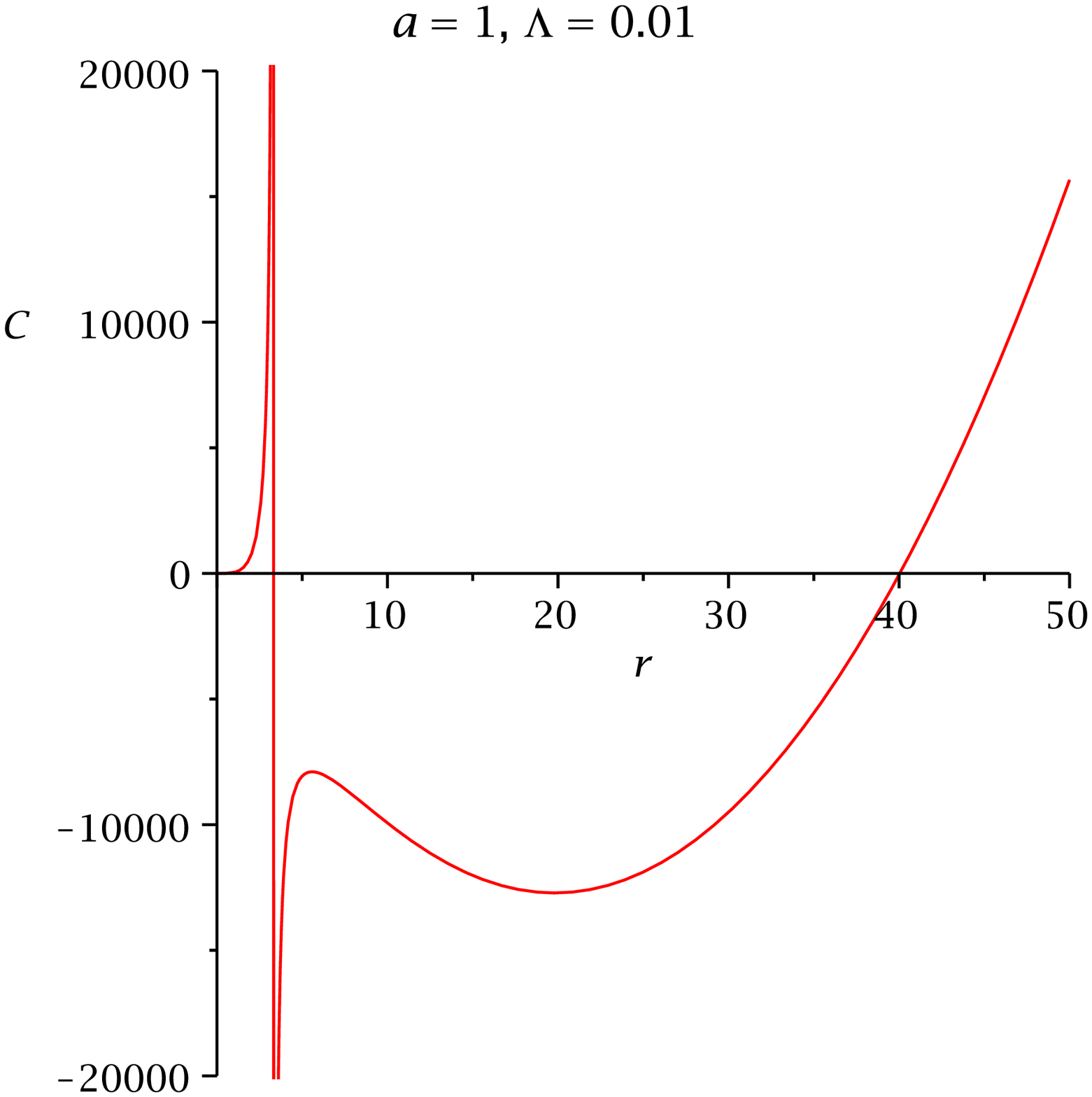}}
 \subfigure[]{
 \includegraphics[width=2.1in,angle=0]{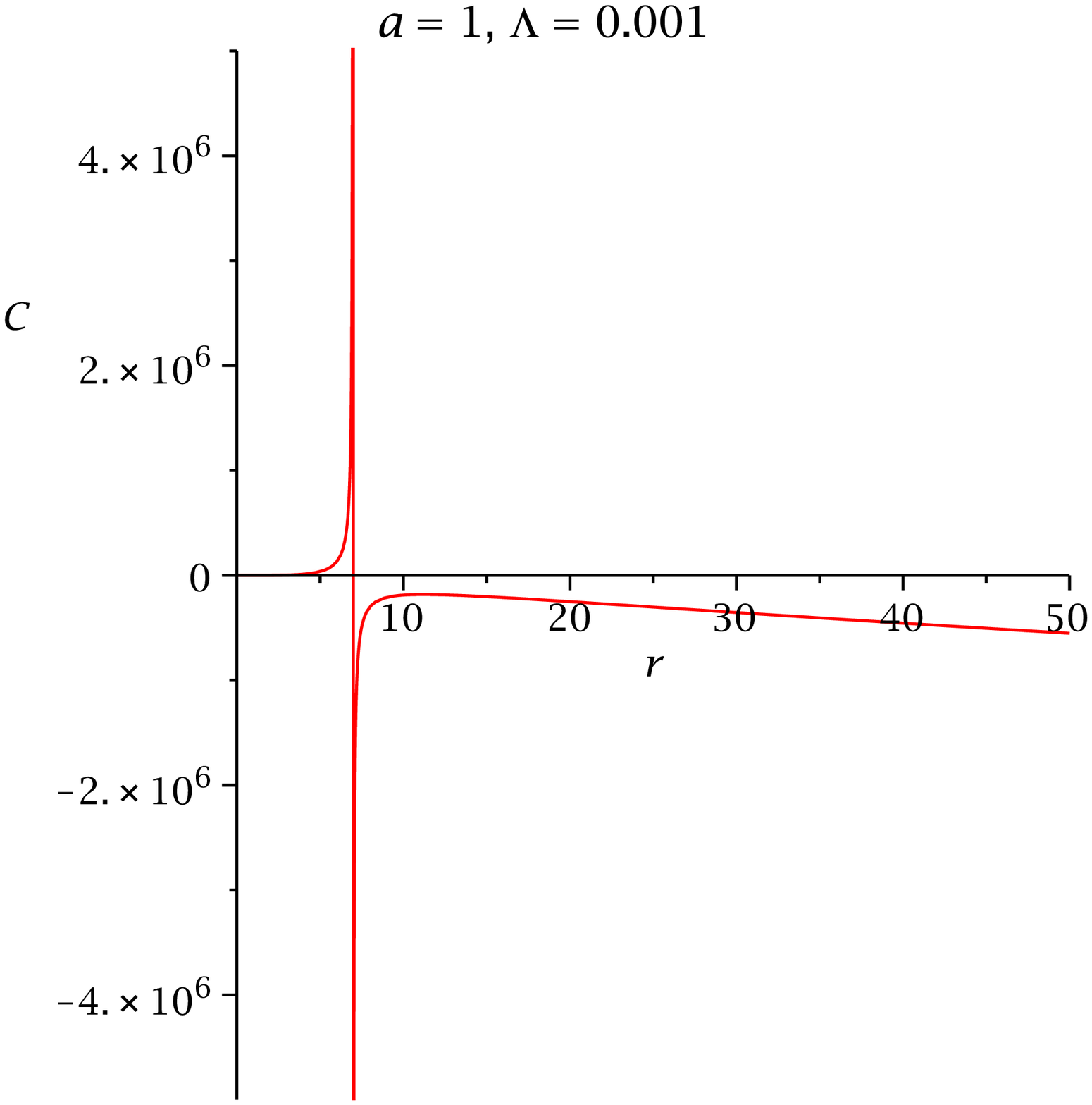}}
 \subfigure[]{
 \includegraphics[width=2.1in,angle=0]{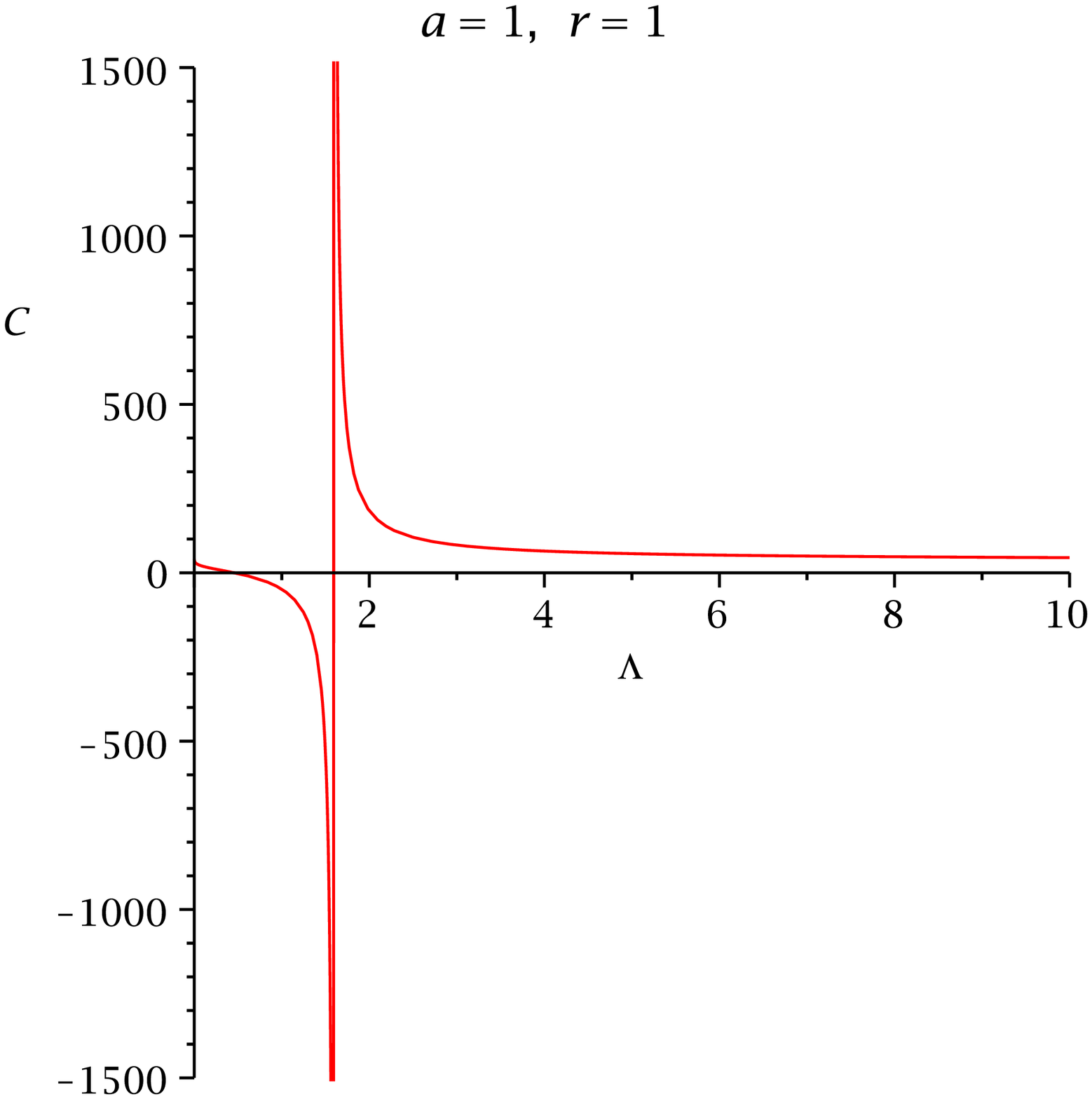}}
\subfigure[]{
\includegraphics[width=2.1in,angle=0]{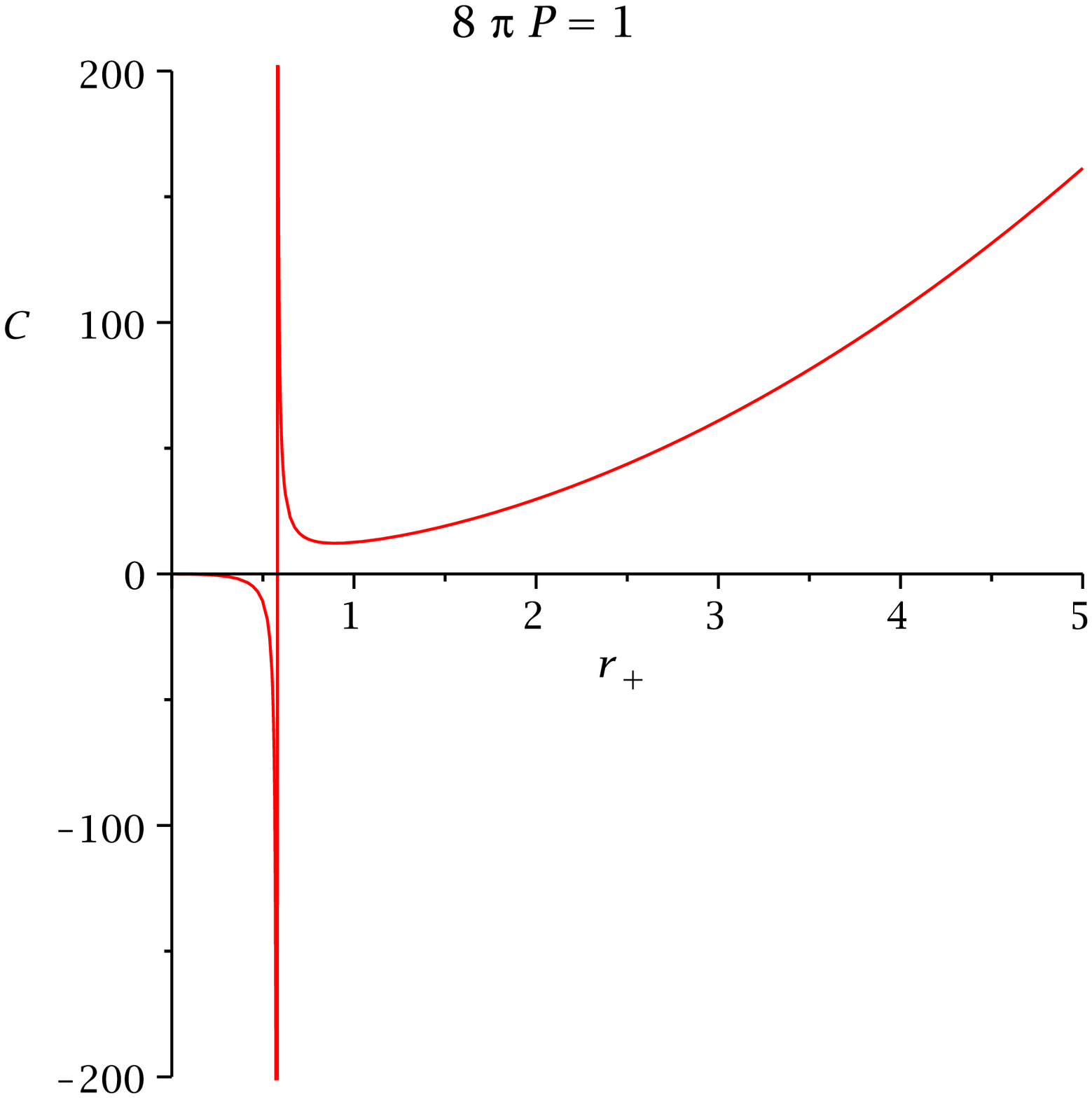}} 
  \caption{\label{f3}\textit{The figure depicts the variation  of $C$  with $r_{+}$ }}
\end{center}
\end{figure}

\begin{figure}[h]
\begin{center}
\subfigure[]{
\includegraphics[width=2.1in,angle=0]{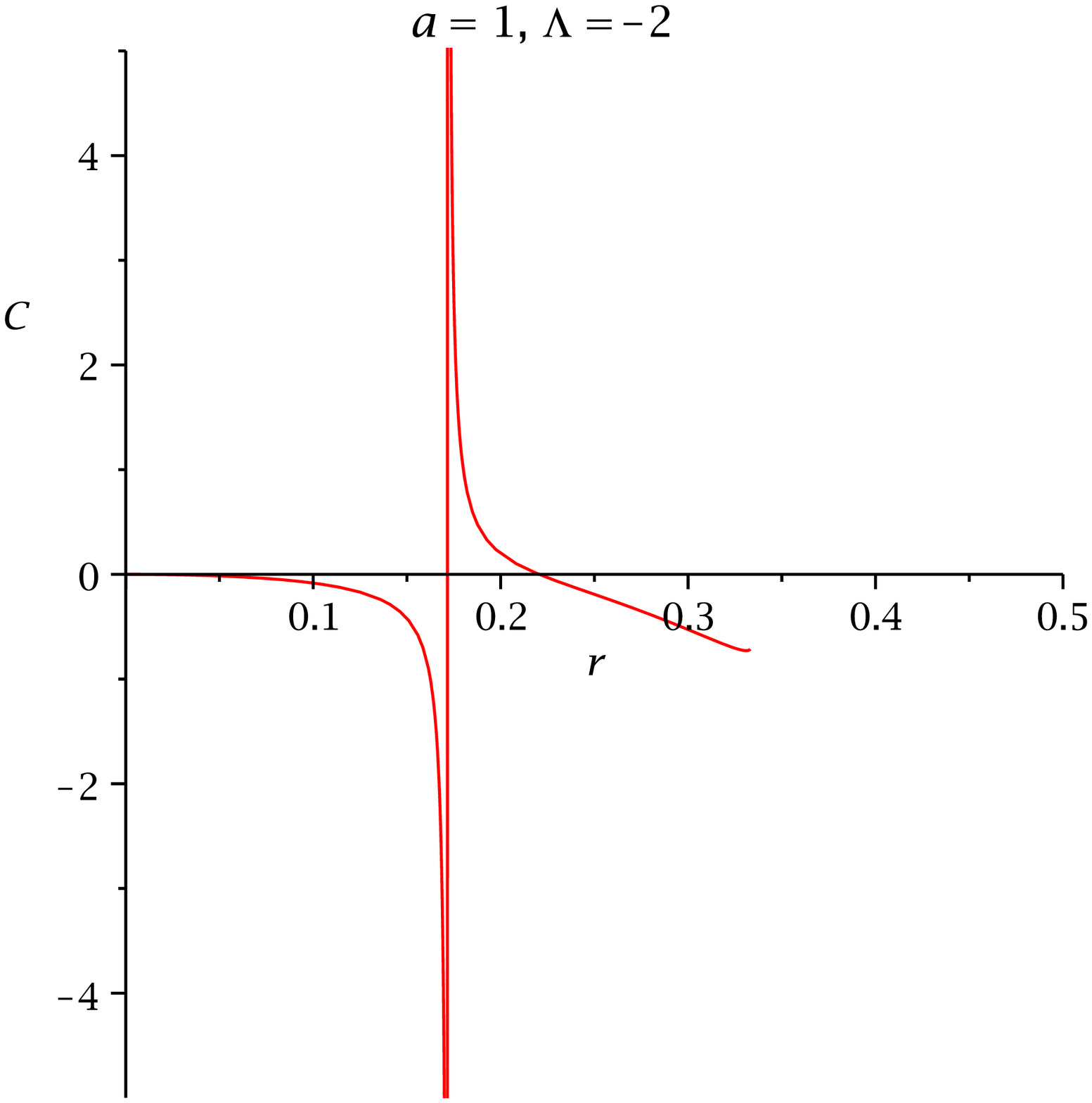}} 
\subfigure[]{
 \includegraphics[width=2.1in,angle=0]{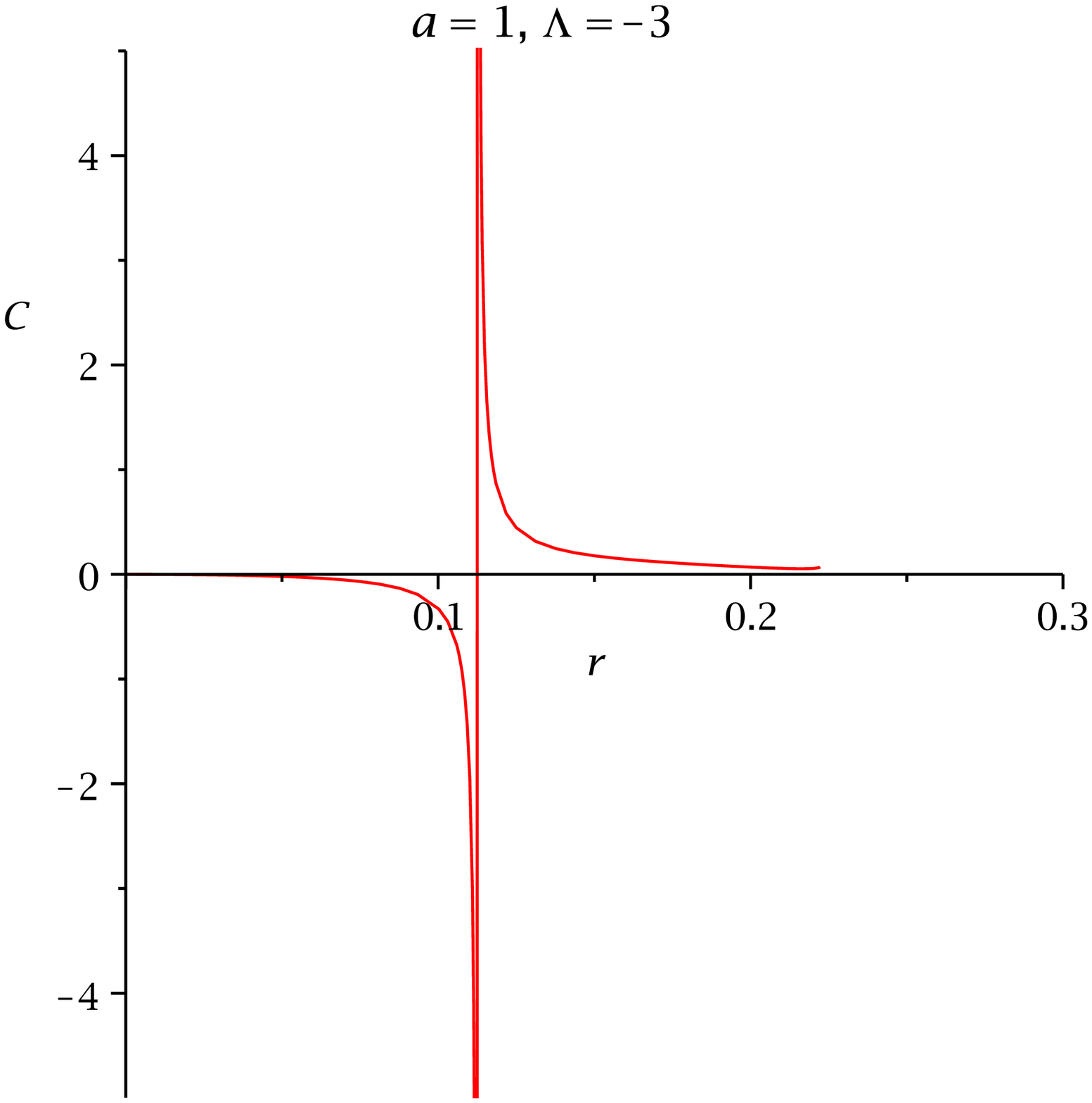}}
 \subfigure[]{
 \includegraphics[width=2.1in,angle=0]{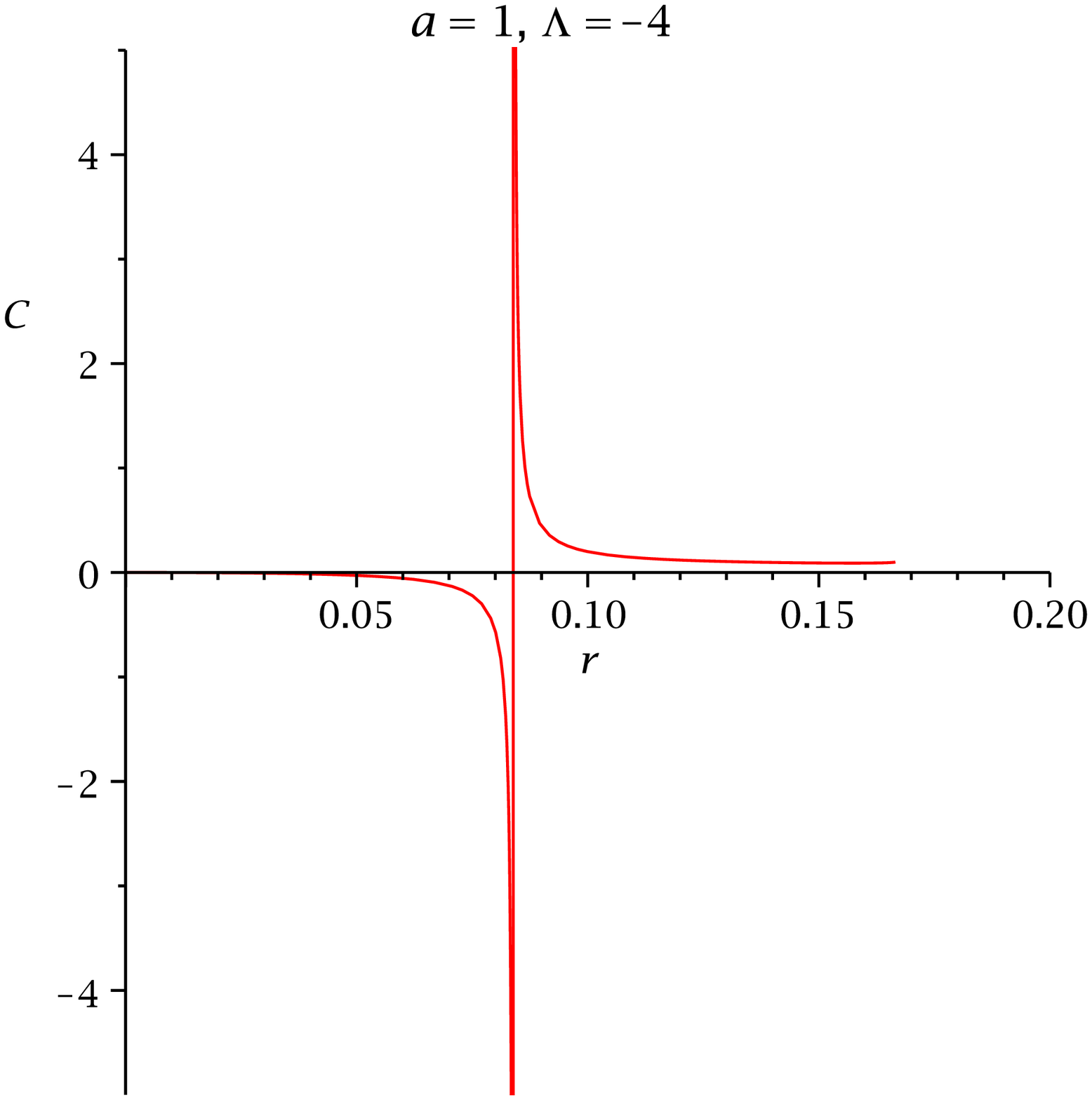}}
 \subfigure[ ]{
 \includegraphics[width=2.1in,angle=0]{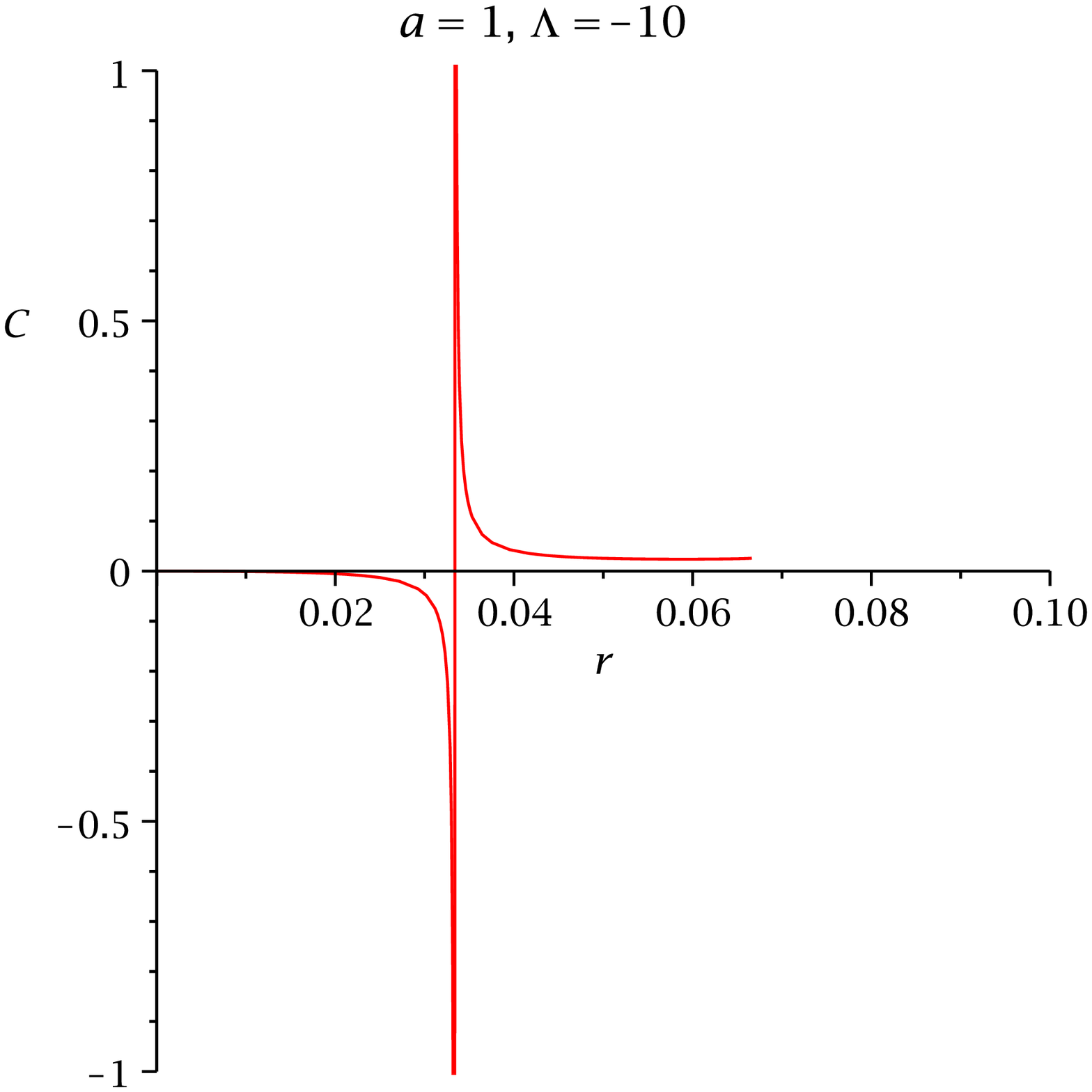}}
  \subfigure[]{
 \includegraphics[width=2.1in,angle=0]{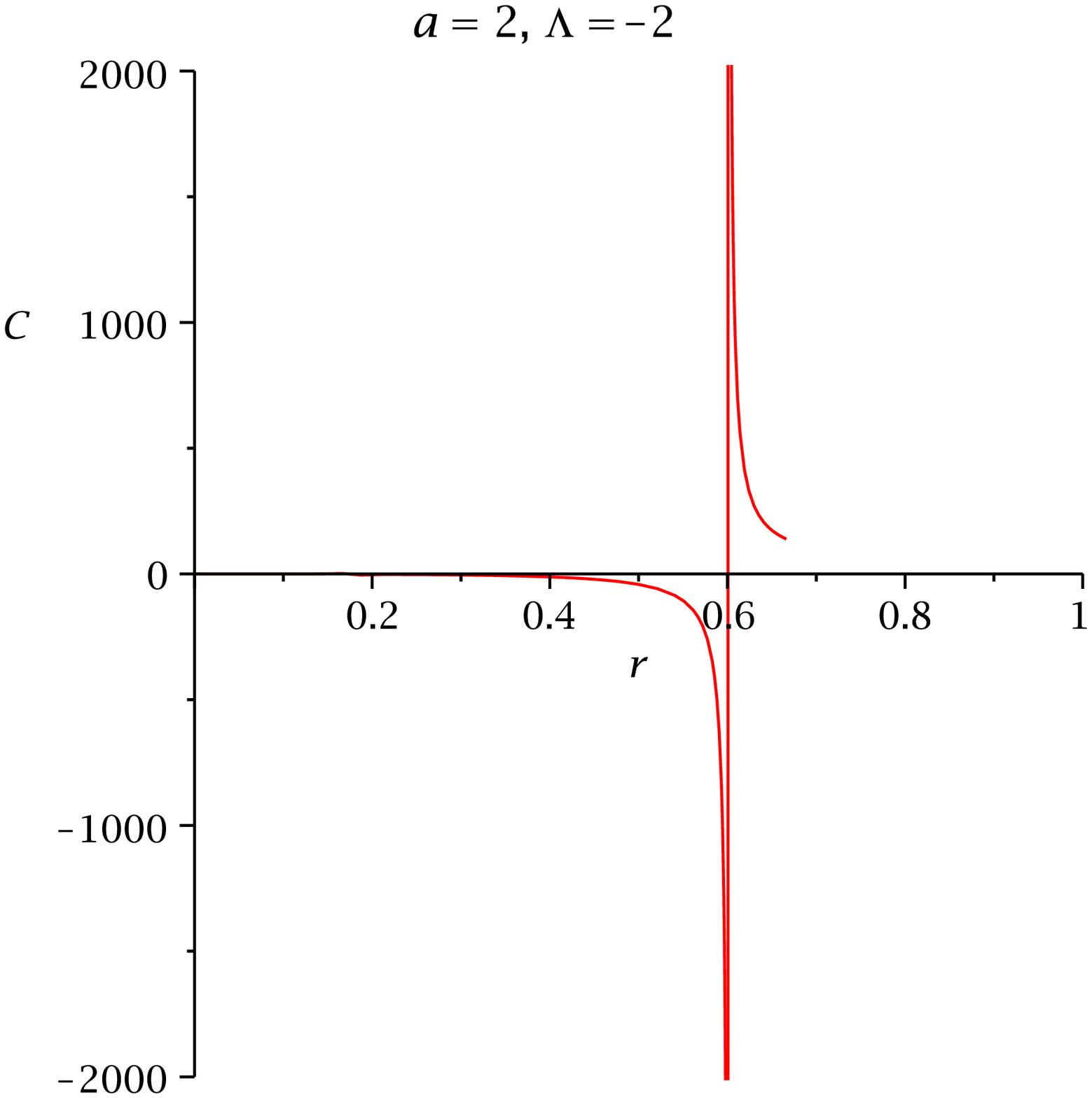}}
 \subfigure[ ]{
 \includegraphics[width=2.1in,angle=0]{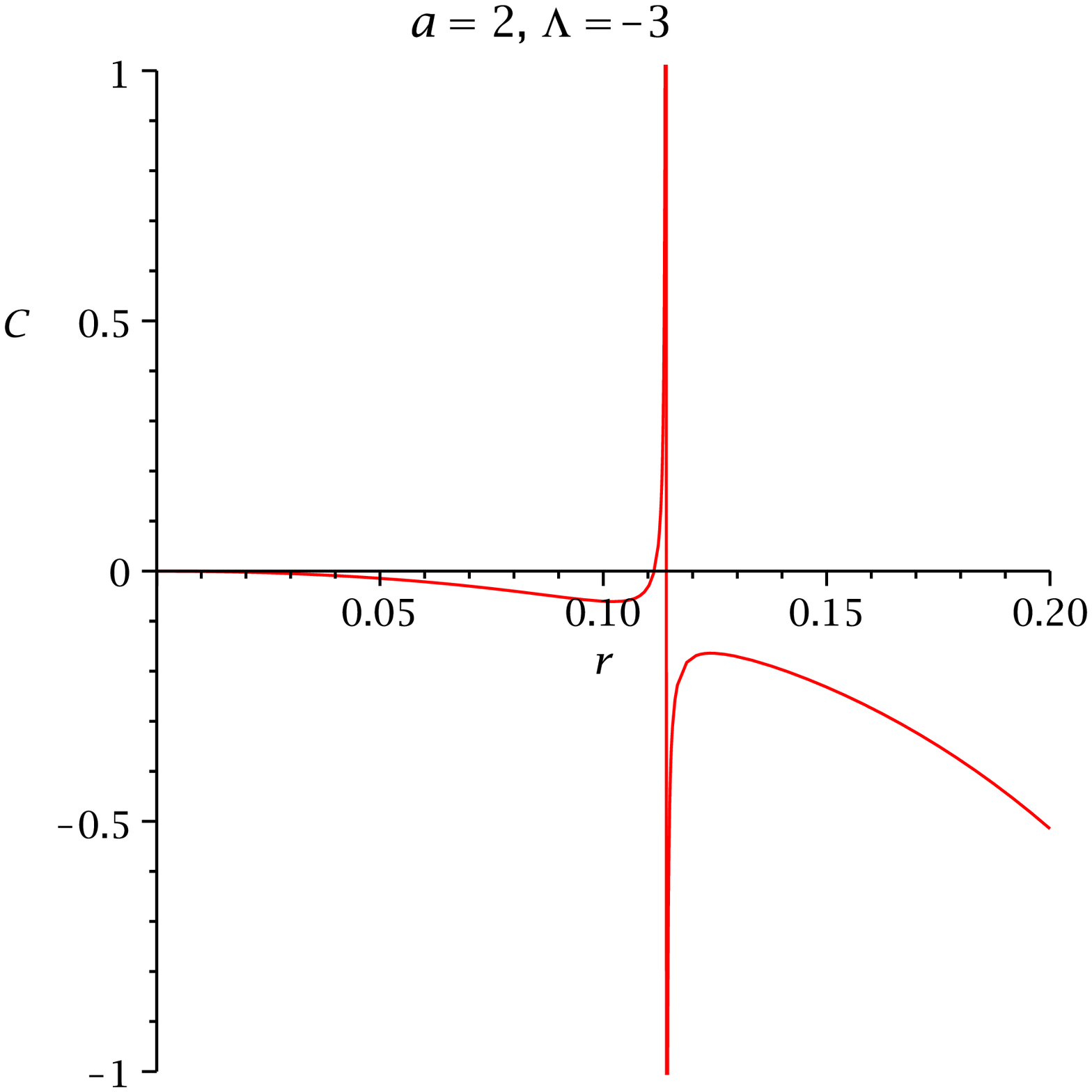}}

  \caption{\label{fg4}\textit{The figure depicts the variation  of $C$  with $r_{+}$ }}
\end{center}
\end{figure}

\begin{figure}[h]
\begin{center}
 \subfigure[]{
 \includegraphics[width=2.1in,angle=0]{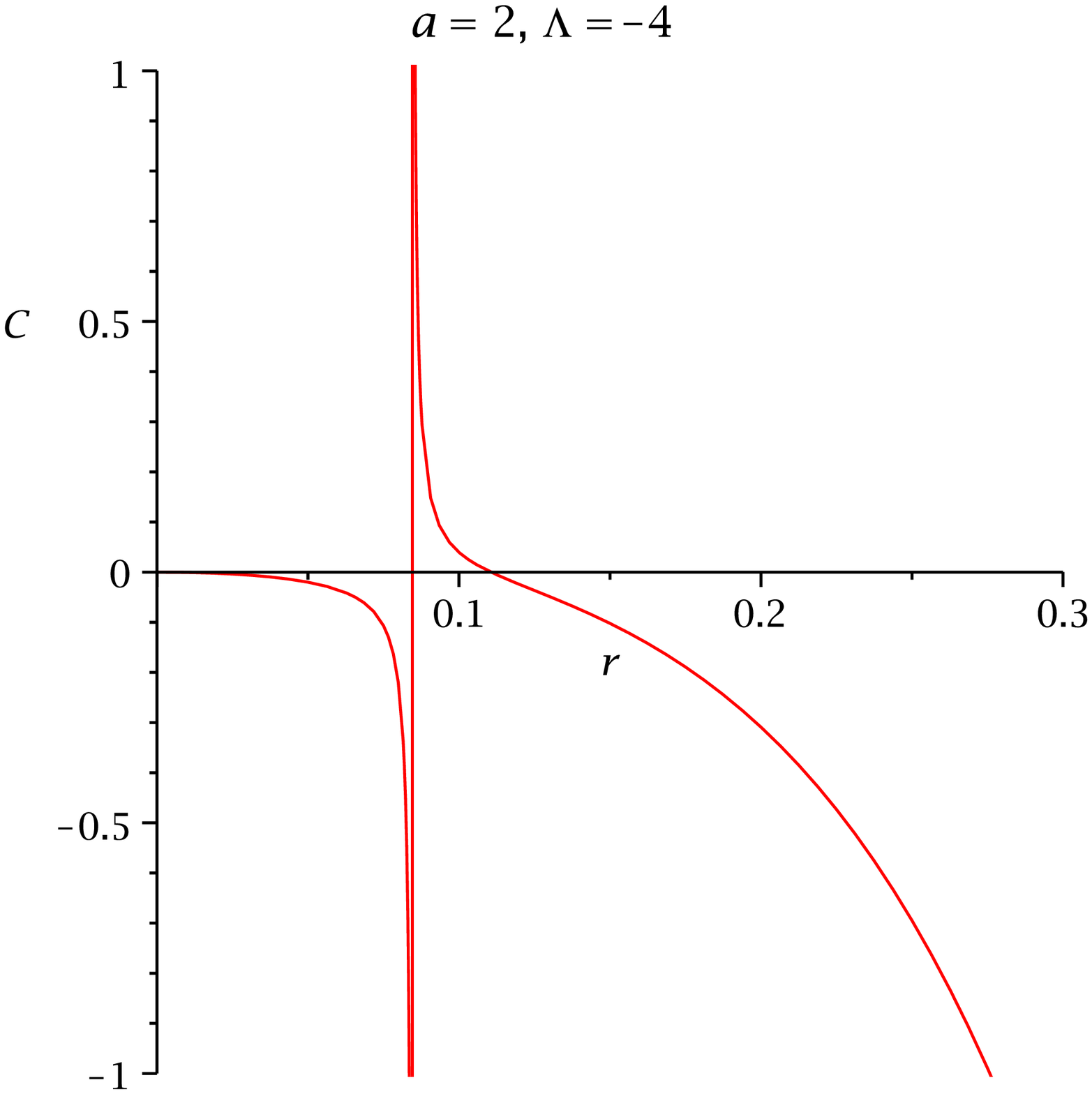}}
 \subfigure[]{
 \includegraphics[width=2.1in,angle=0]{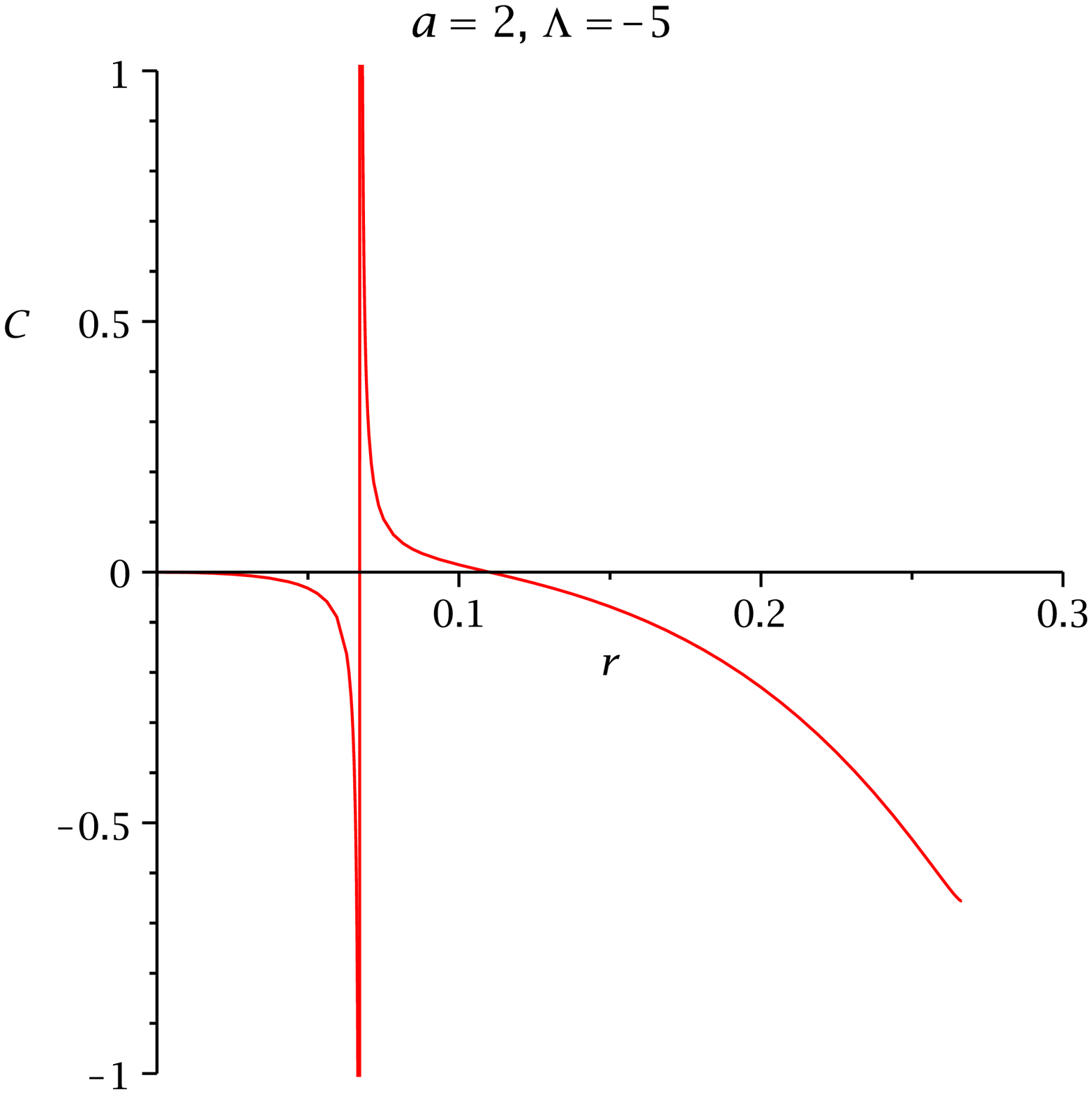}}
\subfigure[]{
\includegraphics[width=2.1in,angle=0]{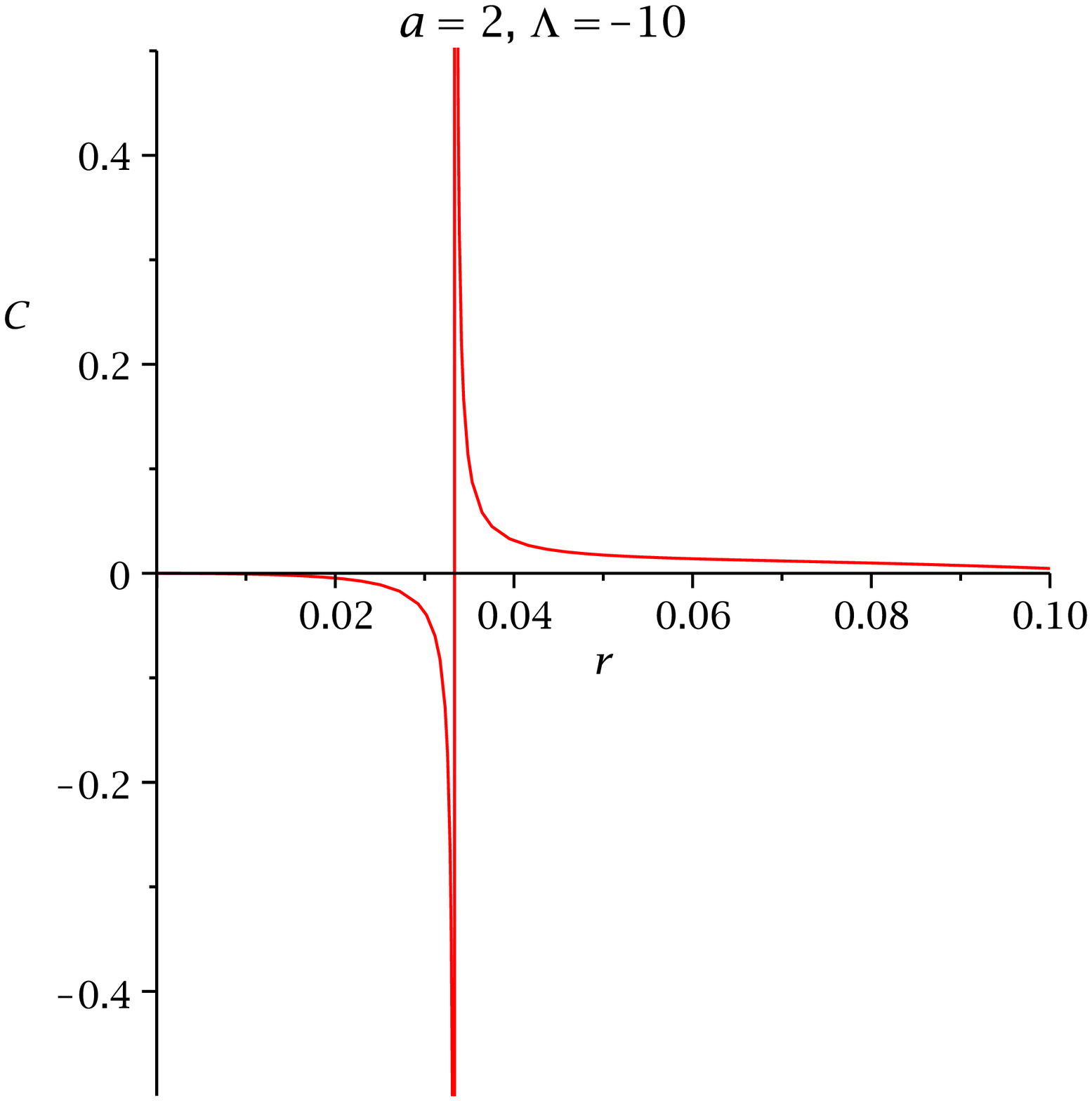}} 
\subfigure[]{
 \includegraphics[width=2.1in,angle=0]{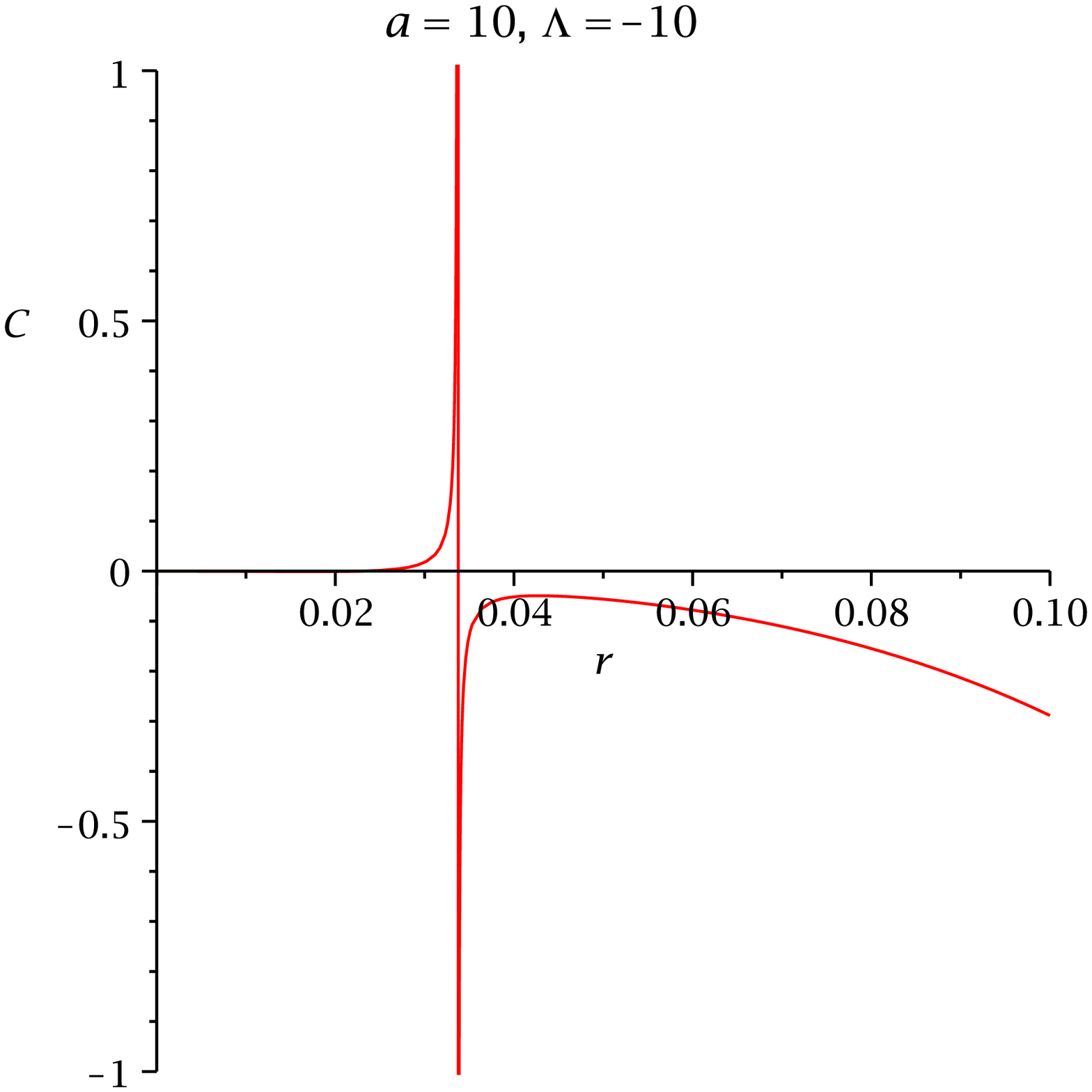}}
 \subfigure[]{
 \includegraphics[width=2.1in,angle=0]{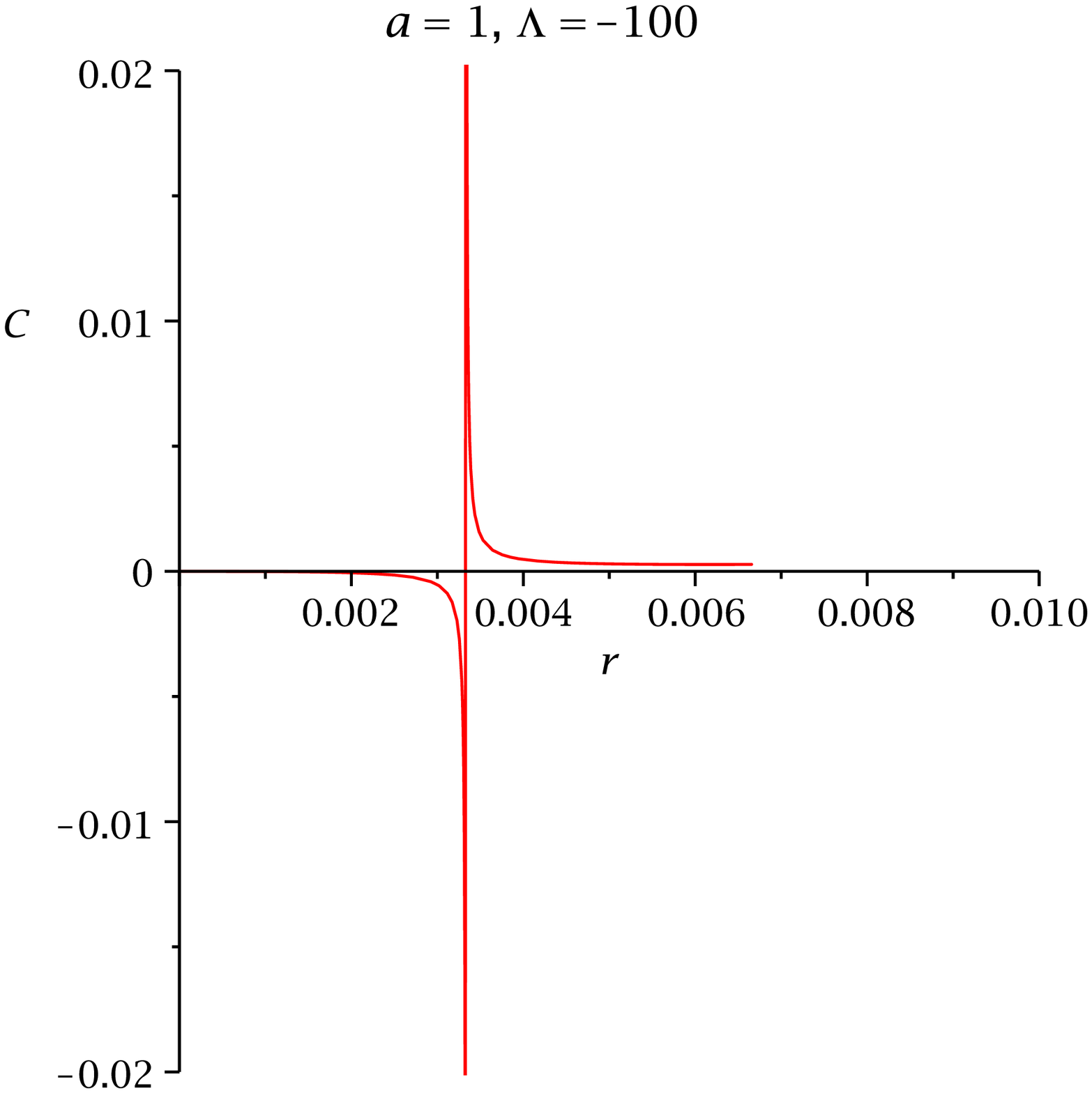}}
 \subfigure[ ]{
 \includegraphics[width=2.1in,angle=0]{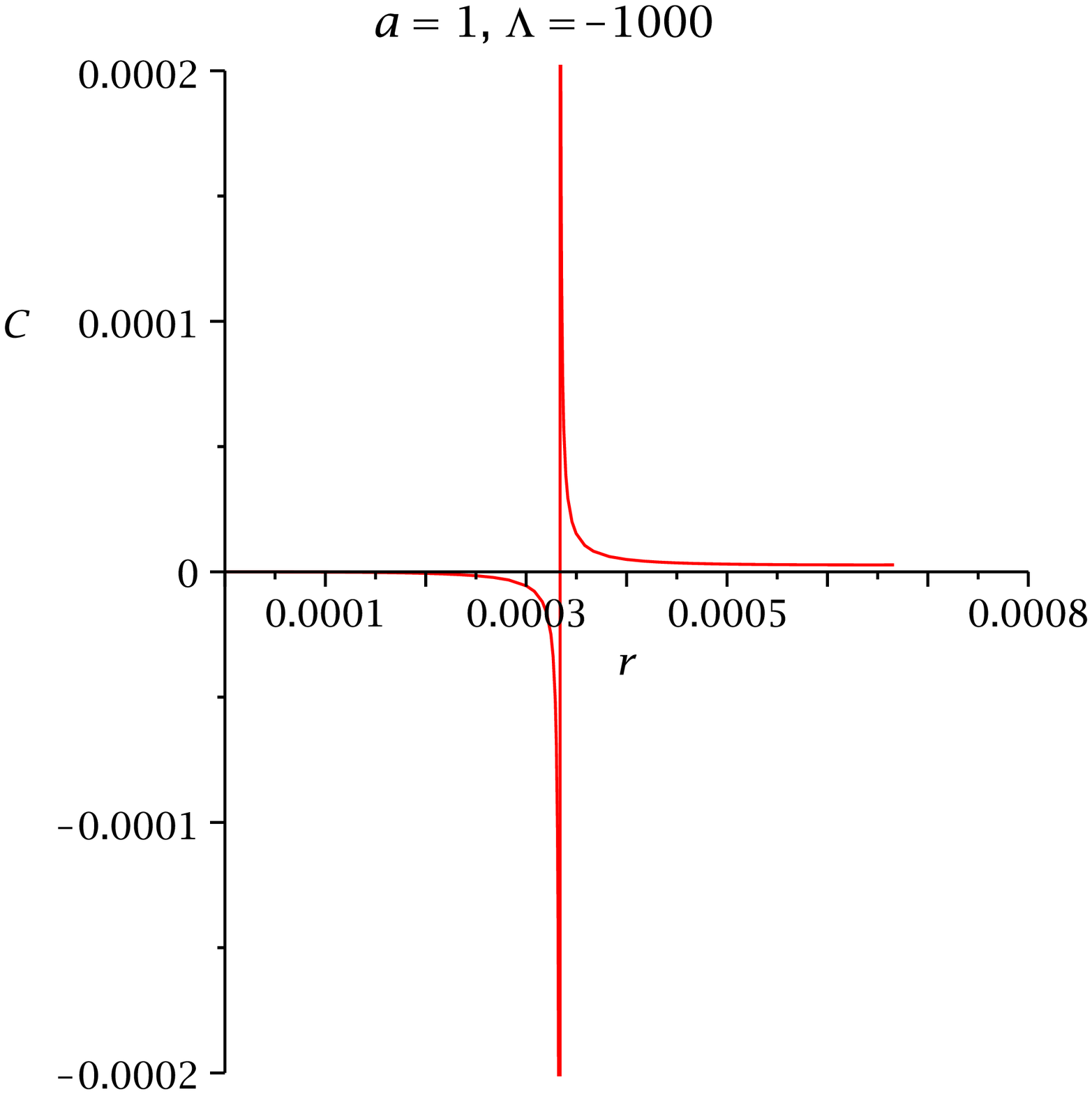}}
  
 \caption{\label{fg5}\textit{The figure depicts the variation  of $C$  with $r_{+}$ }}
\end{center}
\end{figure}

\begin{figure}[h]
\begin{center}
\subfigure[]{
 \includegraphics[width=2.1in,angle=0]{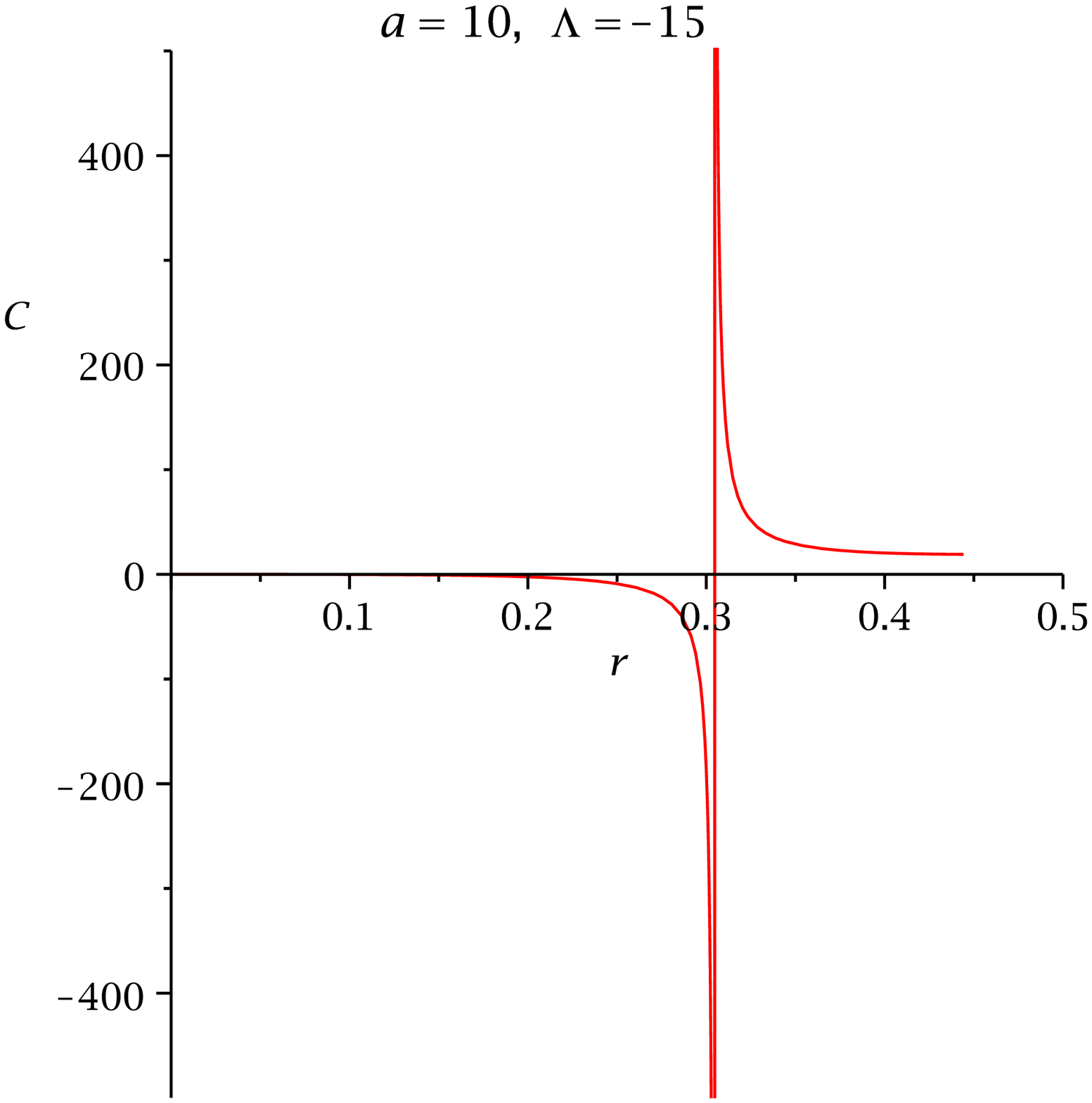}}
 \subfigure[]{
\includegraphics[width=2.1in,angle=0]{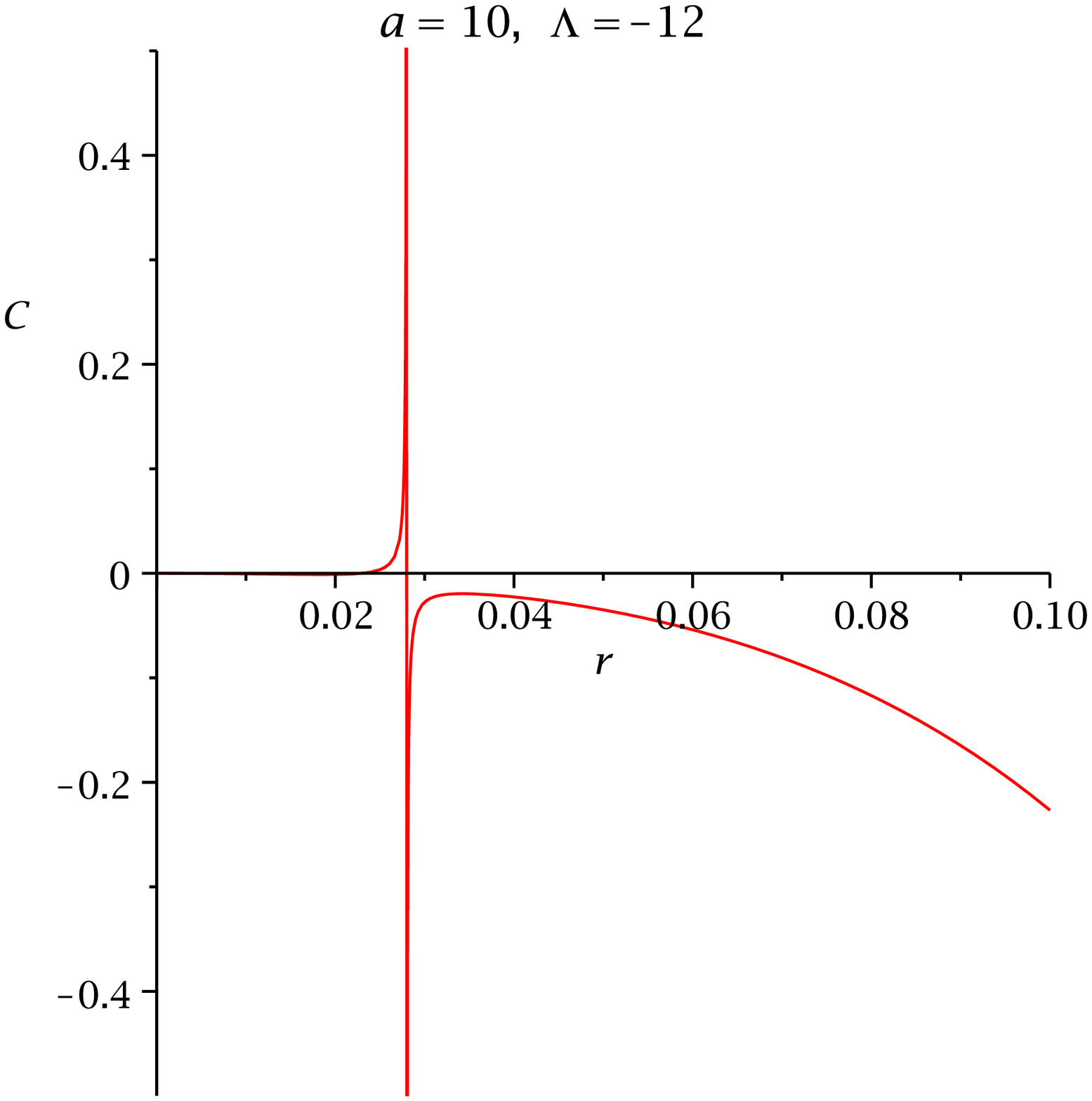}} 
\subfigure[]{
 \includegraphics[width=2.1in,angle=0]{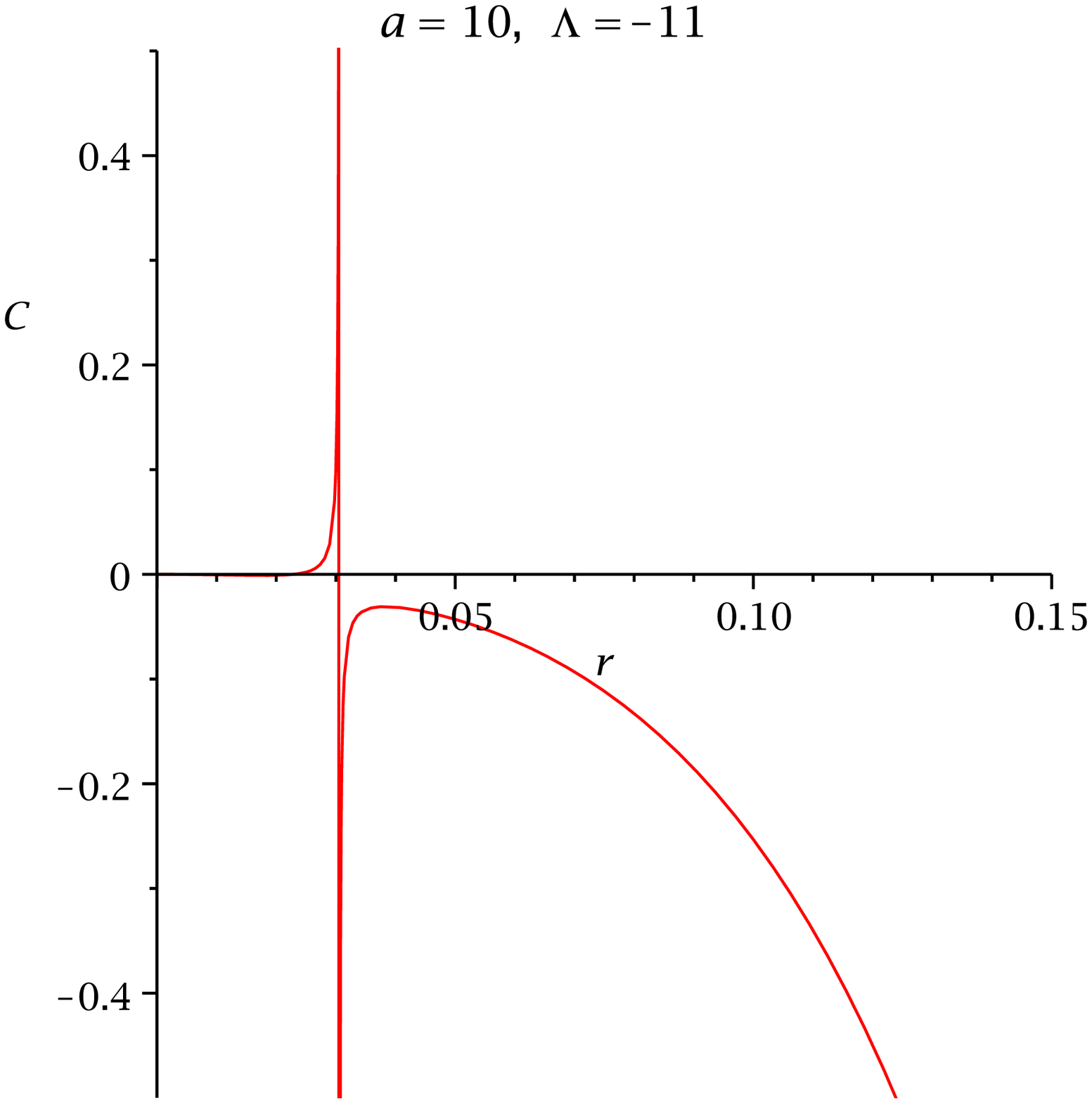}}
 \subfigure[]{
 \includegraphics[width=2.1in,angle=0]{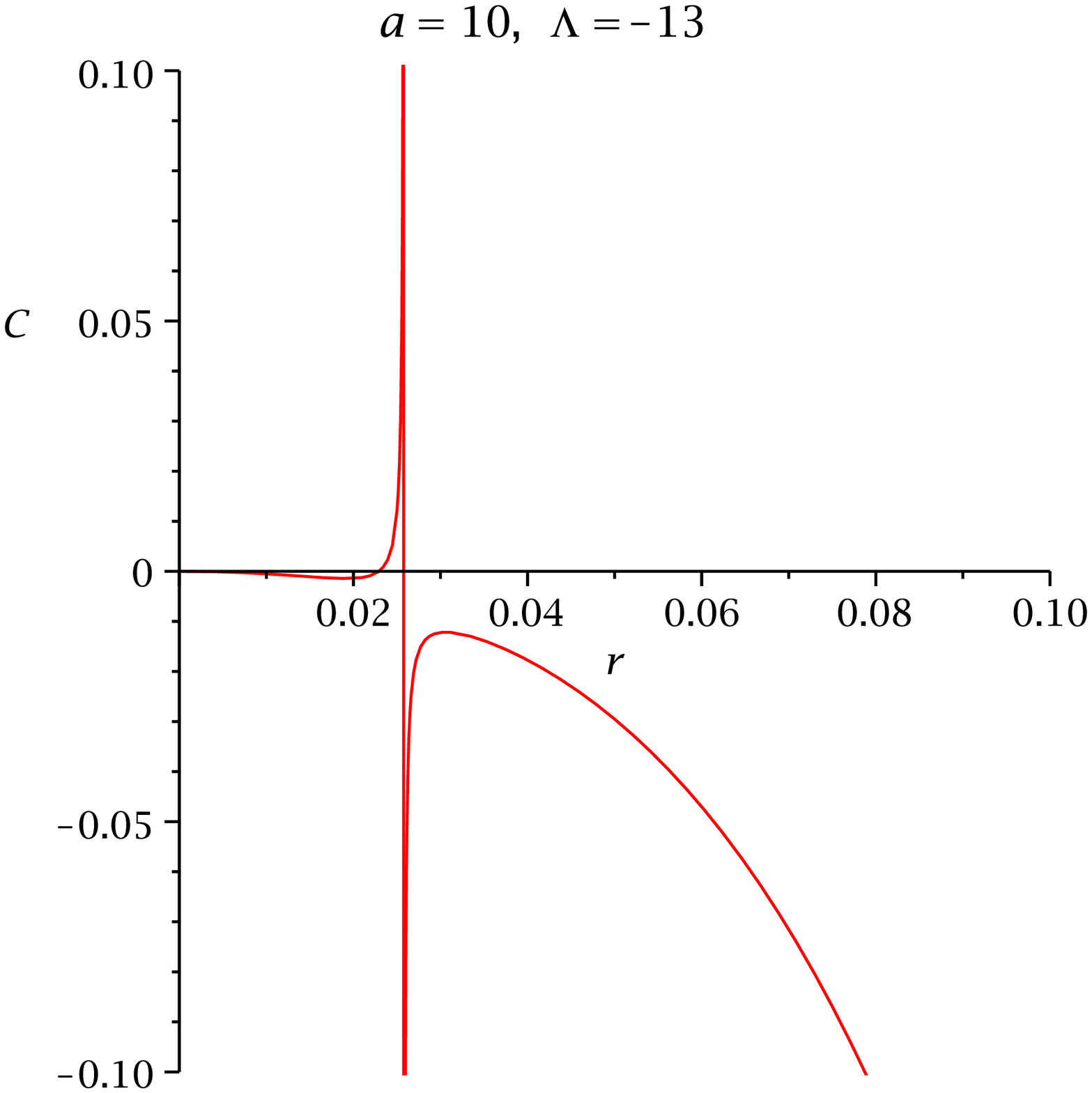}}
\subfigure[ ]{
 \includegraphics[width=2.1in,angle=0]{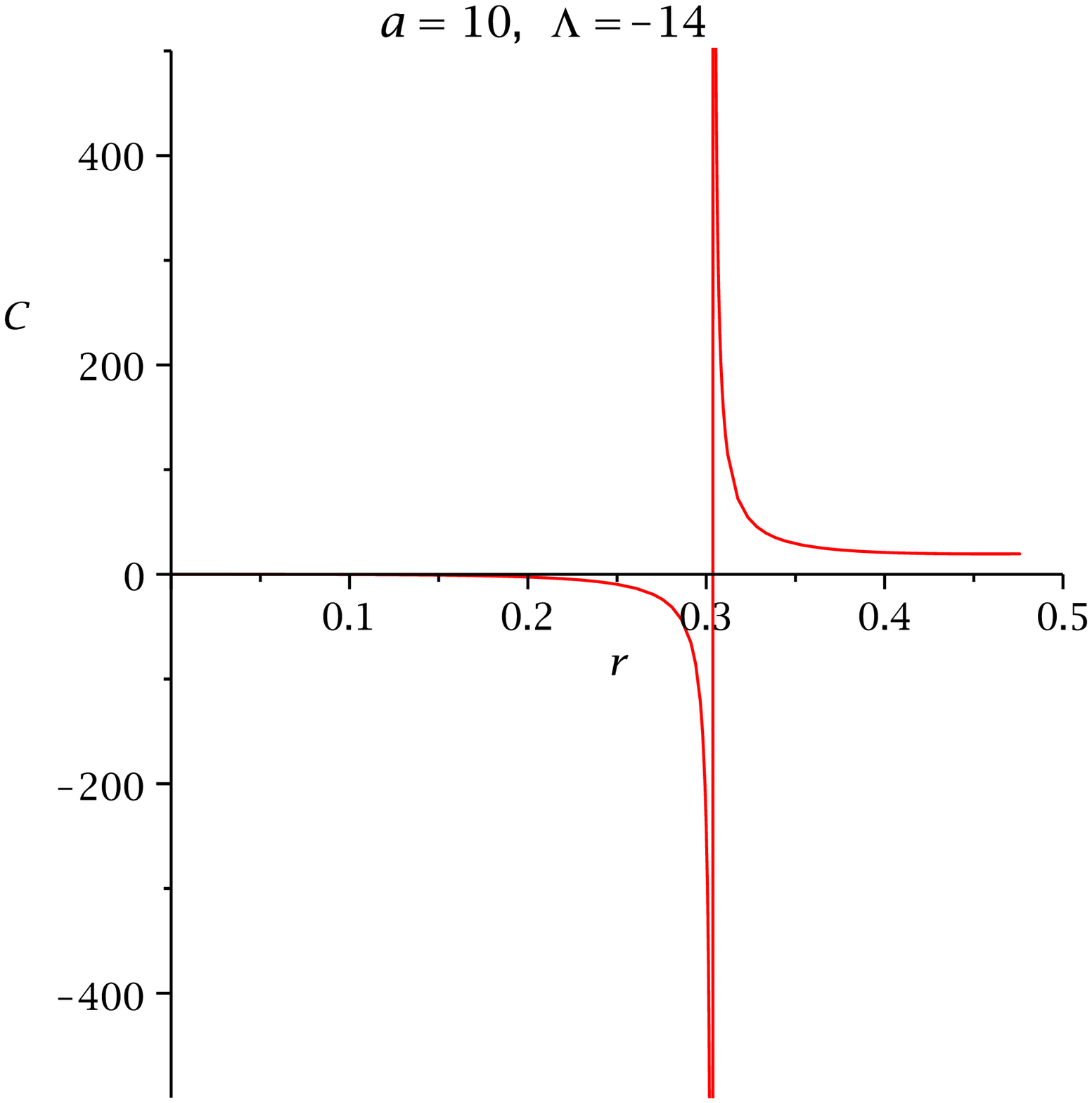}}
 \subfigure[]{
 \includegraphics[width=2.1in,angle=0]{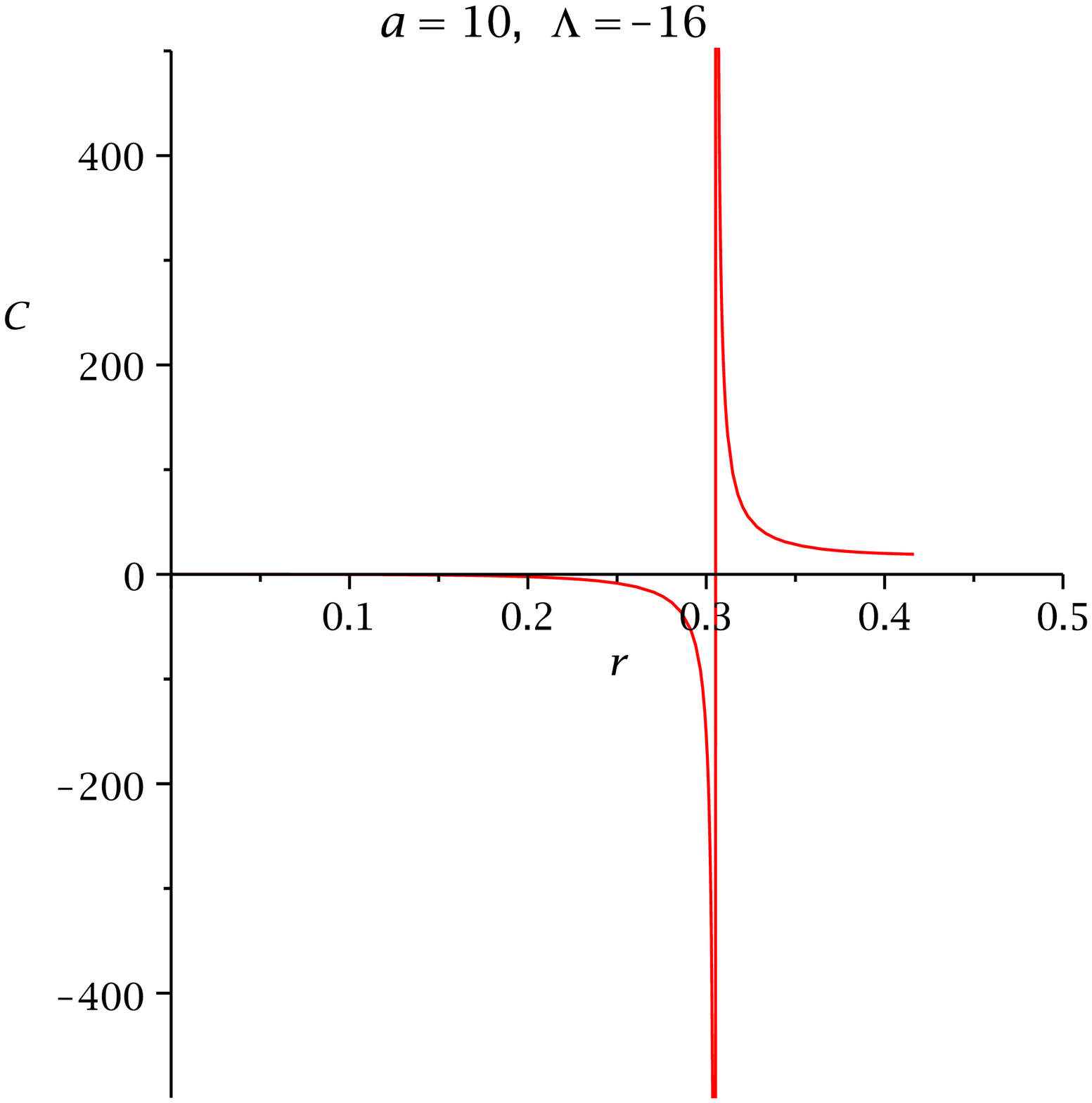}}
 \caption{\label{fg6}\textit{The figure depicts the variation  of $C$  with $r_{+}$ }}
\end{center}
\end{figure}

\begin{figure}[h]
\begin{center}
 \subfigure[]{
 \includegraphics[width=2.1in,angle=0]{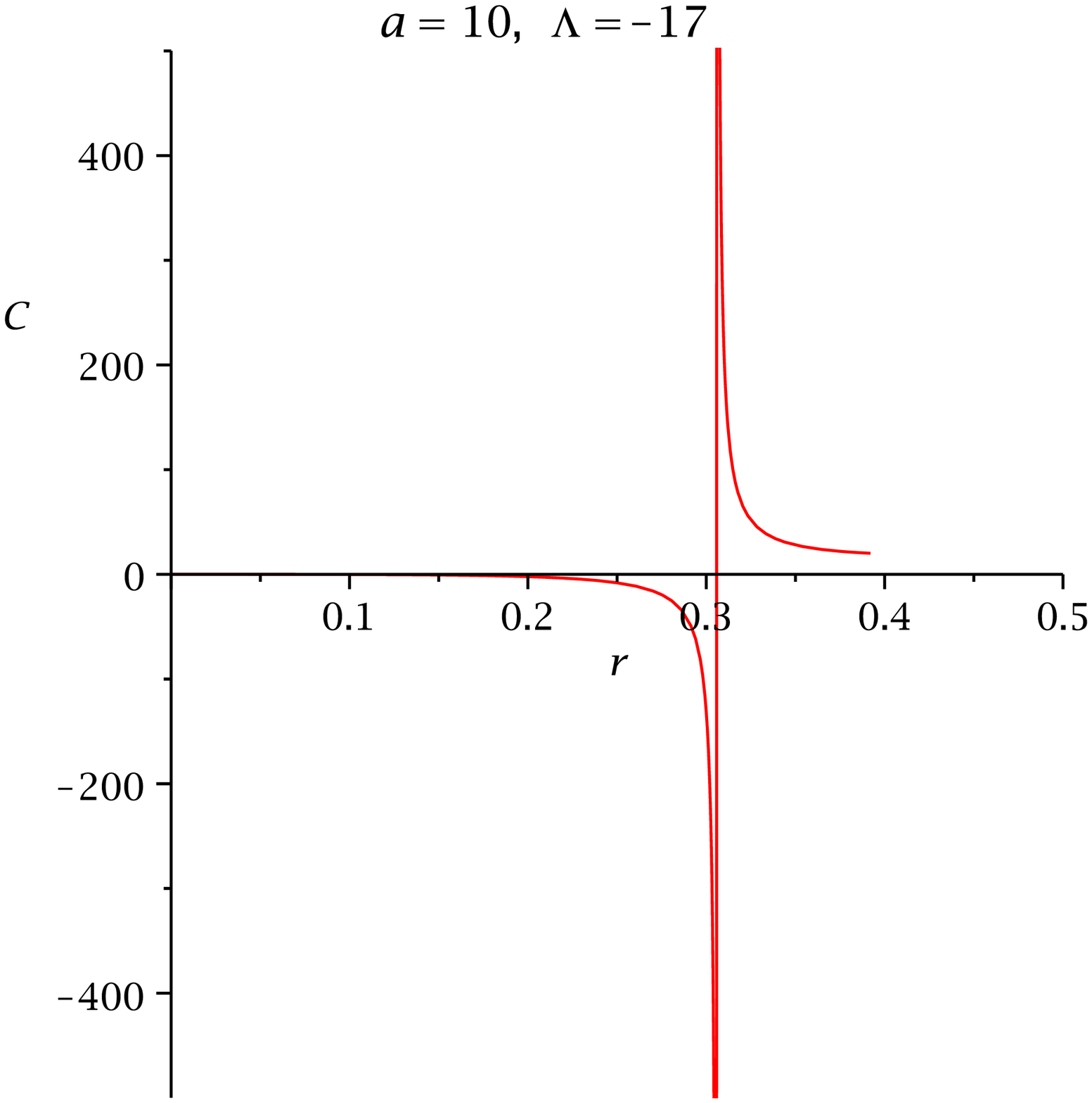}}
 \subfigure[]{
 \includegraphics[width=2.1in,angle=0]{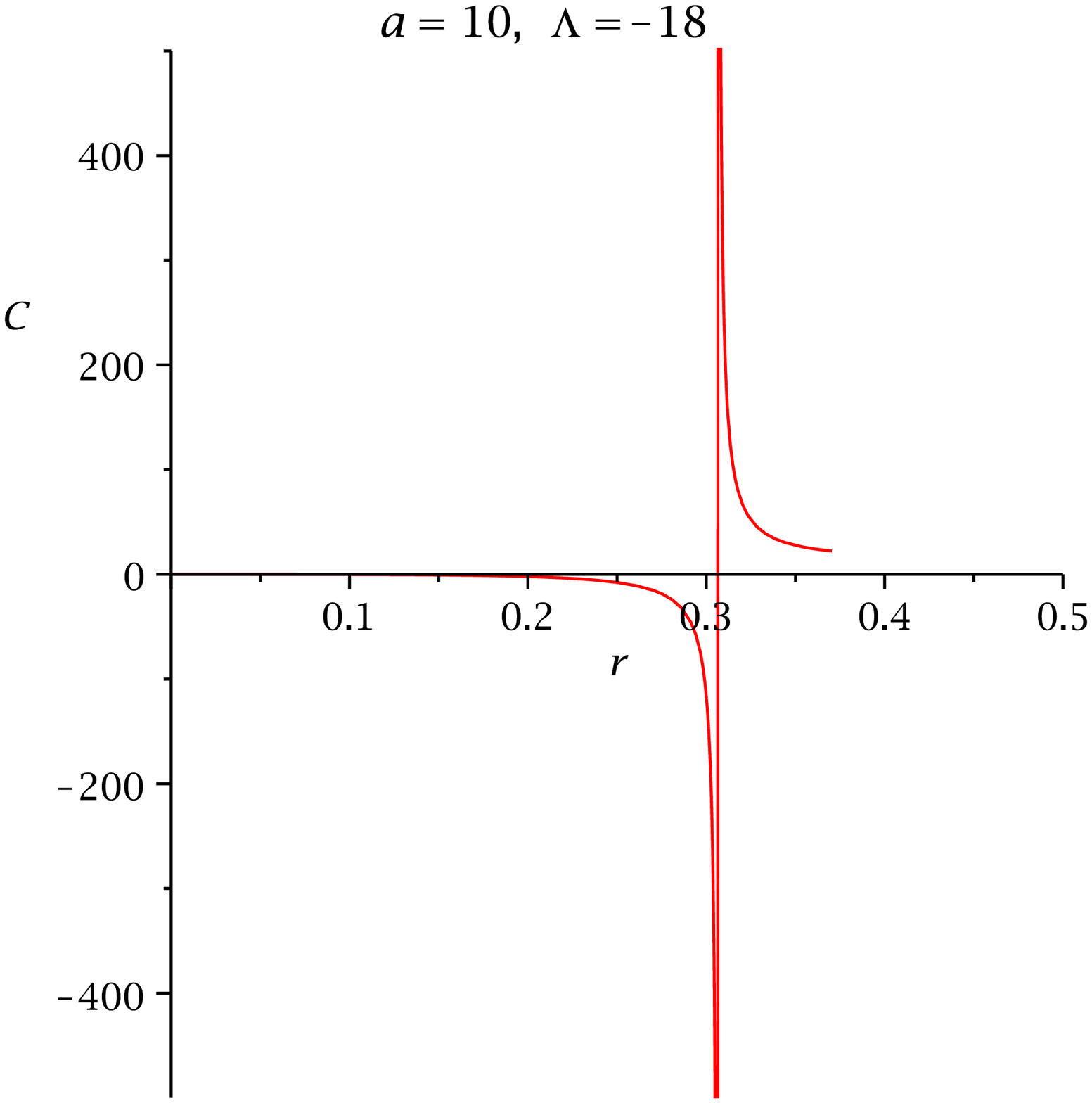}}
 \subfigure[]{
 \includegraphics[width=2.1in,angle=0]{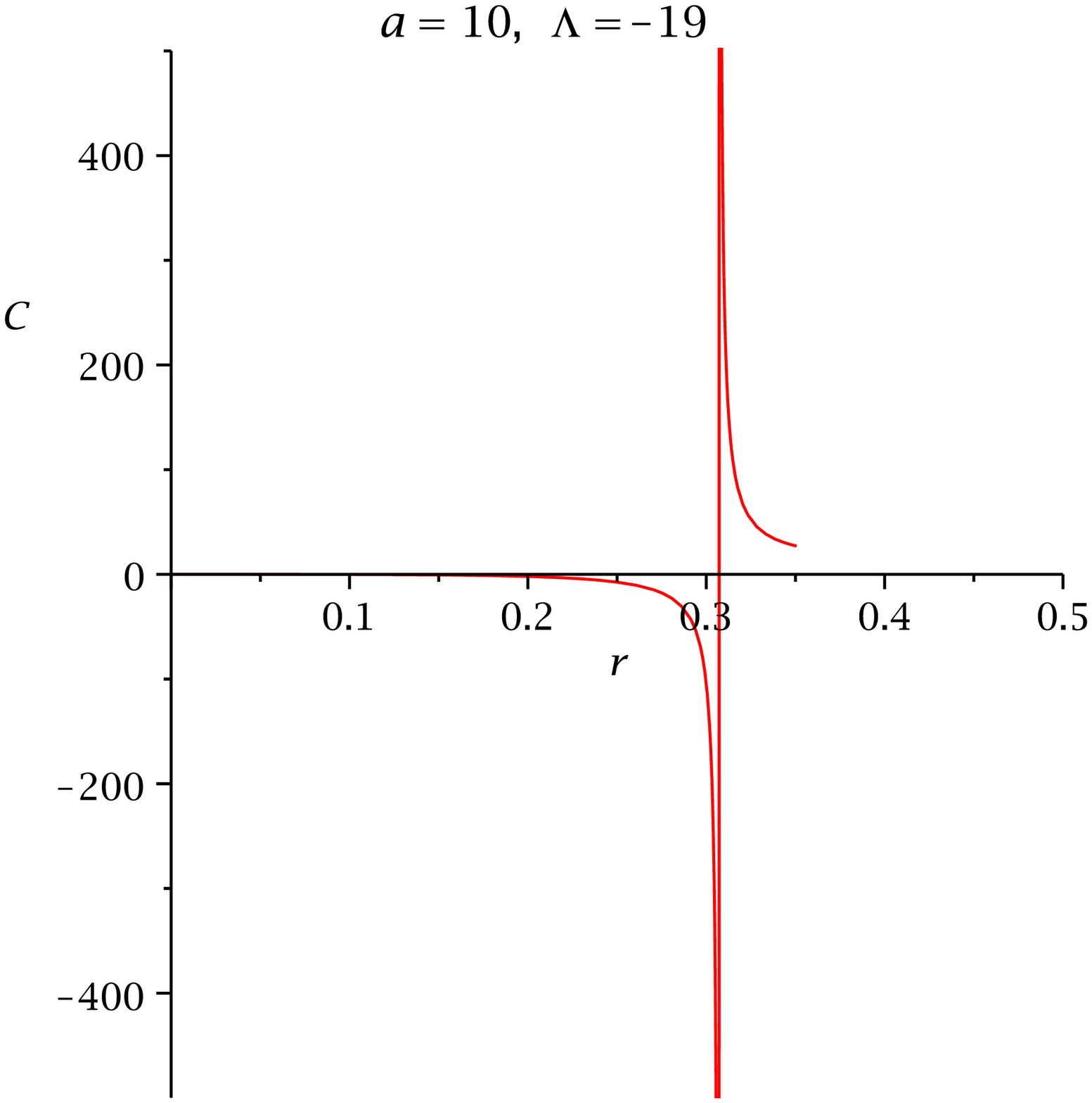}} 
 \subfigure[]{
 \includegraphics[width=2.1in,angle=0]{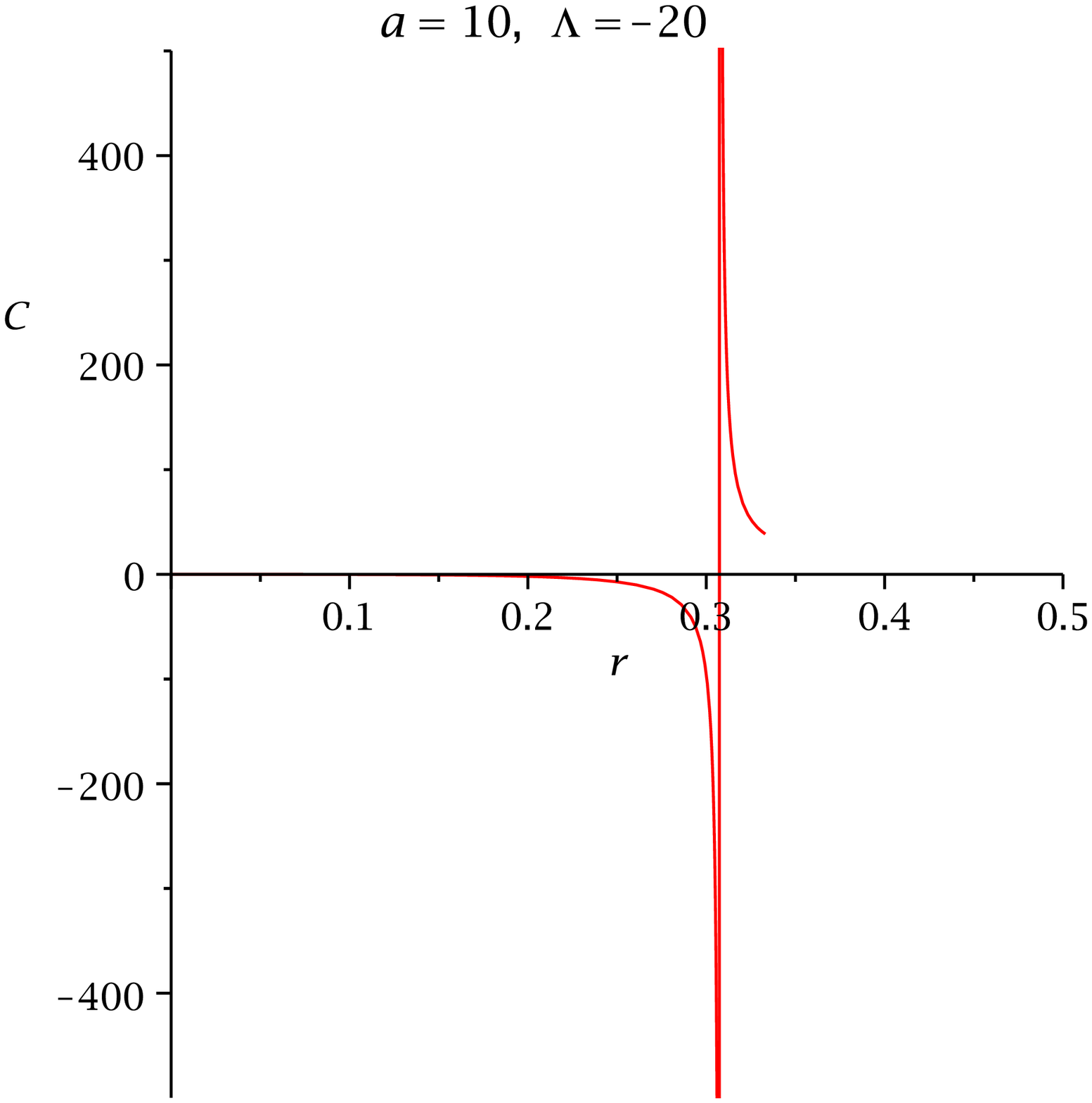}}
 \subfigure[]{
 \includegraphics[width=2.1in,angle=0]{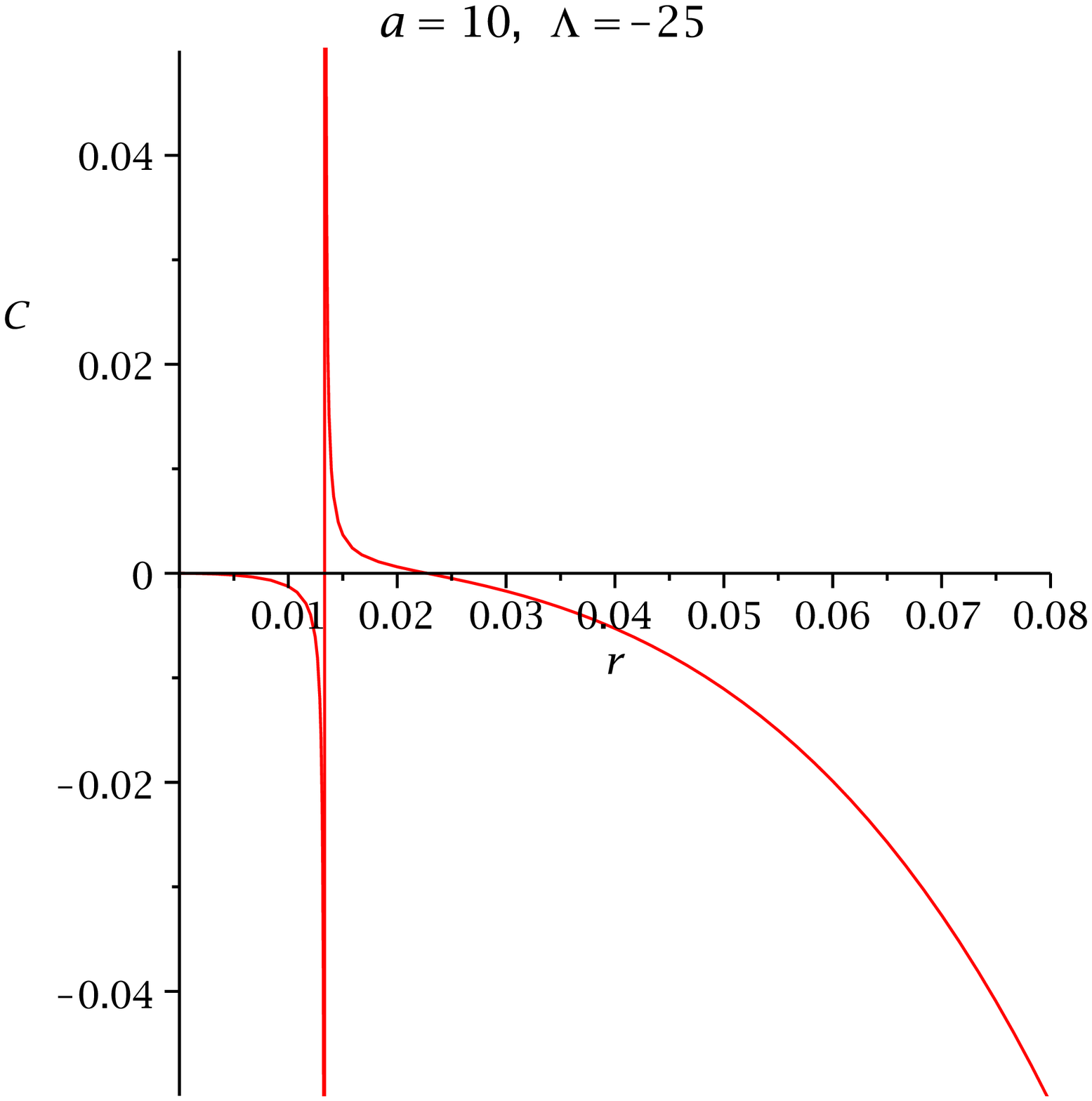}}
 \subfigure[]{
\includegraphics[width=2.1in,angle=0]{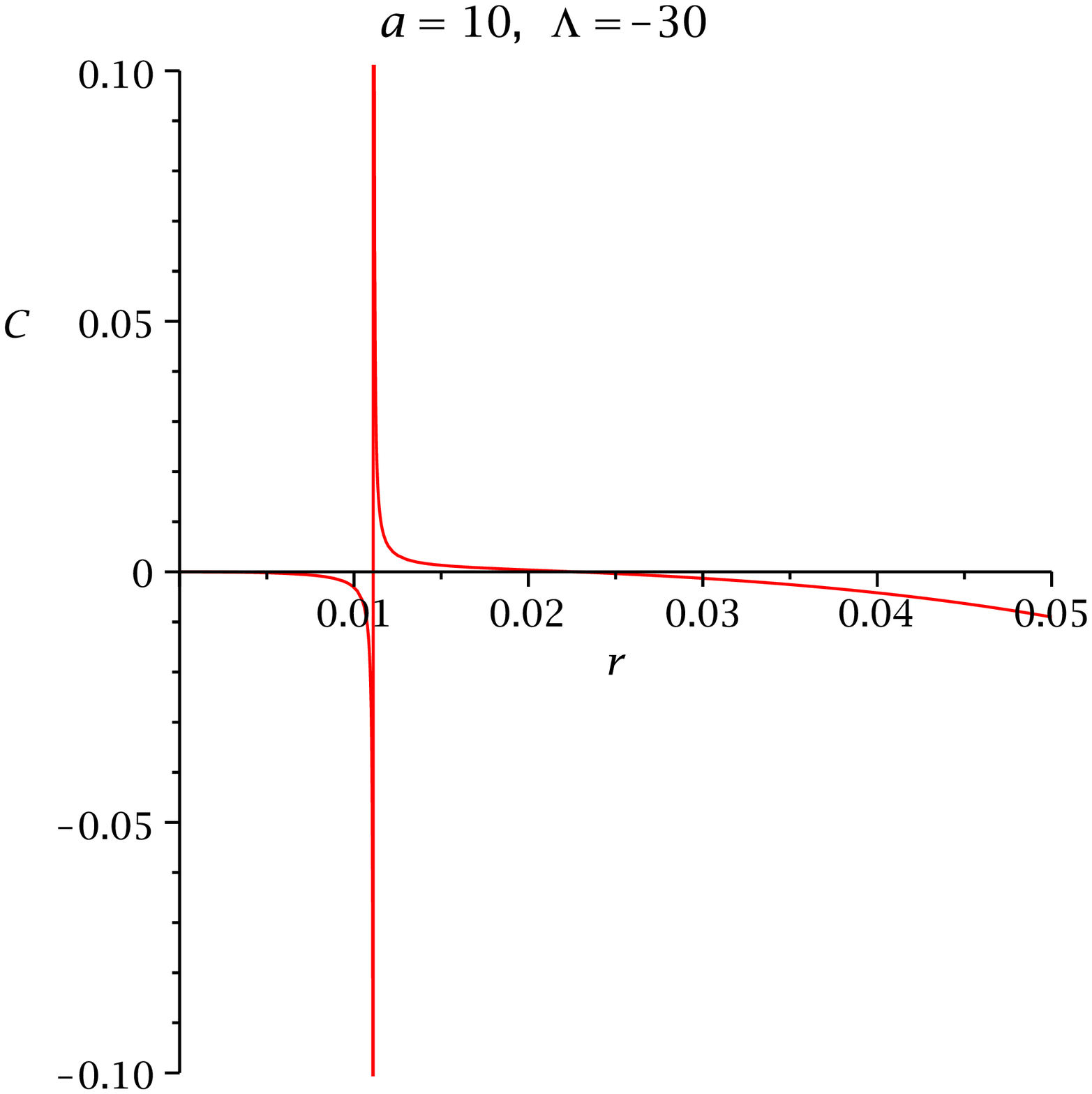}} 
  \caption{\label{g6}\textit{The figure depicts the variation  of $C$  with $r_{+}$ }}
\end{center}
\end{figure}

\begin{figure}[h]
\begin{center}
\subfigure[]{
 \includegraphics[width=2.1in,angle=0]{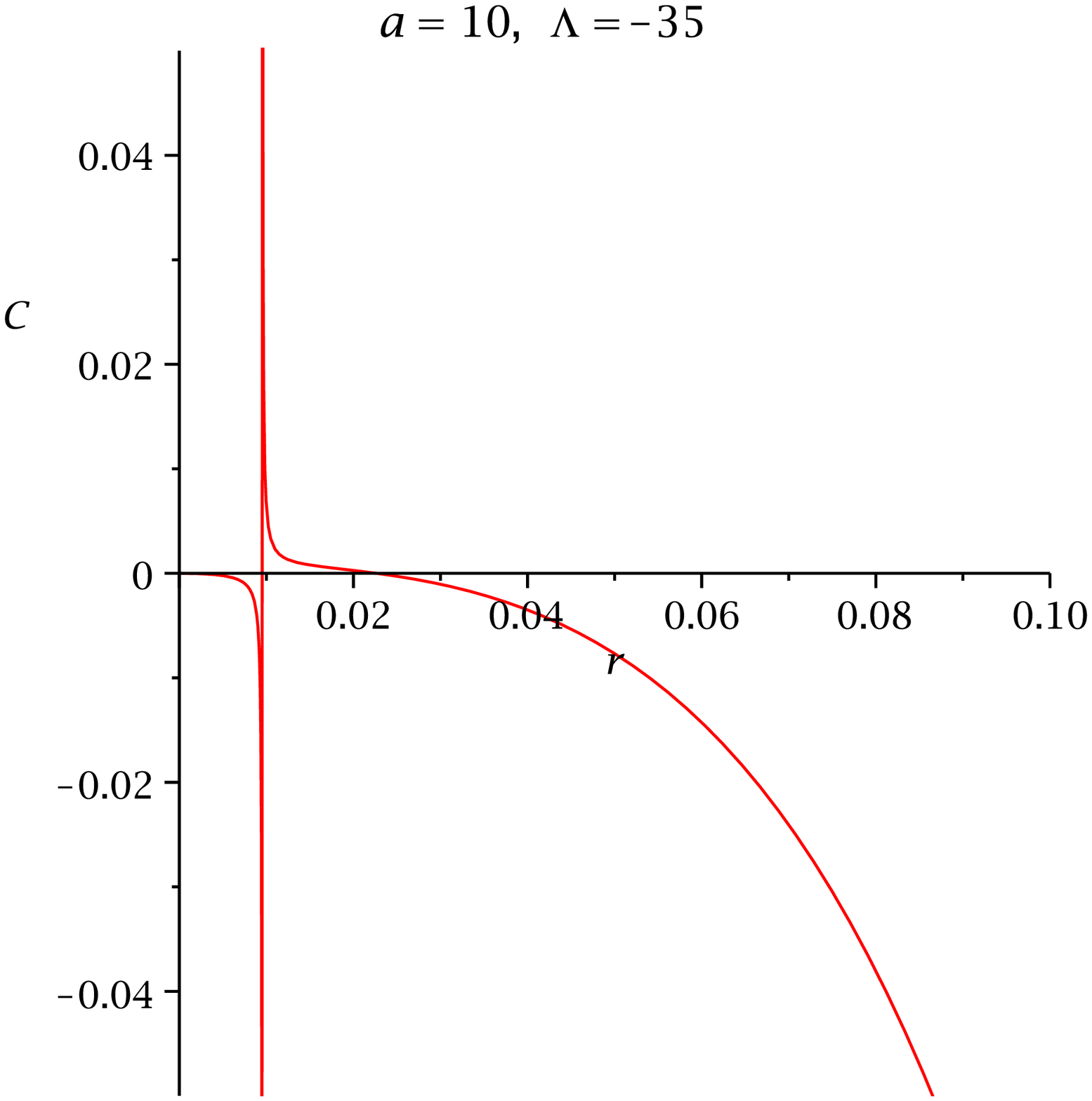}}
 \subfigure[]{
 \includegraphics[width=2.1in,angle=0]{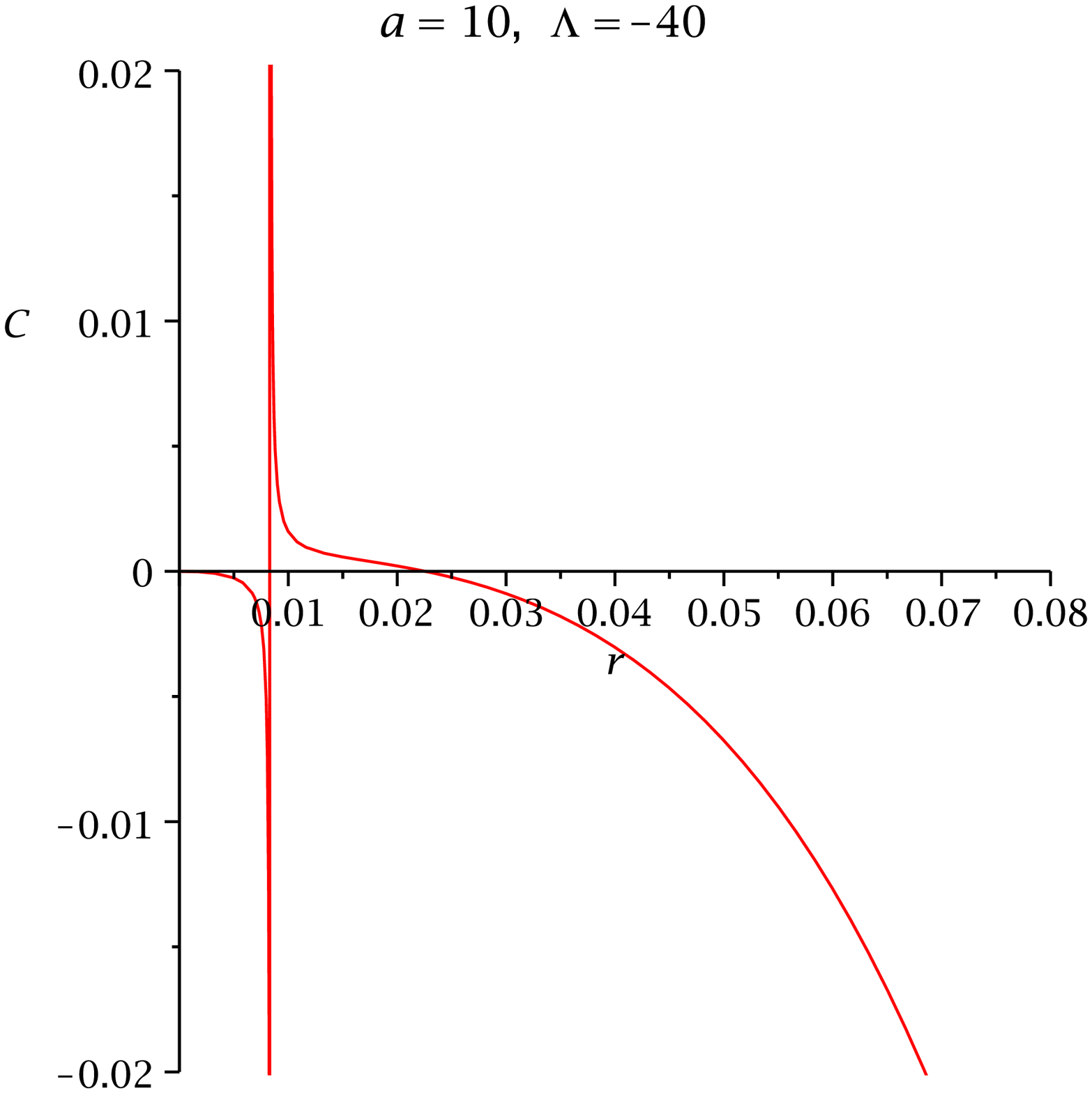}}
 \subfigure[ ]{
 \includegraphics[width=2.1in,angle=0]{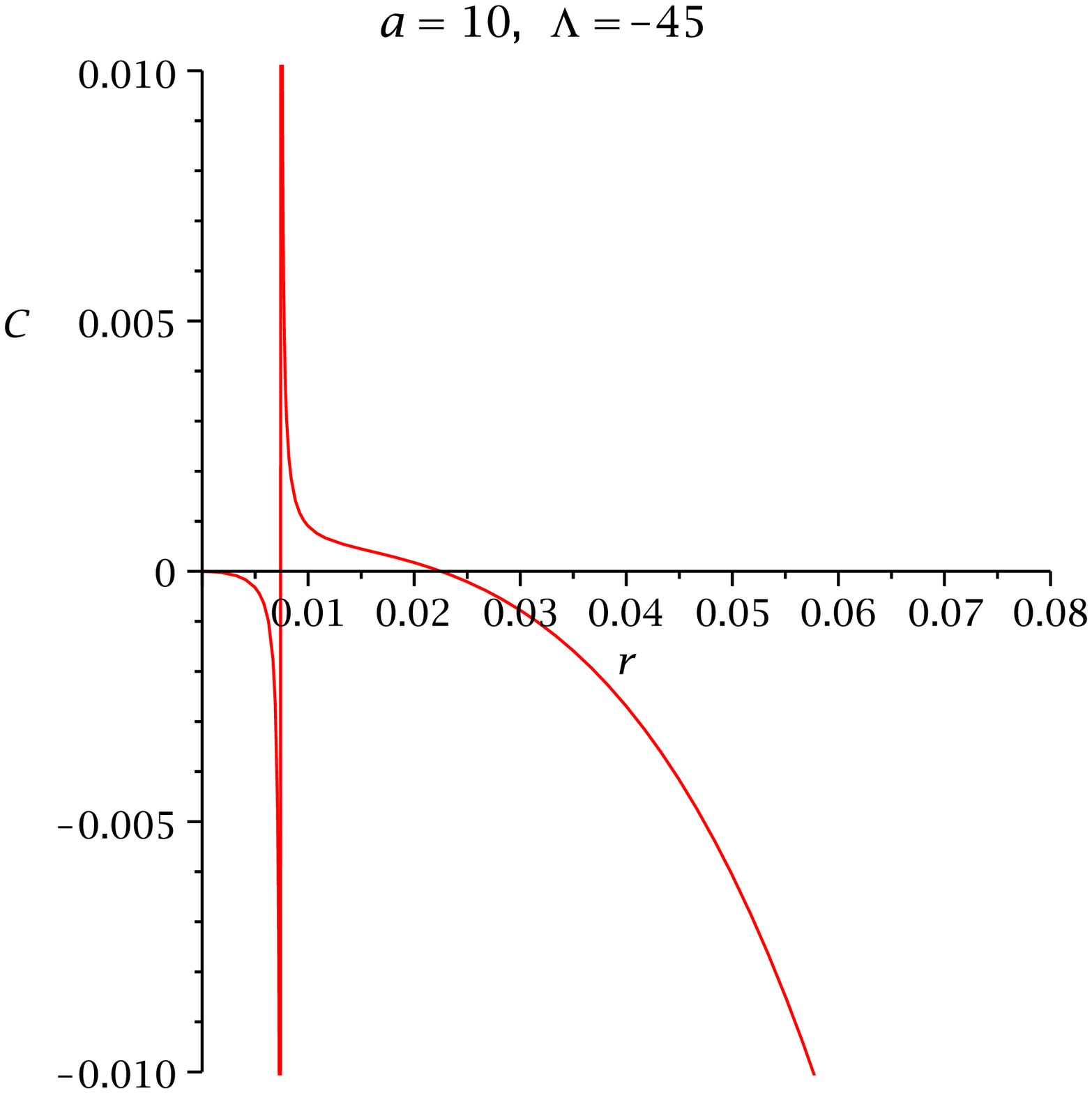}}
 \subfigure[]{
 \includegraphics[width=2.1in,angle=0]{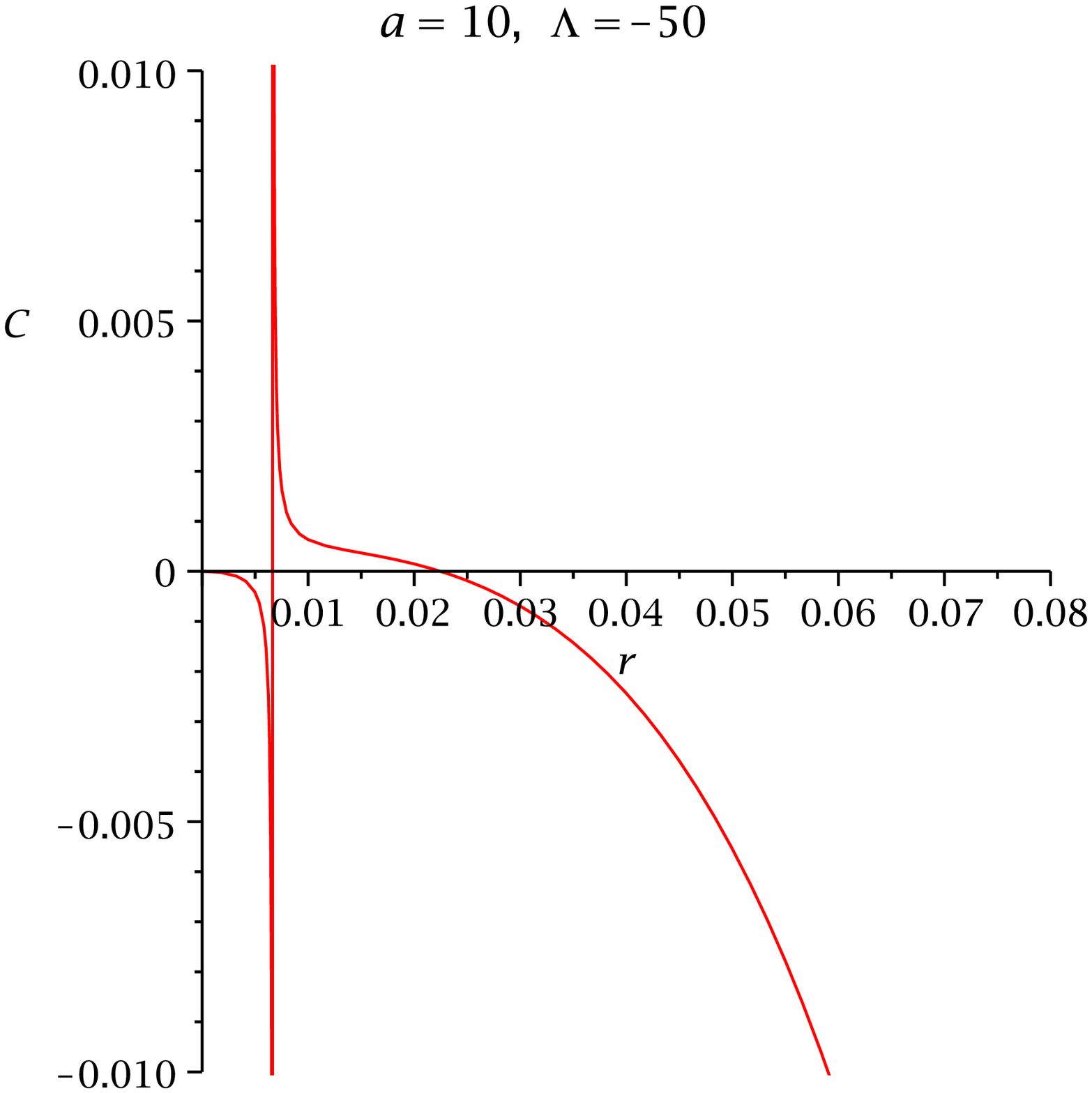}}
 \subfigure[]{
 \includegraphics[width=2.1in,angle=0]{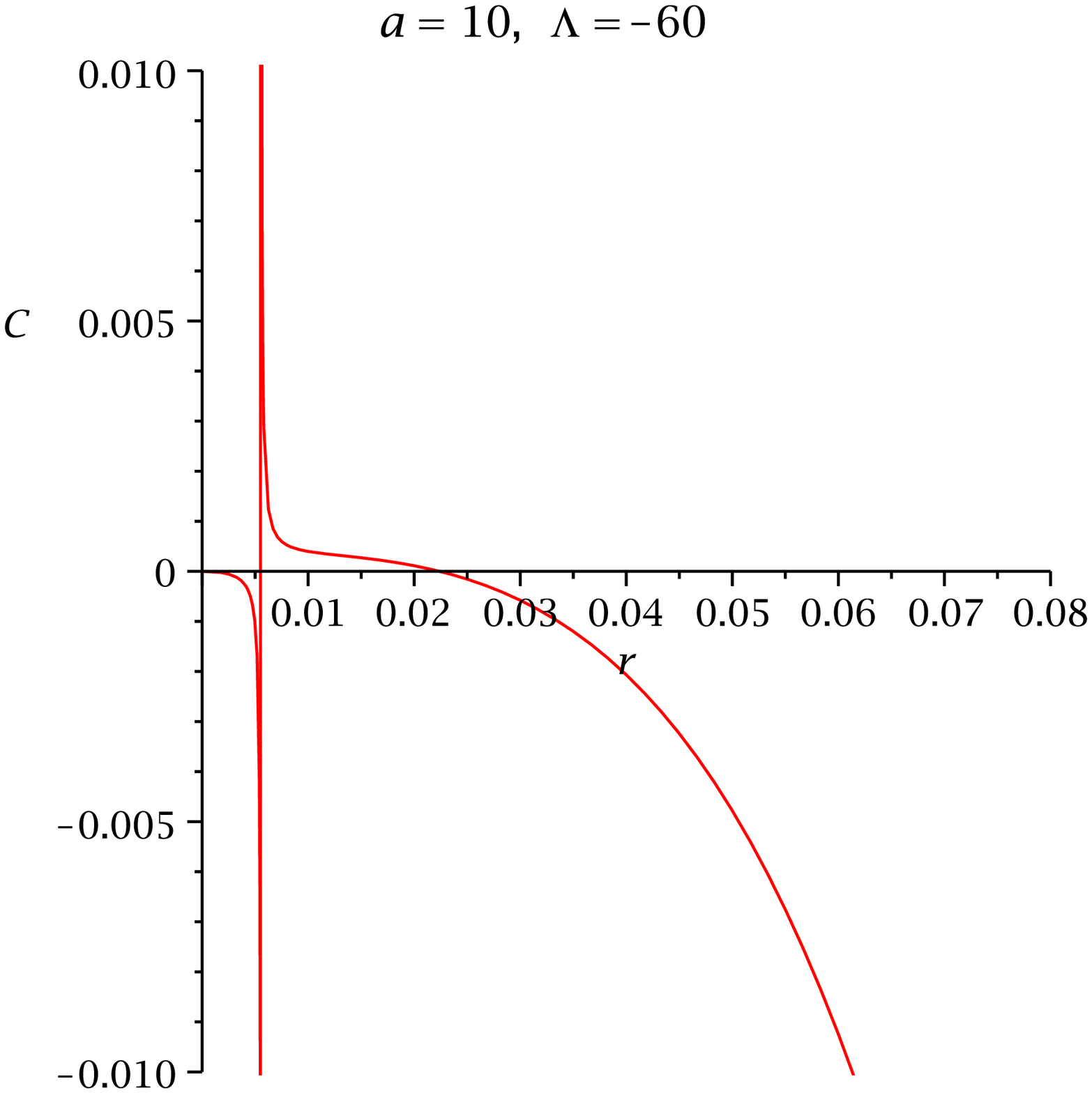}}
 \caption{\label{fg7}\textit{The figure depicts the variation  of $C$  with $r_{+}$ }}
\end{center}
\end{figure}

\end{document}